\documentclass[useAMS,usenatbib]{mn2e}
\usepackage{graphicx}
\usepackage{epsfig}
\newcommand{\HI}{H\,{\sc i}}
\newcommand{\HII}{H\,{\sc ii}}
\newcommand{\SII}{[S\,{\sc ii}]}

\newcommand{\OIII}{[O\,{\sc iii}]}

\newcommand{\NII}{[N\,{\sc ii}]}
\newcommand{\HeII}{He\,{\sc ii}}

\def\p0{\phantom{0}}

\title[New PNe in the Large Magellanic Cloud (III): The PNLF ]{A New Population of Planetary Nebulae Discovered in the Large
Magellanic Cloud (III): The Luminosity Function}
\author[Warren A. Reid and Quentin A. Parker]{Warren A. Reid$^{1}$\thanks{e-Mail:
warren@ics.mq.edu.au; war@aao.gov.au } and Quentin A.
Parker $^{1,}$$^{2}$\thanks{e-Mail: qap@ics.mq.edu.au}\\
$^{1}$Department of Physics, Macquarie University, Sydney, NSW 2109, Australia\\
$^{2}$Anglo-Australian Observatory, PO Box 296, Epping, NSW 1710
Australia}
\begin{document}

\date{}

\pagerange{\pageref{firstpage}--\pageref{lastpage}} \pubyear{2002}

\maketitle

\label{firstpage}

\begin{abstract}
Our previous identification and spectroscopic confirmation of 431 faint, new planetary nebulae in the central 25 deg$^{2}$ region of the LMC permits us to now examine the shape of the LMC Planetary Nebula Luminosity Function (PNLF) through an unprecedented 10 magnitude range. The majority of our newly discovered and previously known PNe were observed using the 2dF, multi-object fibre spectroscopy system on the 3.9-m Anglo-Australian Telescope and the FLAMES multi-object spectrograph on the 8-m VLT. We present reliable \OIII5007\AA~and H$\beta$~flux estimates based on calibrations to well established PN fluxes from previous surveys and spectroscopic standard stars. The bright cutoff (M$^{\ast}$) of the PNLF is found by fitting a truncated exponential curve to the bright end of the PNLF over a 3.4 magnitude range. This cutoff is used to estimate a new distance modulus of 18.46 to the LMC, in close agreement with previous PNLF studies and the best estimates by other indicators. The bright end cutoff is robust to small samples of bright PNe since significantly increased PN samples do not change this fiducial. We then fit a truncated exponential curve directly to the bright end of the function over a 6 magnitude range and test the curve's ability to indicate the position of M$^{\ast}$. Because of the significant increase in the number of LMC PN, the shape of the PNLF is now examined in greater detail than has previously been possible. Newly discovered features include a small increase in the number of PNe over the brightest 4 magnitudes followed by a steep rise over 2 magnitudes, a peak at 6 magnitudes below the bright cutoff and an almost linear drop-off to the faint end. Dips at the bright end of the PNLF are examined in relation to the overall shape of the PNLF and the exponential increase in the number of PNe. Through cumulative functions, the new LMC PNLF is compared to those from the SMC and a new deep local Galactic sample revealing the effects of incompleteness. The new \OIII5007\AA~LMC PNLF is then compared to our new H$\beta$~LMC PNLF using calibrated and measured fluxes for the same objects, revealing the effects of metallicity on the \OIII5007\AA~line.

\end{abstract}


\section{Introduction}

The PN luminosity function (PNLF) describes the number density of
PNe within a given system over any given luminosity range based on a prominent emission line
such as \OIII.
Its most frequent use is as an extragalactic distance indicator where it is the only
standard candle that can be applied to all the large galaxies in the
Local Supercluster (Jacoby, 1980, Ciardullo \& Jacoby et al. 1989, Ciardullo \& Jacoby, 1992, Ciardullo, 2006). In this role, the distance estimate relies on the shape created by the binning of \OIII\,5007\AA~fluxes at the bright end of the PNLF, which appears to be essentially invariant across all galaxies tested thus far, regardless of their morphological type or metallicity
(M\'{e}ndez et al. 1993; Ciardullo et al. 2004). There is a consistent \OIII\,5007\AA~high luminosity cut-off beyond which no
PNe are observed. The bright cut-off point is identified by fitting an exponential curve to the bright end of the PNLF and is believed to be invariant for all PN systems in all galaxies (Stanghellini 1995, Ciardullo 2006).

The concept of using PNe as a cosmological distance indicator was first
suggested in the 1960's (Henize et al. 1963; Hodge 1966). It was slow to gain acceptance because
individual PNe are not standard candles, distances to
local Galactic PNe were very difficult to estimate and extragalactic PNe were severely under-sampled.
The first study using an \OIII\,5007\AA-emission-line based PN luminosity function
as a distance indicator was performed in 1990 by Jacoby et al. (1990). Today, with large numbers of PNe being discovered in external galaxies (Kniazev et al. 2005, Ciardullo 2006, Jacoby 2006) it is
recognised as one of the most important and resilient distance
indicators in extragalactic astronomy (Ciardullo 2006). The PNLF method has been proven to be an accurate standard candle (Jacoby 1989; Ciardullo et al. 1989; Jacoby, Ciardullo,
Ford, 1990; Jacoby, Walker, Ciardullo, 1990b) and appears to be
extremely insensitive to parent stellar population (Ciardullo \&
Jacoby, 1992).

The PNLF has other uses apart from distance determination. Establishing the
consistent shape of the function provides a basis for estimating the number of PNe in any galaxy, given only the number of PNe in the
brightest 3 or 4 magnitudes. A PNLF which comprises a near complete sample of PNe across a galaxy can be used to estimate the
luminosity-to-mass and dynamical age relations (M\'{e}ndez et al. 1993). It has also been suggested as a
valuable tool for studies of the initial-to-final mass relation and
related mass loss processes in stars of early-type galaxies
(M\'{e}ndez et al. 1993). Lastly, the PNLF provides a unique probe into a galaxy's chemical and dynamical evolution. The mass and metallicity of the progenitor star largely determines the maximum \OIII~line luminosity of the PN (Dopita et al. 1992). This makes the shape of the \OIII~PNLF an important diagnostic for galactic chemical evolution.

%The determination of the number of PNe in our Galaxy is frustrated by
%the dust concentration which increases toward the galactic plane.
%The uncertainty of Galactic PN distances has lead to large
%disparities in number estimates for the whole Galaxy. In early work,
%indirect estimates ranged from 5000 (Minkowski 1965, Khromov 1977,
%Alloin, Cruz-Gonzalez and Peimbert 1976) to more than 40,000
%(Cahn and Kaler 1971, Smith 1976, Cahn
%and Wyatt 1976). A new approach was initiated by Ford and Jacoby
%(1978)  which relied on the assumption that there exists a constant
%ratio of PN to galactic luminosity. Estimating the total number of
%PN in M31 to be about 10,000, they computed the total number for our
%Galaxy to be about 5,000. This method however relied on further
%estimates for the number of faint PNe in M31 which were not able to
%be observed.

%A new deep survey within a 2kpc radius of the sun was initiated by Frew (2008) in order to create a PNLF within a constrained Galactic volume. The resulting function included 180 previously known and newly discovered PNe and spanned a range of 10 magnitudes. Adopting a column density of $\mu$ = 23.4 $\pm$ 2.3 kpc$^{-2}$ led to a total estimate of 19,800 $\pm$ 4000 PNe in the Galactic disk (Frew 2008).

%PNe were first proposed as standard candles by Henize & Westerlund (1963) who postulated that there is a maximum brightness for all PNe.

The first luminosity function for PNe in the LMC and SMC was
constructed by Jacoby (1980). The advantage of using PNe in these
local, external galaxies to create a luminosity function was clear since the
distance to all the PNe was essentially equal, faint PNe could still
be identified and each galactic system could be studied in its
entirety. The majority of bright PNe in each galaxy had already been identified so the bright end of the PNLF was able to be modeled and successfully used as a distance indicator. However, with a faint cutoff only 6 magnitudes below the brightest, the overall shape of the PNLF, especially at the faint end, was still unknown.

%In the LMC, previously known PNe were limited to the brightest 6 magnitudes, meaning that the overall shape of the PNLF was not known.

The overall shape of the PNLF has previously been difficult to determine for any galaxy due to sample incompleteness at the faint end. With 431 medium to faint PNe together with 162 previously identified PNe now uncovered within the central 25 deg$^{2}$ of the LMC (Reid \& Parker, 2006b) [RPb] hereafter, we can now update the LMC PNLF and provide measured estimates to compare with previous theoretical simulations and other observed PNLFs. The RP PNe were discovered using an AAO/UKST H$\alpha$ (+ effectively \NII) 40 field mini-survey of the entire LMC, SMC and surrounding regions. The details are provided in Reid \& Parker (2006a) hence [RPa] and will not be repeated here. For details regarding the overall H$\alpha$ survey, see Parker et al. (2005) and for the candidate selection technique and followup spectroscopic object confirmations, please see [RPb].

The new RP sample will enable both the shape and structure of the LMC PNLF to be analysed in detail for the first time over a much wider 10 magnitude range. The distance has always been estimated using an empirical (or theoretical) luminosity function and fitting this curve to the data using a statistically robust technique such as $\chi$$^{2}$ or the maximum likelihood method to give the appropriate bright-end position. The bright-end intersection of this curve with the magnitude plane gives the position of M$^{\ast}$ (the brightest possible PN in the system) and the distance modulus. In this work, we use a cumulative function to determine the position of M$^{\ast}$. We also fit the truncated exponential curve directly to the very complete bright end of the PNLF over a 6 magnitude range to test it's ability to depict PN evolution and indicate distance. Our ability to identify PNe to magnitudes as faint as 25 in \OIII, 10 magnitudes below M$^{\ast}$, gives us confidence that we are largely complete at the bright end of the PNLF. It also largely does away with the need to simulate and extrapolate the PNLF to account for unobserved, faint PNe with (M$^{\ast}$--M) $>$5.

A brief description of the PN spectra used in this study is provided in section~\ref{section2} along with the data reduction procedure. The new method of flux calibration for fibre spectra together with the de-reddening procedure is outlined in section~\ref{section3}. In section~\ref{section4} we present the new \OIII~PNLF for the LMC using 574 bright to faint PNe and compare the shape to the previous best LMC PNLF from Jacoby et al. (1990), and the empirical predictions of Ciadullo et al. (1989) and M\'{e}ndez et al. (1993). In section~\ref{section5} the PNLF is constructed using the H$\beta$ line, a good predictor of central star temperature, and compared to our new determination of the traditional \OIII5007\AA~based PNLF.

\label{section 1}

%%%%%%%%%%%%%%%%%%%%%%%%%%%%%%%%%%%%%%%%%%%%%%%%%%%%%%%%%%%%%%%%%%%%%%%%%%%%%%%%%%%%%%%%%%%%%%%%%%%%%%%%%%%%%%%%%%%%%%%%%%%%%

\section{2dF and FLAMES observations}
\label{section2}

\subsection{Brief description}

A five night observing run on the AAT using 2dF (Lewis et al.
2002) was undertaken in December 2004 to spectroscopically confirm LMC emission candidates as PNe and to eliminate contaminants such as \HII~regions and emission line stars. 2dF was an ideal choice of
instrument for the spectroscopic followup of large numbers of
candidate emission objects due to its unique ability to simultaneously observe
400 targets (including objects, fiducial stars and sky positions) \textbf{with 2 arcsec fibres} over a wide 2 degree diameter field area. The large corrector lens incorporates an atmospheric dispersion compensator, which is essential for wide wavelength coverage using small diameter fibres.

The observations provided $\sim$4,000 spectra. Individual exposure times were mostly 1200s using the 300B grating
with a central wavelength of 5852\AA~and wavelength range
3600-8000\AA~at a dispersion of 4.30\AA/pixel. These low-resolution
observations, at 9.0\AA~FWHM, were used as the primary means of
object classification and provided the bright \OIII5007 fluxes for this study. All fields were then re-observed using the 1200R grating for kinematic studies [RPb] and improved resolution of the \SII~6716,6731\AA~lines for determination of electron densities.

The 2dF raw data were processed using the AAO 2dF data reduction
system, 2dFDR\footnote[1]{http://www.aao.gov.au/AAO/2df/software.html\#2dfdr}. This software
can probe the multifibre spectra creating the
necessary calibration groups (eg. BIAS, DARK, FLAT, ARC
etc). As a calibration exposure is reduced it is inserted into
the appropriate group.

The software was instructed to perform a subtraction of background
scattered light prior to the extraction. The background is
determined by fitting a function through the `dead' or unused fibres in the
image.

A bias frame was obtained for each observed field. The mean of the
bias frame has the bias strip removed. This strip is subtracted from
the data and the bias strip is trimmed from the data frame. A
variance array is then derived from the data with values determined
from photon statistics and readout noise. The resulting frame was
then used throughout the reduction process.

The data reduction system performs a wavelength calibration using
the information from the spectrograph optical model. This is then
refined using data from the arc lamp exposure. The lines in the arc
lamp exposure are matched against a line list. A cubic fit for each
fibre is then performed to the predicted and measured wavelengths of
all lines which are non-blended, not too wide and not too weak. This
fit is then used to further refine the wavelengths.

To perform the sky subtraction, the data is first corrected for the
relative fibre throughput, based on a throughput map derived from
the dedicated sky fibres.  The relative intensities of the skylines
in the object data frame are used to determine the relative fibre
throughput. This method saves time, as no offset­sky observations
are required. The median
sky spectrum was calculated from the median of all the sky spectra
normalized by their mean flux. Each of these spectra are continuum
subtracted, using a boxcar median smoothing with a 201 pixel box to
define the continuum. A robust least­squares fit of the counts was
performed in the differential or continuum subtracted data fibres
versus the counts in the differential or continuum subtracted median
sky. Assuming that the sky was the dominant source of emission, the
slope of this fit gives the fibre throughput. The robust fit is
especially important when dealing with faint objects. The sky fibre
spectra in the data were then combined and subtracted from each
fibre. Cosmic rays were
rejected automatically during the process of combining two or more
observations.

Our data includes spectroscopic observations of a sub-sample of 21 PNe in dense regions of the LMC main
bar, undertaken using the multi-object fibre spectroscopic
system, FLAMES (Pasquini 2002) on the VLT UT2 over three nights in
December 2004. Gratings LR2 and
LR3 allowed us to cover the most important optical diagnostic lines
for PNe in the blue including \OIII\,4363\AA, \HeII\,4686\AA,
H$\beta$ and \OIII\,4959 \& 5007\AA. Grating LR6 covered the
H$\alpha$, \NII\,6548 and 6583\AA~lines as well as the \SII\,6716 \&
6731\AA~lines for electron densities. Details are provided in [RPa] and [RPb] but we will produce a separate paper examining all 420 objects (including PNe, \HII~regions, and WR stars) observed using FLAMES.

The VLT FLAMES spectra were reduced using {\scriptsize IRAF} tasks
{\scriptsize IMRED, SPECRED} and {\scriptsize CCDRED} for multi-spec
files. Cosmic rays were rejected when combining frames. Using the
weighted intensity of the continuum, the {\scriptsize IRAF}
{\scriptsize SCOMBINE} task was then used to combine the three
different wavelength portions of the spectrum into one spectral
image.

\label{section2.1}

\subsection{Continuing Object followup}

\begin{table*}
\begin{center}
\caption{A list of 26 RP objects which have now been re-classified or had their status changed following new high resolution, higher S/N spectral observations and multi-wavelength analysis.  ELS = Emission-line star; PN = Planetary Nebula; SNR = Supernova Remnant; Prob. refers to classification probability where T = True; L = Likely; P = Possible. Under the heading `Reason', (1) refers to analysis using a 2.3m telescope optical spectrum, (2) refers to analysis using follow up observations on 2dF, `IR' refers to false infrared colours at 3.6$\mu$m and 4.5$\mu$m 5.8$\mu$m and 8$\mu$m indicating \HII~regions or hot stars, `Radio' strong radio source ($>$3mJy). Please see the text (section~\ref{section2.2}) for more details.}
\begin{tabular}{llllclcl}
\hline\hline \noalign{\smallskip}
RP cat. & RA $^{(h~m~s)}$ &  DEC $^{(\circ~_{'}~_{''})}$ & Previous & Previous & New    & New & Reason  \\
     No.   &  J2000   &  J2000     &   ID    &   Prob.          &  ID   &   Prob.  &       \\
  \hline\noalign{\smallskip}
 5	&	05 40 28.18	&	-70 56 07.1&	PN	&	T	&	\HII~region	&	T	& weak \OIII5007 (1)	\\
105	&	05 40 45.26	&	-70 28 06.7&	PN	&	T	&	\HII~region	&	T	& Large angular size, Radio, IR 	 \\
148	&	05 37 17.35	&	-70 07 49.3&	PN	&	L	&	SNR	&	L	& large size, strong \SII~lines	(1)\\
352 & 05 37 14.33 &  -66 26 54.61&   PN   &  P   &   ELS     &    L    & Low \OIII5007/H$\beta$ (1)  \\
490 &  05 37 31.73  &  -71 10 48.8&   PN  &   P   &   symbiotic   &   T   & continuum peaking at 7500\AA (1)\\
641	&	05 37 06.38	&	-69 47 17.3&	PN	&	P	&	\HII~region	&	T	& \OIII5007/H$\beta$ = 60\%	(1)\\
667  &   05 30 26.20  &  -70 15 01.5&  PN  &   P   &    ELS     &    P    &Strong continuum (2)  \\
782	&	05 28 18.80	&	-69 28 15.3&		&		&	PN	&	L	& Good PN lines, morphology	(2)\\
798  & 05 26 13.74  &  -69 25 45.1&  PN  &  L   &   ELS   &  L   & Strong continuum  (1) \\
841 & 05 28 08.09  &  -69 10 21.9&   PN  &   P   &   ELS  &  L   & Strong continuum  (1) \\
856	&	05 24 38.82	&	-69 04 13.7&	PN	&	T	&	PN	&	P	& \OIII5007/H$\beta$ =160\%, crowded environ. (2)	 \\
872	&	05 24 25.18	&	-69 39 06.3&	PN	&	L	&	\HII~region	&	T	& Strong continuum	(1)\\
993	&	05 30 54.58	&	-68 34 22.4&	PN	&	L	&	\HII~region	&	T 	&Strong radio source, IR 	 \\
1079 &  05 30 33.09  &  -66 57 41.5 &  PN  &   L   &    ELS   &  L   & Strong continuum (1)\\
1113	&	05 22 49.80	&	-66 40 55.5&	PN	&	P	&	\HII~region	&	T	&\OIII5007 = H$\beta$ (2)	\\
1192  &  05 19 56.77 &   -70 39 03.6&  PN     &   T  &   symbiotic  &  L  & Red continuum (1)\\
1434  &   05 20 16.84 &   -68 45 10.1 &    PN  &   L   &   symbiotic  &  L  & Red continuum  (1)\\
1495	&	05 19 06.88	&	-68 21 34.4&	PN	&	P	&	\HII~region	&	T	& Radio	\\
1534	&	05 21 29.65	&	-67 51 06.6&	PN	&	T	&	PN + \HII	&	L	&Radio, small PN 3" SE  \\
1541	&	05 21 22.00	&	-67 47 28.6&	PN	&	L	&	\HII~region	&	P	& Radio, IR, \OIII5007/H$\beta$ = 60\% 	 \\
1691	&	05 00 32.23	&	-70 00 49.0&	PN	&	P	&	ELS	&	T	& Mostly H emission  (2)	\\
1716	&	04 54 24.49	&	-69 29 42.5&	PN	&	P	&	\HII~region	&	T	& Large size (2)	\\
1760    &  05 03 51.47 &   -68 57 23.7&   PN  &   L   &   ELS  &   L    & Strong continuum (1) \\
1783	&	04 54 33.78	&	-69 20 35.7&	PN	&	P	&	ELS	&	T	& Low \OIII5007/H$\beta$ (2)	\\
1933	&	05 04 47.34	&	-66 40 30.3&	PN	&	P	&	\HII~region	&	T	& Strong radio source (2)	\\
2194	&	05 19 18.44	&	-69 47 17.2&	PN	&	P	&	\HII~region	&	T	& IR  	\\
\hline\noalign{\smallskip}
  \end{tabular}\label{table 1}
   \end{center}
  \end{table*}

Every PN in the LMC sample, both previously known and new, has been re-analysed in order to identify and exclude interlopers which may effect the PNLF. This is particularly important at the bright end for distance determination. H$\alpha$ and off-band red images were examined alongside high and low resolution optical spectra. In addition we have searched the SAGE (Meixner et al. 2006), Spitzer maps and overlaid false colours at 3.6, 4.5, 5.8 and 8$\mu$m to assist verification. Finally, with mosaic radio maps of combined LMC data from ATCA and Parkes we were able to re-classify 3 ultra-bright `True' PNe as contaminants due to their strong radio fluxes $>$3mJy (see Filipovic et al. 2009). This is despite optical spectra which would otherwise strongly indicate a PN. These compact emission objects are now classified as \HII~regions. For a full list of re-classified objects, please see Table~\ref{table 1}.

In order to provide independent checks of our 2dF data and create the most accurate LMC PN sample possible, we have undertaken a deep multi-wavelength study of PN candidates selected from our original discovery list that exhibited unusual characteristics and/or were deemed suspicious. In particular, due to the possibility of minor positioning errors, we have made 78 longslit observations of previous `possible' and `likely' PNe candidates in the RP sample using the MSSSO 2.3m telescope in order to firm-up their status. Using
the double-beam spectrograph (DBS) on this telescope, the visible range (3200-9000\AA), is split by a dichroic at around 6000\AA~and fed into two spectrographs, with red and blue optimized detectors.
 We used the 300B and 316R gratings
to obtain a resolution of 4\AA~for each arm of the DBS. Reduction and extraction of 2.3m spectra were performed
using the standard {\scriptsize IRAF} tasks {\scriptsize IMRED,
SPECRED}, {\scriptsize CCDRED} and {\scriptsize FIGARO}'s task
{\scriptsize BCLEAN}. One dimensional spectra were created and the background sky was subtracted. Final flux calibration used the standard stars, LTT7987, LTT9239, LTT2415 and LTT9491. Results are given in Table~\ref{table 1}, where 8 `likely' and 12 `possible' PNe have been re-classified.

\label{section2.2}

%%%%%%%%%%%%%%%%%%%%%%%%%%%%%%%%%%%%%%%%%%%%%%%%%%%%%%%%%%%%%%%%%%%%%%%%%%%%%%%%%%%%%%%%%%%%%%%%%%%%%%%%%%%%%%%%%

\section{Flux Calibration and De-reddening of the 2dF Fibre Spectra}

\begin{figure}
\begin{center}
  \includegraphics[width=0.48\textwidth]{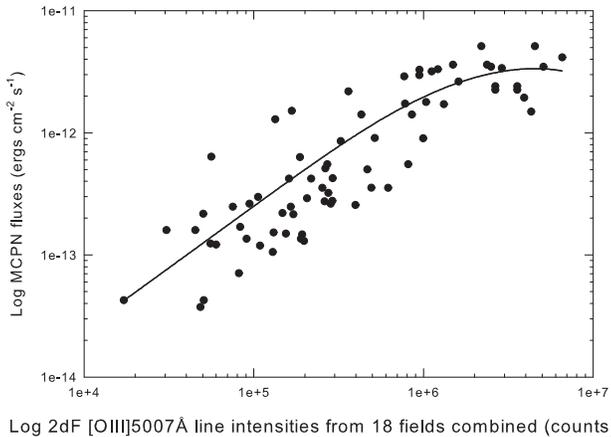}\\
  \caption{The uncorrected MCPN \OIII\,5007\AA~fluxes for 65 PNe are plotted against the \OIII\,5007\AA~2dF 300B PN line intensities from all field exposures. General agreement is evident through an essentially log-linear relation with a turn-over at the bright end but the scatter is too wide ($\sim$1 dex) to provide a reliable flux calibration. The line represents a least squares fit to the data.}
  \label{Figure 1}
  \end{center}
 \end{figure}

Decent flux calibration of the 2dF LMC PN spectra was required in order to combine PN spectra from different 2dF fields and to make meaningful comparisons between fibre spectroscopy and long-slit observations of individual objects. A reliable flux calibration permits quick conversion to magnitudes for
luminosity studies. It also permits a study into other nebula characteristics such as excitation class, temperature, density and mass. Since the majority of the spectra were gained by
multi-fibre spectroscopy, it was necessary to verify the reliability
of the measured line intensities. A comparison of the same objects
observed in two or more overlapping 2dF field plate exposures revealed
variations in line strengths of up to 35\% from field to field,
clearly rendering the standard reduced spectra unsuitable for combination in their raw state. A further comparison was made between combined 2dF line intensities and published fluxes from the Magellanic Cloud Planetary Nebulae
spectral catalogue ([MCPN] thereafter) (Stanghellini et al. 2002)\footnote{Magellanic Cloud Planetary Nebula data is available from http://archive.stsci.edu/hst/mcpn/MCPN}. The results,
using all 65 MCPN which correspond with the RP sample, are shown in
Figure~\ref{Figure 1}. General agreement is evident through an essentially log-linear relation but the scatter is too wide ($\leq$1 dex) to provide a reliable flux calibration. The variations in the \OIII\,5007\AA~line
intensities shown are mainly due to the different relative strengths
in the \OIII\,5007\AA~line from one 2dF field exposure to another in
the raw, reddened spectra. A flux calibration of the spectra was required but to
achieve this, a reliable method had to be
found. This has proved quite difficult to do previously with
fibre-based spectra (eg. Mortlock et al. 2001; Georgakakis et al. 2004). Below, an effective new method is described,
which, in its careful application, is shown to be able to produce reliable flux
calibration for the 2dF spectra. %This is extremely important as
%decent flux calibration enables a whole range of diagnostic and
%physically meaningful line measurements to be undertaken.

\label{section3}

\begin{figure*}
\begin{minipage}[t]{8.5cm}
\psfig{file=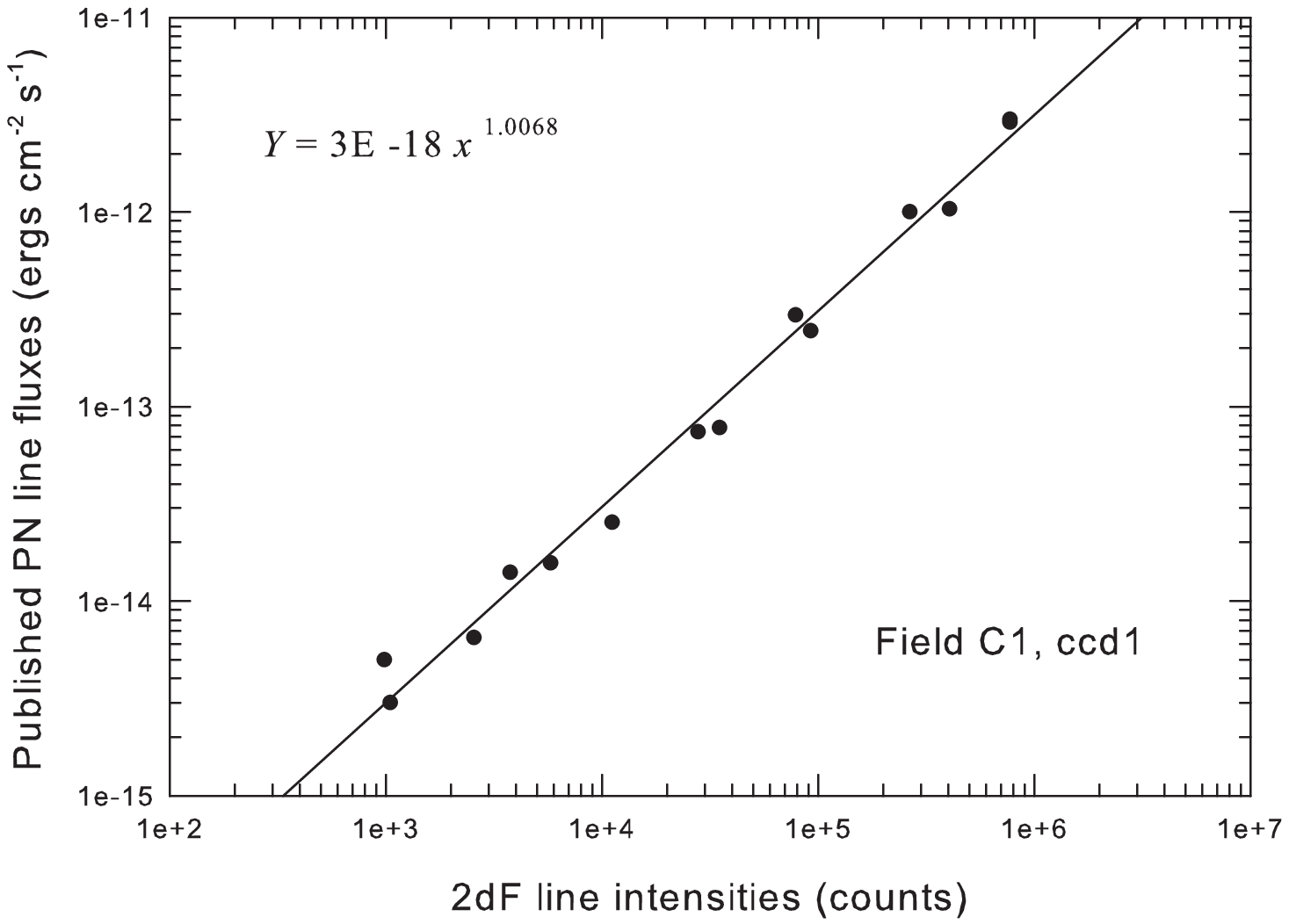,angle=0,height=5.5cm,width=7.9cm}
 %{\small Calibration graph for 2dF field C on ccd1.}
\label{figure2}
%\hfill
\end{minipage}
\begin{minipage}[t]{8.5cm}
\psfig{file=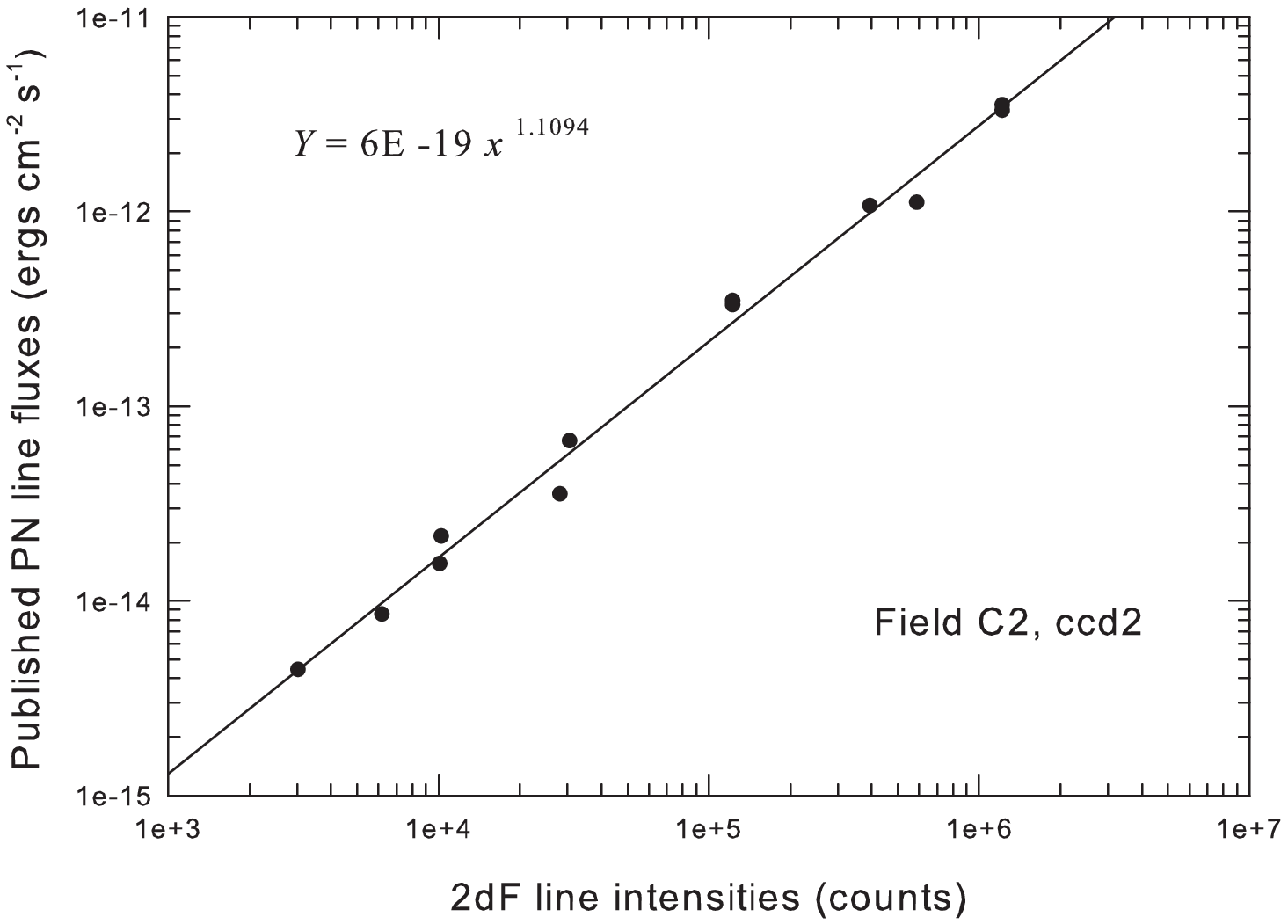,angle=0,height=6.4cm,width=9.3cm}
%{\small Calibration graph for 2dF field C on ccd2.} \label{figure3}
\end{minipage}
\end{figure*}
\begin{figure*}
\begin{minipage}[t]{8.5cm}
\psfig{file=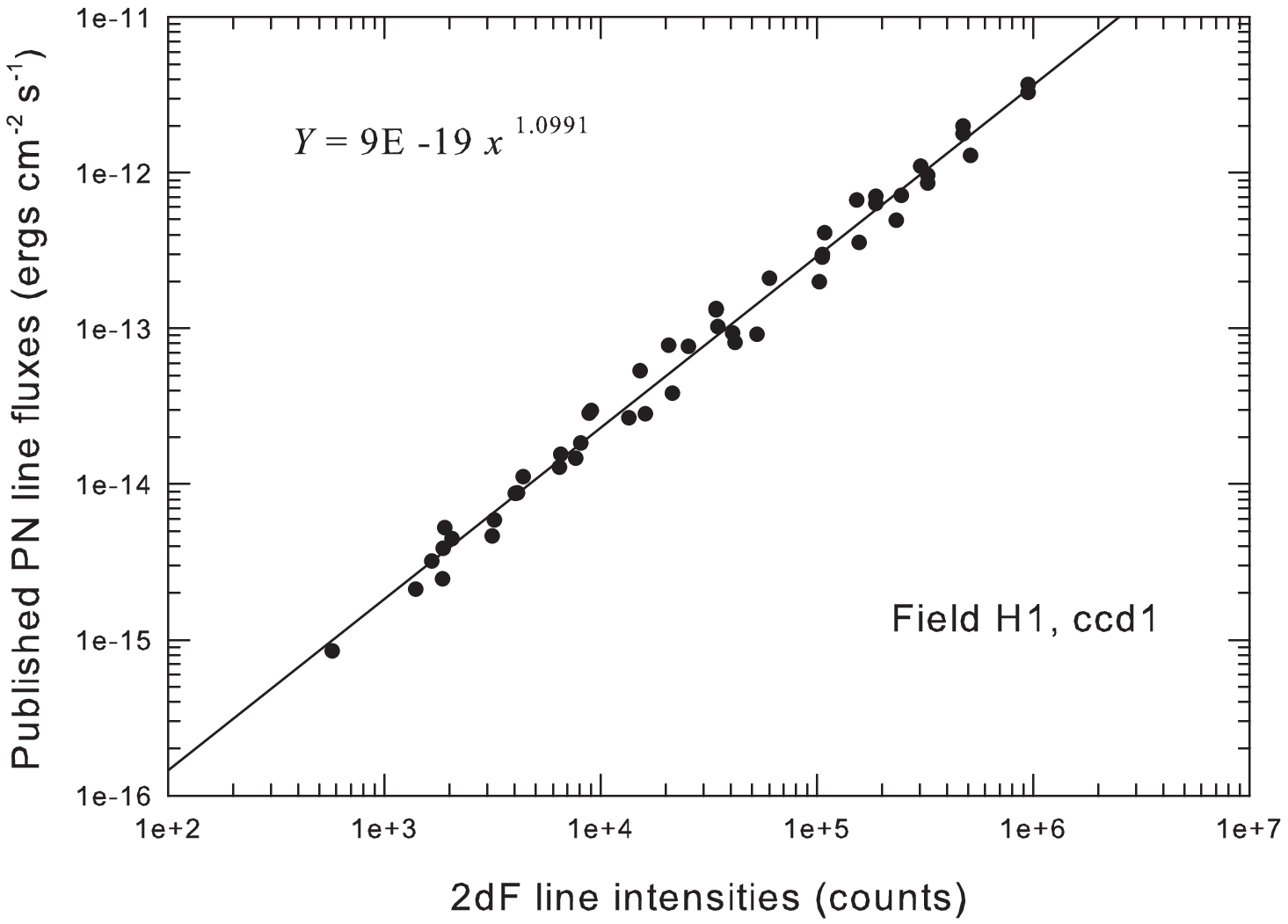,angle=0,height=6.4cm,width=9.3cm}
 %{\small Calibration graph for 2dF field H on ccd1.}
\label{figure2}
%\hfill
\end{minipage}
\begin{minipage}[t]{8.5cm}
\psfig{file=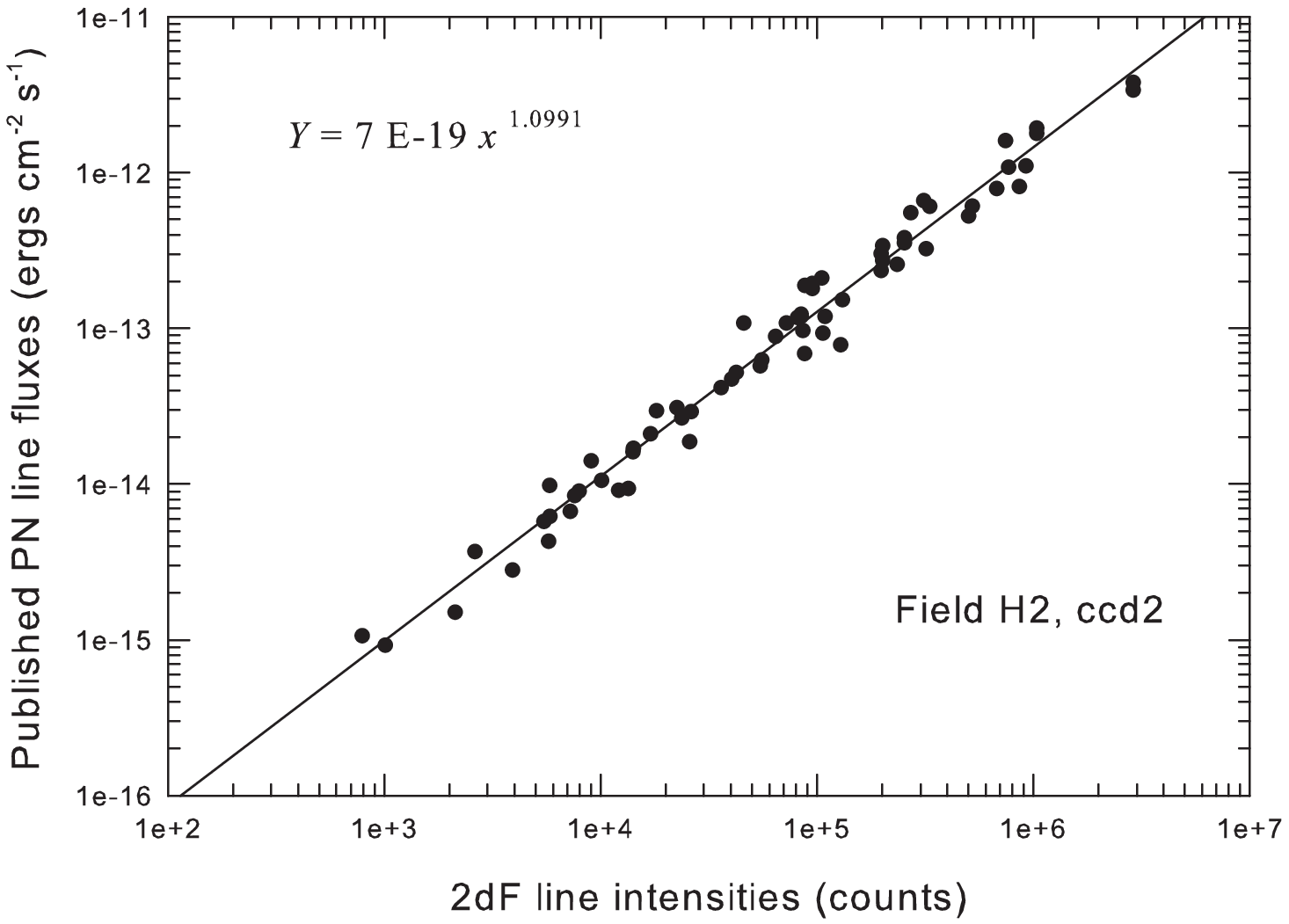,angle=0,height=6.4cm,width=9.4cm}
%{\small Calibration graph for 2dF field H on ccd2.} \label{Figure 5}
\end{minipage}
\caption{Four examples of the flux calibration procedure. In each case, an excellent agreement has been found between previously published \OIII~and H$\beta$~line fluxes and 2dF line intensities. The graphs on the left provide the calibration for {\scriptsize CCD}1 and the graphs on the right provide the calibration for {\scriptsize CCD}2 from the same field-plate observation. Field C1 is calibrated to 7 PNe, C2 to 6 PNe, H1 to 20 PNe and H2 to 28 PNe. The equation shown in the top left of each graph has been applied to all \OIII~and H$\beta$~line intensities for spectra fed to that particular {\scriptsize CCD} and field-plate combination.}
\label{Figure 2}
\end{figure*}

\subsection{Flux calibration technique}

Altogether, 18 overlapping 2dF field-plate exposures were taken in order to cover the entire central 25deg$^{2}$ survey region of the LMC. A comparison of line intensities from overlapping areas revealed that each field plate/observational setup resulted in its own individual spectrum intensity scale. In addition, the 400 spectra from each field were fed to 2 identical spectrographs. Although they were well matched, small instrumental differences were noticeable. Overall, the combined use of 2
different field-plates, each sending spectra to 2 different spectrographs and CCD cameras resulted in the observed deviations
seen when comparing previously published fluxes with line intensities for the same objects from multiple 2dF fields (Figure~\ref{Figure 1}).

Experiments were undertaken in order to find the most reliable method of flux calibration
using the observed line intensities. The best results were obtained by
individually calibrating each spectral line on each field plate from each {\scriptsize CCD} camera
to raw MCPN fluxes gained from HST exposures. The known PNe included and observed on
each field plate were used as flux calibrators for each individual
measured field. In order to extend the reliability of the \OIII5007\AA~flux calibration to magnitudes $>$19, the \OIII4959\AA~and H$\beta$4861\AA~lines (Shaw et al. 2006) were included in the calibration. All three lines are close enough in the optical spectrum to produce reliable flux ratios. The individual H$\beta$ and \OIII4959,5007\AA~2dF line intensities for known PNe observed for each {\scriptsize CCD} and each field plate exposure were
 graphed against the individual published fluxes for the same lines (eg. Figure~\ref{Figure 2}). The same method was applied to our multi-fibre spectra from VLT FLAMES. All 2dF and FLAMES fields purposefully contained sufficient PNe with well calibrated fluxes to enable this process. The agreement of flux calibrated PNe from each spectrograph/field plate combination was considered strong enough (within 0.2 dex, see Figure~\ref{Figure 2}) to allow its application to all the H$\beta$~and \OIII4959,5007\AA~emission lines for other PNe observed in the same field.
 In each case, a line of
 best fit was derived and the underlying linear equation extracted. This
 equation became the calibrator for each emission line in each uncalibrated PN from the {\scriptsize CCD} and individual 2dF field
 plate exposure used. The equation was applied to
 each spectrum with a detectable line intensity in that field.

Fluxes for LMC PNe from other
catalogues (Jacoby et al. 1990, Leisy et al. private communication,
Meatheringham et al. 1988) were also included in order to build up
the number of calibrators per field. Where only a de-reddened flux
value was published, a relative reddening was applied in order to
make a better correlation. Where 2 or more published fluxes were found for the same PN, stronger comparisons could be made. Each independent survey showed a mean agreement to within 0.2 dex, slightly increasing as the magnitude grew fainter. A comparison between calibrated \OIII5007\AA~fluxes from 2dF and published fluxes from the MCPN, Jacoby et al. (1990) and Leisy et al. are shown in Figure~\ref{Figure 3}. The agreement ($\sigma$$<0.2$ dex) is very good across all surveys; especially since they represent space vs ground based observations in some cases with a variety of detectors and measurement methods. It also shows that the 2dF calibrated fluxes remain within the $\sigma$$<0.2$ dex agreement. The mean slope of the fit (2dF/MCPN) is 1.01, where 1 would  represent a perfect match. The median is exactly 1 and the standard
 deviation $\sigma$ = 0.12 dex.

 A spread in
published line fluxes for the same PN is common but helps us to establish the uncertainties. Since the MCPN catalogue contained the largest
number of PNe and the raw fluxes were `as measured' (not
de-reddened) these were given preference where irregularities became obvious. The MCPN set also
includes some ground-based fluxes from ELCAT (Kaler et al. 1997) available at http://stsdas.stsci.edu/elcat/
where the spectra are uncorrected for extinction. The intention was to provide self consistent fluxes for all the 2dF observations.

A calibrated quality and consistency check was then performed where fluxes for several PNe observed on more than one field plate exposure
(due to the overlaps
in 2dF pointings across the survey) were compared for
consistency once the new flux calibration had been applied to each
field. Figure~\ref{Figure 4}, shows an excellent match in the high
as well as the low flux regimes, proving that reliable and
consistent fluxes can in fact be achieved this way. The equation to convert line intensities to fluxes in each field was then
applied to the 2dF and FLAMES \OIII\,5007\AA~line intensities for each
corresponding field. %Within the database the equations can undergo
%fine adjustments to achieve the best possible correlation.
Further line-to-flux calibration equations were then derived for
other spectral lines such as H$\alpha$ which, together with H$\beta$, controls the
de-reddening. A similar procedure was then carried out for each of the
spectral lines to provide their individual flux calibration.

\label{section 3.1}

\subsection{Reliability of the line intensities and fluxes}

At the distance of the LMC, most PNe are compact and point-like in ground-based observations allowing most of the flux to fall within the 2 arcsec diameter 2dF fibre. This is despite the larger reported angular diameters,
which suffer from point spread function growth as a function of source intensity and include the very faint surrounding
halos. The extended outer halos typically represent about 1,000th of the flux
of the inner PN (Corradi, 2003). Clearly, the central region of LMC
PNe contains the majority of the light. There are 70 PNe out of our sample of 589 with inner
shell diameters which exceed 2 arcsec by more than 1 arcsec.
These are among the brightest LMC PNe and, despite the anticipated loss of light by 2dF spectroscopy, are still able to be accurately flux calibrated.
For example, the loss of light from bright PNe is evidenced by the bright turnover in Figure~\ref{Figure 1}.
However, individual spectrograph calibration which includes the brightest PNe in each field (Figure~\ref{Figure 2}),
is able to largely correct for the assumed loss of light at the highest magnitude level. It does this by
calibrating the increasing brightness received by the fibre directly to a well established flux level.
This level will vary for each spectrograph/plate observation.
For this reason, calibration of 2dF spectra to a
single standard star would produce erroneously low fluxes for bright and
extended LMC objects. In the same way, if a longslit is too narrow, flux levels will be underestimated (Jacoby \& Kaler, 1993). This may be the cause of some small
deviations in previously published fluxes for LMC PNe.

Flux ratios of doublet or close lines from the central shell of the PN, such as
the \SII~ratio used for the estimation of nebula densities, are largely consistent across the nebula and do not require flux calibration. The \SII6716,6731 lines were measured from the
high resolution 1200R spectra. All other lines, into the blue, were measured from the 300B spectra.

\subsubsection{Error estimations}

The line intensities measured from 2dF 300B and 1200R spectra have
two main sources of error. The first is due to an uncertainty
in where to place the continuum level for line intensity
measurement. Since the author performed all the line measurements,
they were all performed in the same manner, at the same base
positions for any line, relative to the continuum and/or noise.
Repeated measurements however indicated an increased error estimate
should apply to lines with a S/N of 5 or less. The error estimate increases from 10\% at S/N = 5 to 40\% at S/N = 3. About 5\% of H$\beta$ intensities are low enough that the increased measurement error should be applied. This is never applied to the \OIII5007~line used for the PNLF in this study. These lines mostly have intensities $>$4 $\times$ H$\beta$ and have repeat error estimates no greater than 3\% for the faintest PNe. This system however provides a general error
estimate for the line measurements in all 631 PNe.

\begin{figure}
\begin{center}
  \includegraphics[width=0.48\textwidth]{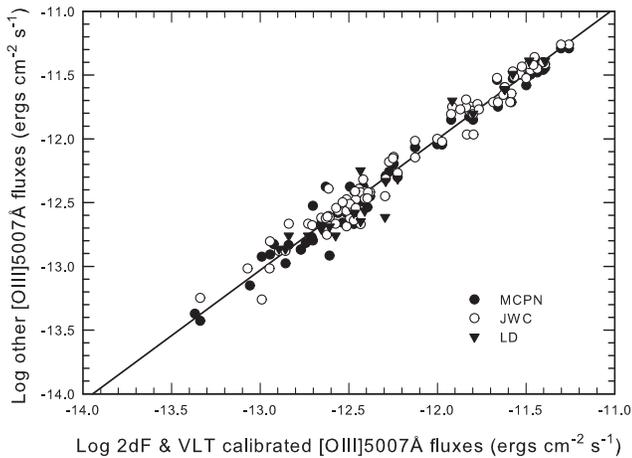}\\
  \caption{The calibrated \OIII\,5007\AA~2dF 300B line fluxes are plotted against the published \OIII\,5007\AA~fluxes from the MCPN on-line catalogue [MCPN], Jacoby, Walker, Ciadullo 1990b, [JWC] and Leisy \& Dennefeld 2006, [LD]. The sample of 85 calibrated data points from 2dF and 13 from VLT FLAMES are drawn from 21 different fibre field exposures. Extremely good agreement can be seen between all the combined data points and published fluxes. Through this calibration method, 2dF and VLT FLAMES line intensities for all the remaining PNe observed in a given 2dF or FLAMES field exposure are reliably calibrated to fainter magnitudes. The
  mean agreement $\Delta$flux (RP-MCPN) is 0.03$\pm$0.10 dex. A similar mean agreement is found between RP and JWC (-0.004$\pm$0.11) and RP and LD (0.02$\pm$0.13).}
  \label{Figure 3}
  \end{center}
 \end{figure}

\begin{figure}
\begin{center}
  \includegraphics[width=0.47\textwidth]{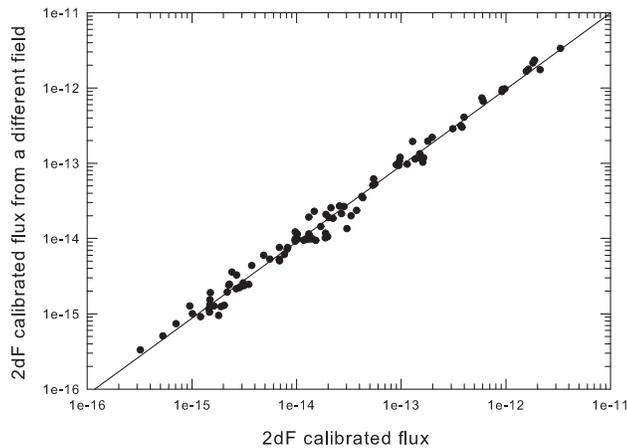}\\
  \caption{The \OIII\,5007\AA~calibrated fluxes for 81 PNe with multiple observations (due to overlapping fields) are plotted in order to check the integrity of the calibration across different observations including different {\scriptsize CCD} and field plate combinations. A good match ($\sim$0.2 dex) along the line of equality has been found, where there is a mixture of fields and flux intensities represented over a 4 dex range.}
  \label{Figure 4}
  \end{center}
 \end{figure}

Secondly, there
are difficulties in placing exact errors on the
individual lines, due to the inherent variations in detector
response with wavelength. Luckily, this is the same for all objects
on any given field plate exposure and can therefore be estimated
from the response curve. The exposure times are also exactly the
same across any given field and most of the exposures for the
different fields are also of the same duration. With a small error included for these two
variables, the remaining difficulty is the
different absolute fluxes which cause a variation in the signal to
noise. One fortunate aspect of having a relatively large number of
PNe re-observed on different 2dF fields is that a comparison between
two or more observations of the same object is possible. A
comparison of line intensity ratios for the same PN, observed on two
or more fields (Figure~\ref{Figure 4}), gives an agreement to within $\pm$0.2 dex with $\sigma$=0.15 using a sample of 81 repeated PNe.
This discrepancy includes ratios between strong lines and between
strong to weak lines.

In addition to the estimates for spectral line errors, the flux
calibration adds an extra level of error. With a very good
calibration of the 300B spectra to the MCPN fluxes
(Figure~\ref{Figure 2}), the average difference to the published
fluxes following calibration is $\pm$0.18 dex with $\sigma$ = 0.5. These published fluxes
will themselves have both measurement and systematic errors and
these have been estimated at 0.2 dex through direct comparison. An allowance is made for these
errors to increase with decreasing flux. This estimated increase is
based on the standard deviation in the calibrated fluxes, increasing
to $\pm$0.3 $\sigma$ = 0.2 dex.

With ground-based spectroscopy, atmospheric dispersion can
produce an extended object on the slit where blue light is separated
from red light, also received by the detector (Filippenko, 1982).
This effect can cause a systematic difference when comparing
standard stars observed at a lower air mass. It can also produce larger images in the blue than in the red due to more blue light being captured by the fibre. The increase is estimated as (1/$\lambda$)$^{0.2}$ (Fried 1966; Boyd
1978). To overcome this problem, 2dF has an `Atmospheric Dispersion Compensator' (ADC) which largely compensates for this. With our flux calibration technique, however, lines from different regions of the optical spectrum are individually calibrated to HST-based fluxes, largely circumventing problems of atmospheric dispersion and the differential air mass of standard stars.  A correction for extinction is applied using the
Balmer decrement, where the fixed ratio 1:2.86 applies between
H$\alpha$ and H$\beta$ in the absence of extinction. This corrected
ratio is discussed in detail in the following subsection.

In addition to the photometric and flux calibration errors, an error
estimate for the inclination of the LMC to the line of sight may be included. This could
be as much as 0.04 mag (Jacoby et al. 1990) based on a 1$\sigma$
depth uncertainty.

One further problem can cause errors in flux estimations for LMC
PNe. The LMC is filled with faint stars and diffuse
emission of varying intensity. For individual slit spectroscopy, it can be extremely
difficult to find a region of blank sky along the slit. In the case
of 2dF fibre spectroscopy, we dedicated $\sim$40 fibres to specially
selected sky positions. These positions represented average values of background sky in each field observed. The H$\alpha$ map provided the most appropriate positions. For each field, the
2dfdr reduction program sampled and averaged these regions before applying the sky subtraction to the objects. The resulting
error estimates for \OIII5007 as a function of line measurement and flux calibration are based on results from
Figures~\ref{Figure 3} and~\ref{Figure 4}. Line measurement errors increase from 2\% to 10\% and flux calibration errors increase from 6\% to 20\%. The line
measurement error also includes an estimated error based on a function of
the discrepancy between repeated, field to field and line ratio
measurements.

\label{section3.2}

\subsection{Corrections for extinction and reddening}

Extinction of light from distant objects is mainly the result of
interstellar dust. The light is both scattered and absorbed, which
increases the interstellar extinction towards shorter wavelengths.
The amount of extinction will differ for each object and needs to be
corrected to gain true fluxes. In the optical regime, the
H$\alpha$/H$\beta$ ratio was used to determine the extinction
constant $\textit{c}$H$\beta$ (i.e., the logarithmic extinction at
H$\beta$ for each nebula. These hydrogen transitions are the strongest and easiest to accurately measure in the nebula spectrum and therefore provide a better
estimate than other H lines. %For the reddening law, the interstellar
%extinction curves of Nandy et al. (1981) for the LMC were used for
%the extinction functions ($\textit{f}$~($\lambda$)). This law and
The observed H$\alpha$/H$\beta$ ratio, when compared to the
recombination value of 2.86 (Aller 1984), gives a logarithmic
extinction at H$\beta$ of:

\begin{equation} \textit{c}(\textrm{H}\beta) = (log(\textrm{H}\alpha /
\textrm{H}\beta) - log(2.86)) / 0.34
\end{equation}

%corresponding to a visual extinction of
%\begin{equation}
%A_{V} = 2.1 \textit{c}(H\beta).
%\end{equation}

This estimation is based on the relationship between observed and
intrinsic intensities:

\begin{equation}
\frac{I_{obs}(\textrm{H}\alpha)}{I_{obs}(\textrm{H}\beta)} =
\frac{I_{int}(\textrm{H}\alpha)}{I_{int}(\textrm{H}\beta)}
10^{-c(\textrm{H}\beta)[f(\textrm{H}\alpha)-f(\textrm{H}\beta)]},
\end{equation}

where[$\textit{f}$~(H$\alpha$)-$\textit{f}$~(H$\beta$)] = --0.34
from the standard interstellar extinction curve given in Osterbrock
(1989) and $\textit{c}$(H$\beta$) is the logarithmic extinction of
H$\beta$. The intrinsic ratio is mildly dependent on temperature:
for T$_{e}$ between 2,500 and 20,000 K, the ratio varies from 3.30
to 2.76 (Osterbrock, 2006).

The value of $\textit{c}$ and $\textsl{I}$($\lambda$) for all other
lines $\lambda$ was then used to correct for interstellar extinction
using the galactic extinction law from Whitford (1958) in the form
of Miller \& Mathews (1972):
\begin{equation}
I_{cor}(\lambda) = I_{obs}(\lambda) 10^{c[1+f(\lambda)]}.
\end{equation}
Reworking this equation for the H$\beta$ line all other lines could
be corrected for reddening using:
\begin{equation}
\frac{I_{cor}(\lambda)}{I_{cor}(H_{\beta})} =
\frac{I_{obs}(\lambda)}{I_{obs}(H_{\beta})} 10^{cf(\lambda)}.
\end{equation}

Application of the Balmer decrement ratio 2.86 between the H$\alpha$ and H$\beta$ lines is able to correct the spectrum in terms of the required ratios however the degree of c is dependent on flux calibration for each of these lines and any internal inconsistencies. For fibre-based observations, the uncertainty in $\textit{c}$(H$\beta$) is extremely difficult to estimate as there may be small inconsistencies between fibres resulting from light transmission and position of the fibre on the plate. These will have an effect on the
reddening corrected fluxes. Calibration of both the red and blue sides of the spectra each show a maximum uncertainty of 0.2 dex. This is consistent with the uncertainty found comparing published fluxes, indicating that internal inconsistencies are low. The reddening law of Howarth (1983) where $\textsl{E}$($\textsl{B-V}$) = 0.689c was also employed. It then follows that A$_{5007}$ = 3.5 $\textsl{E}$($\textsl{B-V}$) and A$_{4861}$ = 3.63 $\textsl{E}$($\textsl{B-V}$).

%The value for the \OIII5007\AA~line extinction
%$\textit{f}$~($\lambda$) is --0.038.
When estimating the errors,
flux calibration is given predominant consideration together with the wide spectral
range between the H$\alpha$ and H$\beta$ lines. This is estimated for H$\alpha$ using the same method
of 2dF flux calibration described above where the H$\alpha$ spectral line in each
nebula was individually compared to published H$\alpha$ fluxes for the same objects. The estimated uncertainty in
$\textit{c}$(H$\beta$) as a result of combined maximum errors in
line measurement and flux calibration estimations is $<$7\%.

%The uncertainty in $\textit{c}$(H$\beta$) has a modest effect on the
%reddening corrected line ratios. The value for the \OIII5007\AA~line extinction
%$\textit{f}$~($\lambda$) is --0.038. With such a small correction, the \OIII5007\AA~line flux estimates are very unlikely to be affected by any error estimates in $\textit{c}$(H$\beta$). When estimating the errors,
%flux calibration is given consideration together with the wide spectral
%range between the H$\alpha$ and H$\beta$ lines. This is estimated for H$\alpha$ using the same method
%of 2dF flux calibration described above where the H$\alpha$ spectral line in each
%nebula was individually compared to published H$\alpha$ fluxes for the same objects. The estimated uncertainty in
%$\textit{c}$(H$\beta$) as a result of combined maximum errors in
%line measurement and flux calibration estimations is $<$7\%.

The optical image of every PN was inspected for the presence of
diffuse \HII~across its immediate field. This was used as a
means of testing the environment of PNe with high values of
$\textit{c}$(H$\beta$) in the range 1.4 to 2.0. The
available 8$\mu$m Spizer SAGE image (Meixner et al., 2006) was
also examined for areas where IR emission extended across individual
PNe. This was added confirmation that foreground dust could play a
significant role in flux determination for some LMC PNe. The intensity of the ambient
emission in H$\alpha$ was then measured in counts, where, for a
measure below 3 $\sigma$, the noise begins to have a significant
effect on the observed PN flux. Velocity measurements
(RPb) for PNe and the velocity for the HI disk
in the vicinity of each PN were then examined to estimate whether
the PN was foreground, within, or
background to the bulk of observed emission. %Using the new velocity
%measurements (section~\ref{section 6}) it was determined as to
%whether each PN lie within or behind the \HI~disk and vertical area
%within which \HII~regions occur for their particular image cell
%(Figure~\ref{figure 7b}).
From the list of 292 newly discovered `true' PNe, 11 were found in
areas of relatively strong \HII~emission. Of these, only 2 were
identified as lying beyond the visible emission. Of the
newly discovered PNe, all 4 `likely' PNe in areas of strong, dense
\HII~emission ($>$3 $\sigma$ the noise) exist within or beyond the
\HI~disk.  All of these PNe have values of $\textit{c}$(H$\beta$)
$>$ 0.8.

 \begin{figure}
\begin{center}
  \includegraphics[width=0.5\textwidth]{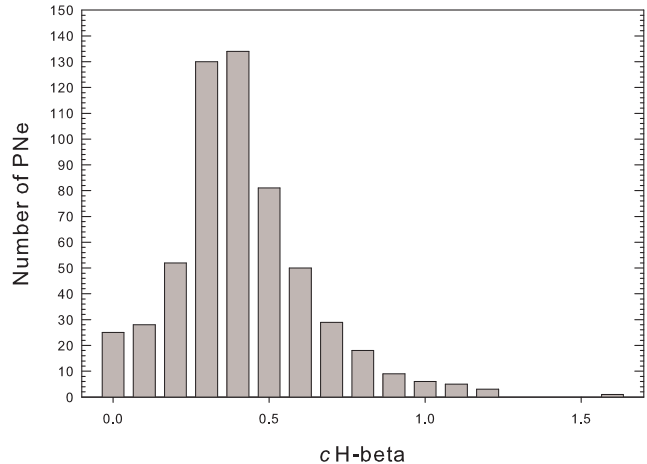}\\
  \caption{The range of derived values for $\textsl{c}$(H$\beta$) using the RP and previously known LMC PNe from 2dF 300B spectra. A peak at 0.4 shows the low overall extinction of most LMC PNe. Most of the bright, previously known PNe occupy the range 0-0.4.}
  \label{Figure 5}
  \end{center}
 \end{figure}

Due to the suspected low metallicity of many of the new, faint PNe
in the RP LMC sample, as well as the generally low metallicity of all
PNe in the LMC compared to the Galaxy (Leisy \& Dennefeld, 1996), we
also consider the effect of collisional excitation on the H$\alpha$
line (Stasi$\acute{n}$ska, 2002) and the effect it may have on the
extinction calculation. Because PNe are excited by power-law
photoionisation or shock-heating from the central star, collisional
excitation and self-absorption can affect the intrinsic Balmer
decrement. This can be caused by high $\textit{T}$$_{e}$ resulting
in the H$\alpha$/H$\beta$ ratio no longer holding the canonical
recombination value of 2.86 (Osterbrock \& Ferland 2006). It is possible that this scenario applies to some PNe in the LMC. Since the de-reddening
procedure involves H$\alpha$ there may be some overestimation of the
extinction $\textit{c}$(H$\beta$). It has been shown that this
error, which is independent of extinction, can be large when the
lines have far separated wavelengths (Stasi$\acute{n}$ska, 2002). In
low metallicity nebulae, there may be a range of evolutionary
sequences where H$\alpha$/H$\beta$ is significantly larger than the
accepted recombination value. The resulting error has been estimated
to be as high as 10\%. Since there is no precise means to estimate
the amount of correction required, care was taken to identify faint,
low mass PNe with high T$_{e}$ and extremely high H$\alpha$/H$\beta$
$>$10 ratios which may have been affected by power-law
photoionisation or shock-heating (see Stasi$\acute{n}$ska, 2002 fig.
2). In the LMC sample, 6 PNe with $\textit{c}$(H$\beta$) $>$ 0.8 and T$_{e}$ $>$ 25,000K have been identified. They are marked with a $\dag$ in column 3 of appendix tables 3 and 4. Where inconsistencies (such as ultra-low N/O abundances) are found
in 1 of these PNe (RP1584), the cause can be traced to collisional excitation or
self absorption affecting $\textit{c}$(H$\beta$).

The derived values for $\textsl{c}($H$\beta$) are shown in the
histogram of Figure~\ref{Figure 5}. Despite low extinction towards the LMC PNe we still find a few PNe
with values of $\textsl{c}$ reaching as high as 1.6. Most of the brighter
previously known LMC PNe generally occupy the lower range from 0 to
0.6 however some fainter Jacoby (1980) PNe in dusty, nebulous areas have $\textsl{c}($H$\beta$) values up to 1.0. Results for bright LMC PNe agree with values
derived using fluxes in the MCPN catalogue. The majority of
these PNe have c(H$\beta$) between 0 and 0.6 but also range as high
as 1.6. For all the RP PNe, there is a peak at 0.4 with the largest proportion of PNe occupying the range 0.3 - 0.5. The mean extinction of 0.4 corresponds to $\textsl{E}$($\textsl{B}$-$\textsl{V}$)=0.31 mag, which is considerably higher than the line-of-sight reddening of 0.074-0.11 for stars in the LMC (Caldwell \& Coulson 1986). It is also higher than a value of 0.2 found for LMC PNe by Herrmann \& Ciardullo (2009) using a pre-RP catalogue sample to produce a basic skeleton of our Figure~\ref{Figure 5} plot. They cover less than half the magnitude range with lower sampling and base the Balmer decrement on estimated H$\alpha$ values. Nonetheless, their sample of the bright end also shows evidence of an increase in extinction beginning at 3 magnitudes below the brightest.

Somewhat higher values of
$\textsl{c}($H$\beta$) in the RP sample are to be expected due to
their faint magnitudes and often dusty environments. It does suggest that average internal dust reddening may be as great as 0.5 mag. Most of the dust is expected to consist of carbon-rich and dredged-up material, the quantity of which will have dependence on the mass and metallicity of the central star (Doptia et al. 1992). PNe at the bright end mainly occupy the medium excitation range (Reid \& Parker 2006c) with central stars ranging in temperature from 80,000K to 126,000K making them very efficient at converting central star luminosity to luminosity in the nebula. Necessary
corrections for $\textsl{c}($H$\beta$) were applied to the whole LMC
sample on an individual object basis. The derived values of $\textsl{c}($H$\beta$) for each
previously known and newly discovered PNe are given in appendix Tables 3 and 4 respectively.

When deriving the distance to an extragalactic source using the PNLF, we use the raw \OIII5007\AA~fluxes corrected only for foreground reddening. This is because none of the measured PNLFs and simulations to date have included the effects of circumstellar dust. Most surveys have been based on the \OIII5007\AA~line. The H$\beta$ line, required to derive c, is extremely difficult to measure from individual objects in external galaxies. Circumstellar dust therefore plays a large part in determining the observed brightness of the PNLF bright end. In order to make the correct comparison, our PNLF for distance determination has only been corrected for foreground reddening. The contribution of Galactic dust was removed assuming a foreground reddening value of E(B-V) = 0.074 (Caldwell \& Coulson, 1985).

\label{section3.3} %%%%%%%%%%%%%%%%%%%%%%%%%%%%%%%%%%%%%%%%%%%%%%%%%%%%%%%%%%%%%%%%%%%%%%%%%%%%%%%%%%%%%%%%%%%%%%%%%%

\section[]{The new PNLF for the LMC}
\label{section4}

The PNLF is a plot of the total number of PNe in a defined volume at a particular magnitude at a given point of time. Since the \OIII\,5007\AA~line is generally the brightest line optically emitted by the nebula, it has traditionally been used to plot the PNLF. The raw fluxes are listed in Appendix Tables 4 and 5 for the
164 previously known and 432 newly discovered PNe respectively. The \OIII5007\AA~fluxes for the observed PN sample, corrected only for foreground reddening
have been transformed to the magnitude system using the relation
provided by Jacoby (1989):

\begin{equation}
\textit{m}_{5007}= -2.5~\textrm{log}~\textsl{F}_{5007}-13.74 .
\end{equation}

\begin{figure*}
\begin{center}
  \includegraphics[width=0.80\textwidth]{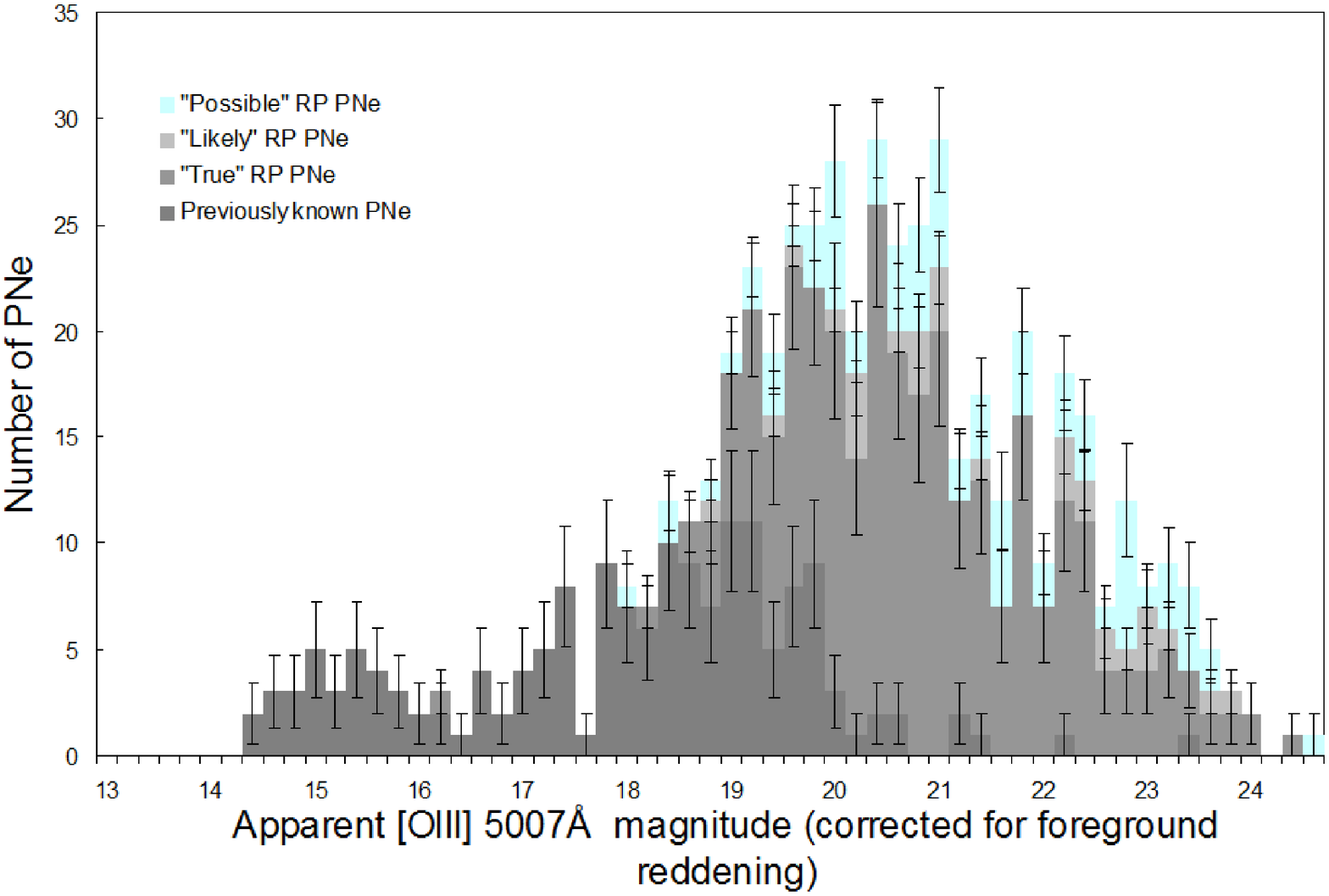}\\
  \caption{The foreground de-reddened planetary nebulae luminosity function for the central 25 deg$^{2}$ of the LMC, with PNe separated into the categories of previously known and true, likely and possible RP PNe in the survey area for this study. The bins are 0.2 mags in width. Poisson error bars are shown. The errors are calculated individually for previously known, true, likely and possible PNe and combined for each bin. }
  \label{Figure 6}
  \end{center}
 % \end{figure*}
%\begin{figure*}
\begin{center}
  \includegraphics[width=0.80\textwidth]{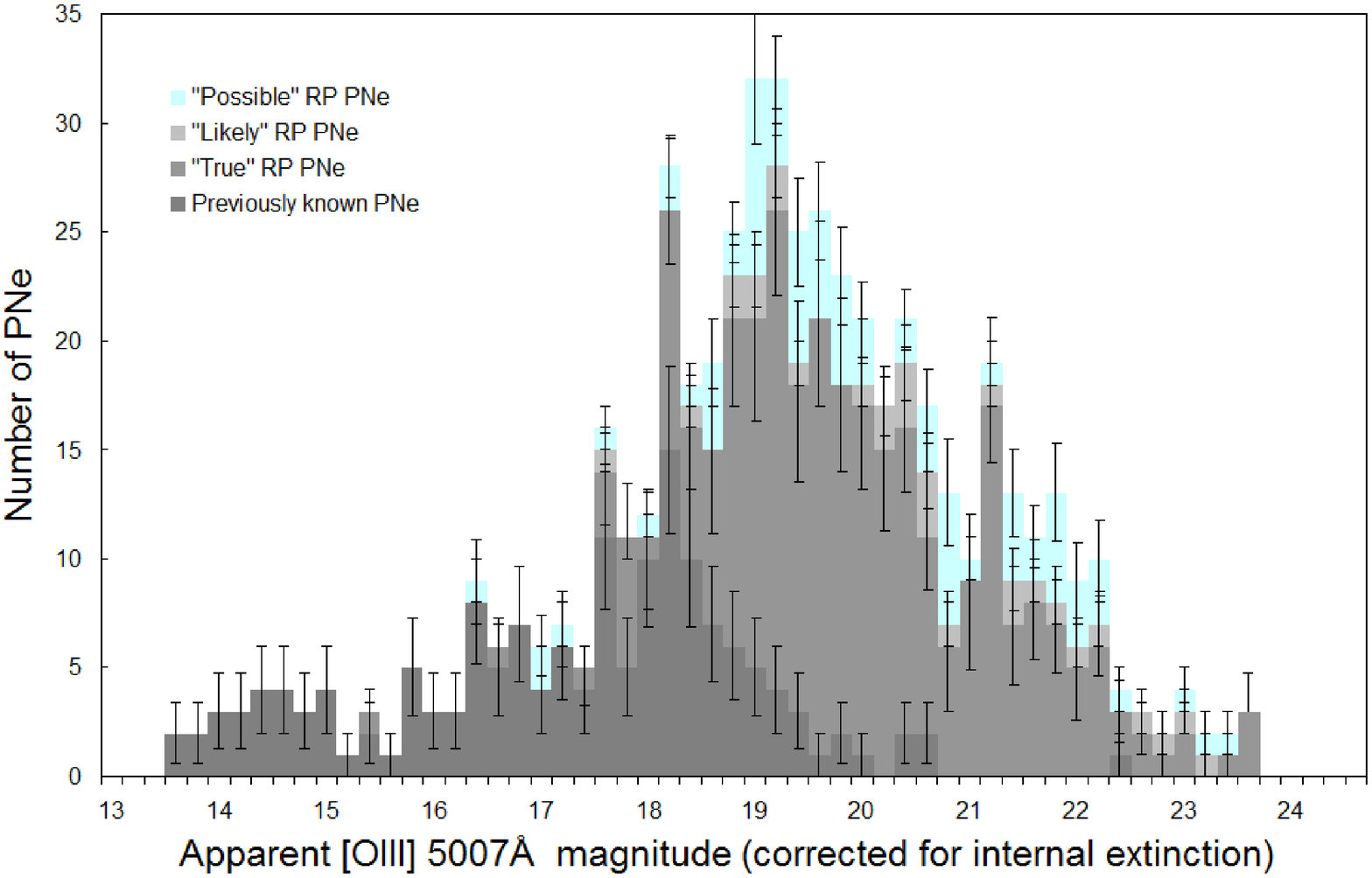}\\
  \caption{The planetary nebulae luminosity function for the central 25 deg$^{2}$ of the LMC, with PNe separated into the categories of previously known and true, likely and possible RP PNe in the survey area for this study. All PNe are individually corrected for internal extinction using the Balmer decrement. The bins are 0.2 mags in width. Poisson error bars are shown. The errors are calculated individually for previously known, true, likely and possible PNe and combined for each bin. }
  \label{Figure 6a}
  \end{center}
  \end{figure*}

Using the
magnitudes, the luminosity function can be displayed through several
methods. In principle we can get an approximation to the luminosity
function $\rho(\textit{l})$ by plotting the number of PNe found within any number of equally spaced luminosity bins. Using this method, all previously known and new RP PNe within the central 25deg$^{2}$ of the LMC have been plotted into one of 54 bins, each 0.2 magnitude wide. This luminosity function, based on the foreground de-reddened flux of the \OIII\,5007\AA~emission line, in magnitudes, is shown as a
histogram in Figure~\ref{Figure 6}. The magnitude assigned to each bin represents the magnitude at the central position of each bin. Poisson error bars have been included.

  \begin{figure}
\begin{center}
  \includegraphics[width=0.48\textwidth]{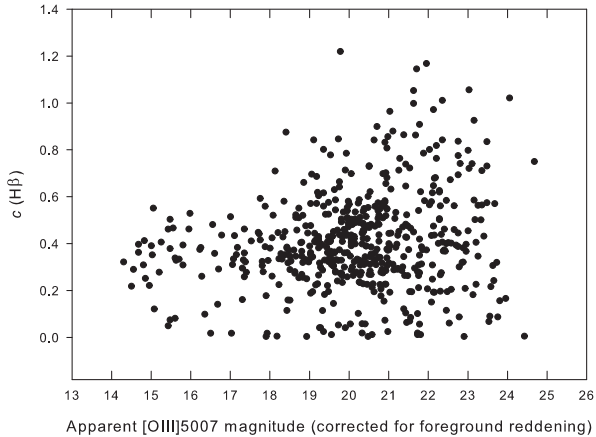}\\
  \caption{In order to observe the effects of internal dust extinction within the LMC PNe, the foreground de-reddened \OIII5007\AA~magnitudes are plotted against extinction. PNe in the brightest 4 magnitudes have low to medium extinction while PNe with medium to high extinction are only found less than 4 magnitudes below the brightest. These highly reddened faint PNe probably have intrinsically fainter cores which have
evolved over a longer timescale. It is therefore possible that the
dust properties of these objects may be rather different to those of the higher-core mass
objects.}
  \label{Figure 6b}
  \end{center}
  \end{figure}

In the first PNLF derived for the Magellanic Clouds, Jacoby (1980) using a sample of 41 PNe, demonstrated the theoretical exponential function of Henize \& Westerlund (1963), whereby a PN is treated as a uniformly expanding homogeneous sphere ionised by a non-evolving central star. It follows, therefore, that the number of PNe in each luminosity bin is proportional to the time those PNe spend at that luminosity (Jacoby, 1980, Ciardullo et al. 1989, Ciardullo et al. 2004). This may be represented by an exponential curve (Henize \& Westerlund, 1963) with a sharp truncation designed to accommodate the bright end (Ciardullo et al. 1989). The resulting curve is described by:

\begin{equation}
N(M) \propto e^{0.307M} \{1-e^{3(M^{\ast}-M)}\}
\end{equation}

In this equation, the key parameter $\textsl{M}^{\ast}$~is the
absolute magnitude of the brightest possible PN. The limit to the high
luminosity of the \OIII\,5007\AA~line is mainly due to the initial
mass of the progenitor star and its evolution to a white dwarf as a
function of time in the ranges of 3-11 Gyrs (Jacoby, 1989; M\'{e}ndez et al., 1993; Stanghellini, 1995; Richer et al., 1997; Jacoby, 1997). This PNLF model has been demonstrated as an excellent standard candle when compared to other distance indicators (Ciardullo, 2006).

It has been shown that there is a weak correlation between the luminosity of $\textsl{M}^{\ast}$ and the metallicity of a galaxy (Ciardullo \& Jacoby, 1992; Dopita et al. 1992; Richer 1993; Ciardullo et al. 2002, 2005). This trend has been modeled by Dopita et al. (1992), showing that $\textsl{M}^{\ast}$ is bright for solar oxygen abundances but fades where metallicity increases or decreases away from solar. The correction for $\textsl{M}^{\ast}$ is given by:

\begin{equation}
\Delta M^{\ast} = 0.928[\textrm{O/H}]^{2} + 0.225[\textrm{O/H}] + 0.014
\end{equation}

where solar metallicity is assumed to be 12 + log\,(O/H) = 8.87 (Grevesse, Noels \& Sauval 1996) (see Ciardullo et al. 2002 for additional details).
The metallicity of the LMC has been determined from the emission lines of \HII~regions. A mean value of (12 + log\,[O/H]) 8.45 $\pm$0.11 from
LMC \HII~regions was found by averaging mean values from Dufour, Shields \& Talbot (1982), Dennefeld (1989), Monk, Barlow \& Clegg (1988) and
Ferrarese et al. (2000a). With this value, using the metallicity dependence of Doptia et al. (1992) we find a metal dependence ($\Delta M^{\ast}$) of 0.1\,mag.

In the LMC, the estimated metallicity is low compared to the Galaxy. The low oxygen abundance in LMC PNe, whether
inherent or through depletion, also raises the electron temperature
thereby increasing the rate of collisional excitations per ion.
Since oxygen is a coolant, a decrease in oxygen abundance only
decreases the \OIII\,5007\AA~flux by roughly the square root of the
difference in abundance (Jacoby 1989). The central star at the core
of a PN affects the luminosity the opposite way to low metallicity.
Low metallicity in the progenitor star produces a more massive
radius with greater UV flux (Lattanzio 1986). %Brocato et al. 1990.
This energy compensates for the decreased emissivity of the nebula.
The total \OIII\,5007\AA~flux then becomes somewhat independent of
metallicity.

 In order to reveal the effects of dust, internal to the LMC and the environment of each individual PN, we include another PNLF (see Figure~\ref{Figure 6a}) where the same objects have been corrected for what we will refer to as `internal extinction' using the Balmer decrement. Both functions have been plotted on the same scale in order to clearly demonstrate the offset caused by extinction. A comparison of the foreground de-reddened (Figure~\ref{Figure 6}) and internally de-reddened (Figure~\ref{Figure 6a}) plots shows a highly consistent shape in the function.

 Unfortunately, with most external
galaxies, it is almost impossible to correct each observed PN for the effects of internal extinction. Remarkably however, Figure~\ref{Figure 6a} implies that if the foreground de-reddened PNLF is simply shifted by applying 1 magnitude of extinction to each object, the internally de-reddened PNLF is recovered, so that the shape of the observed and de-reddened PNLFs are essentially identical. This result is rather surprising since although individual object extinctions vary between 0 $<$ \textit{c}H$\beta$ $<$ 1.3 (0 - 2 mag), from a global perspective, the PNLF responds as though each PN has \textit{c}H$\beta$ = 0.4. The magnitude range and shape of the plot (Figure~\ref{Figure 6a}), corrected for internal extinction, has only shifted $\sim$1 magnitude brighter than that corrected for line of sight reddening. It implies that the one dimensional PNLF of external galaxies can be modeled by simply applying a mean offset.

In order to observe the effects of internal dust extinction within the LMC PNe, the foreground de-reddened \OIII5007\AA~magnitudes are plotted against extinction in Figure~\ref{Figure 6b}. PNe in the brightest 4 magnitudes have low to medium extinction (0 to 0.6) while PNe with medium to high extinction ($>$0.6) are only found $>$\,4 magnitudes below the brightest. These highly reddened faint PNe probably have intrinsically fainter cores which have
evolved over a longer timescale. It is therefore possible that the
dust properties of these objects may be rather different to those of the higher-core mass
objects. Most of the PNe, across the full magnitude range, are located in the low to medium extinction range. This suggests that, regardless of the diminishing intrinsic luminosity as a consequence of age, a large proportion of these PNe with high-mass cores have evolved fairly rapidly and their surrounding circumstellar envelope, which provides
the extinction, have remained of similar opacity.

This is not to say that there has not been evolution within each nebulae such as decreasing electron density against changes in mass. Although the trends shown in Figure~\ref{Figure 6b} are strong, there are selection effects which may come into play. A further study including electron densities, masses and dynamical ages will be required in order to reveal these effects and understand the role played by circumstellar dust.

%\begin{figure}
%\begin{center}
 % \includegraphics[width=0.48\textwidth]{REID_Fig8a.eps}\\
  %\caption{The cumulative bright end of the planetary nebulae luminosity function to mag 17.2 for the central 25 deg$^{2}$ of the LMC, binned into 0.2 mag intervals and plotted in linear space. The linear least squares line of regression points to the limiting magnitude of the brightest possible PN within this system. This PNe is estimated to be at a magnitude of 14.02, indicated by the intersection of the trend line and the $\textit{x}$-axis.}
 % \label{Figure 8a}
  %\end{center}
 % \end{figure}

\begin{figure}
\begin{center}
  \includegraphics[width=0.48\textwidth]{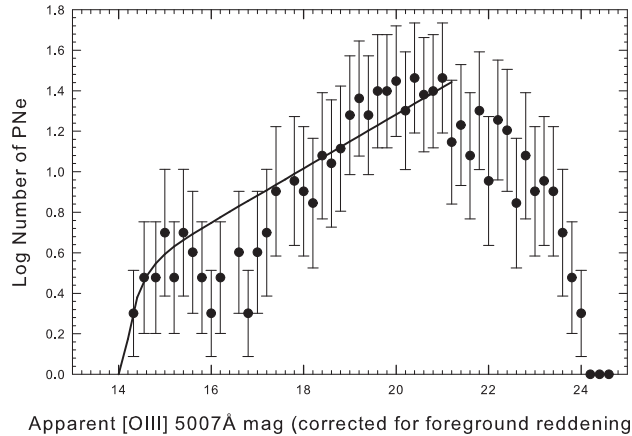}\\
  \caption{The planetary nebulae luminosity function for the central 25 deg$^{2}$ of the LMC, derived from an
  homogeneous sample of previously known and RP LMC PNe. In order to use the PNLF for distance determination,
  the data have only been corrected for line of sight reddening. The data have been binned into 0.2 mag intervals
  and plotted in log space. The solid line is the log of the truncated exponential curve (Equation 6)
  as predicted by Ciardullo et al. (1989) convolved for errors and placed at the best-fit position to the bright end of the
  observed function. %The broken blue line
  %represents the truncated exponential curve (Ciardullo et al., 1989) convolved for the errors shown in Figure~\ref{Figure 6}
  %and placed at the best fit position to the bright end of the PNLF.
  Poisson error bars are included. It is assumed that the decreasing number at magnitudes m$_{5007}$ $>$\,21 is due to incompleteness. This provides the first, direct estimation of $\textsl{M}^{\ast}$ for \OIII5007 in the LMC using a 3.4 magnitude range at the bright end of the LMC PNLF.}
  \label{Figure 7}
  \end{center}
  \end{figure}

 % \begin{figure}
%\begin{center}
 % \includegraphics[width=0.44\textwidth]{REID_Fig9.eps}\\
 % \caption{The previous planetary nebulae luminosity function for the LMC, from Jacoby et al. (1990b). The samples, binned to 2.5 magnitudes are probably complete to $\sim$2 magnitudes below the brightest PN. The PNLF includes 133 bright PNe from across the entire LMC not included at the bright end of the new PNLF (Figures 6 - 8). Also shown is the model curve for the PNLF from Ciardullo et al. (1989). Filled circles represent objects brighter than the completeness limit adopted by Jacoby et al. (1990b). Open circles are beneath this level and so were excluded from the fit.}
  %\label{Figure 8}
  %\end{center}
 %\end{figure}

 \begin{figure}
\begin{center}
  \includegraphics[width=0.48\textwidth]{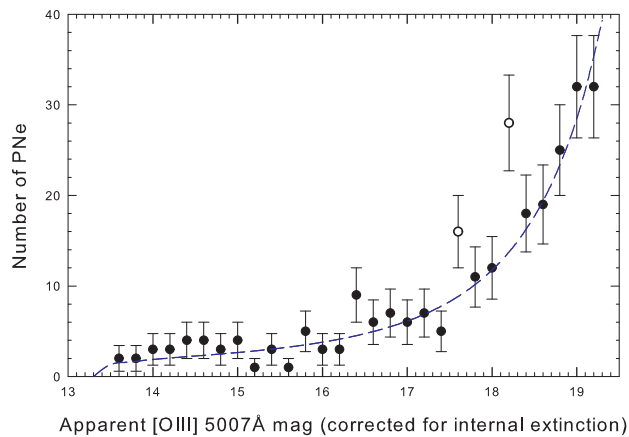}\\
  \caption{The bright end of the planetary nebulae luminosity function to mag 19.3 for the central 25 deg$^{2}$ of the LMC, binned into 0.2 mag intervals and plotted in linear space. In order to examine the true shape of the \OIII5007-based PNLF, the data are fully corrected for internal extinction. The broken line is the theoretical truncated exponential curve (Equation 6) as predicted by Ciardullo et al. (1989). Error bars employ poisson statistics. The curve matches the bright end of the function very well. Data bins that exceed 2$\sigma$ away from the theoretical curve are shown as open circles. This provides the first, direct estimation of the extinction corrected value of $\textsl{M}^{\ast}$ for \OIII5007 in the LMC using the bright end of the LMC PNLF.}
  \label{Figure 9}
  \end{center}
  \end{figure}

  \subsection{New PNLF-based distance estimate to the LMC}

The standard means of determining the distance to a galaxy using the PNLF has been previously confined to the brightest 1 or 2 magnitudes. We now extend this range to the brightest 3.4 magnitudes. With a good estimate of completeness to this magnitude depth, we avoid including the effects of any possible dip in the function, 4 magnitudes below the brightest.

%\textbf{To estimate the distance, we create a cumulative PNLF, covering the brightest 3.4 magnitudes (see Figure~\ref{Figure 8a}). The point at which the extended least squares line of regression crosses the \textit{x}-axis marks the position of the brightest possible PN in the system known as $\textsl{M}^{\ast}$. With this method, we estimate the apparent magnitude of $\textsl{M}^{\ast}$ to be 14.02.}

To estimate the distance, we use the $\chi$$^{2}$ method to fit a model PNLF to the observed PNLF corrected for foreground reddening (Figure~\ref{Figure 7}). The bright cut-off absolute magnitude ($\textsl{M}^{\ast}$)~has been estimated as $\textsl{M}^{\ast}$ = --4.44 $\pm$ 0.05 (Ciardullo et al. 1989, 2002; Jacoby et al. 1992). We adopt this value because it reflects the theoretical dependence of $\textsl{M}^{\ast}$ on metallicity (Dopita et al. 1992) and in order to make a direct comparison to previous LMC PNLF distance estimates using this value. This gives us a distance modulus of 18.46$\pm$0.2 for the LMC where the magnitude errors are used as systematic uncertainties for the distance modulus. The error estimate allows for errors in line measurement and calibration, a 1$\sigma$ error in the fit, uncertainties in the
distance calibrator galaxy, M31 and the shape model of the PNLF. The previous best PNLF for the LMC using real LMC PNe for a distance estimate is shown as Figure 4 in Jacoby et al (1990b). This figure includes the model exponential curve of Ciardullo et al. (1989). Figure~\ref{Figure 6} shows that we have only added one extra PNe to the bright end of the function. This being the case, it is not surprising that the distance modulus does not deviate greatly from the distance modulus of 18.44 previously
determined by Jacoby et al. (1990b).

Speculation has arisen as to whether the shape of the PNLF itself can be used for distance determinations (eg. Mendez \& Soffner, 1997). Only a deep and complete survey can answer this question.

Although it is extremely difficult to place an exact limiting magnitude on our survey data, we are confident that we are complete to
7 mag below the brightest. We are, therefore, in a good position to test if the truncated exponential curve
(Ciardullo et al. 1989) is a good predictor of PNLF shape and can be used as a distance indicator.
%The distance to the LMC is derived by fitting a linear trend line to the cumulative PNLF over the brightest 3.5 magnitudes. The intersection of this line with the x-axis indicates the position of $\textsl{M}^{\ast}$, the brightest possible PN in the system.

In order to test this, we fit the theoretical truncated
exponential from Ciardullo et al. (1989), to the brightest 6 magnitudes of the PNLF using the Levenberg-Marquardt fit method to achieve the best fit position. To check how well the observed PNLF follows the theoretical exponential curve (equation 6) we ran the Kolmogorov-Smirnov (KS), Anderson Darling and $\chi^{2}$ tests since all three work very well with binned data. Three assumed distributions were tested in order to find the best fit. The results represent values of the 0.2 magnitude binning and no poor fitting data points have been smoothed or excluded. The KS test, based on the largest vertical difference between the theoretical and empirical cumulative distribution function, found a statistic of D=0.11 down the brightest 6 magnitudes assuming an exponential distribution.

The threshold value of the significance level (P-value) of 0.677 between the observed and theoretical functions means the null hypothesis (H$_{0}$) that the observed distribution does not match the theoretical distribution must be accepted with a 95\% confidence limit. The major departure from the theoretical curve occurs 1.5 magnitudes below the brightest PN and lasts for 1.5 magnitudes. This may be referred to as a dip in the observed function and will be discussed in subsection~\ref{subsection 4.2}. Despite peaks and troughs in the observed function, if one were to estimate PN populations simply by integrating equation 6 over 6 magnitudes one would over estimate the population by only $\sim$3\%.
\begin{table}
%\small
\caption{Summary of the goodness of fit tests between the observed and empirical PNLFs.}
\begin{tabular}{|l|l|c|c|c|}
  \hline
  Extinction & Distribution & Kolmogorov  &  Anderson  & $\chi^{2}$ \\
             &               &  Smirnov  &  Darling  &             \\
  \hline\hline
  Foreground & Exponential  & 0.119 & 2.262 & 1.147 \\
  " & Normal & 0.229 & 2.072 & 3.684 \\
  " & Johnson SB & 0.074 & 11.119 &  N/A  \\
   Internal & Exponential  & 0.134 & 0.781 & 0.909 \\
   " & Normal & 0.257 & 2.523 & 4.418 \\
   "  & Johnson SB & 0.188 & 14.35  &  N/A \\
  \hline
\end{tabular}\label{Table 2}
\end{table}

%\textbf{If the intermediate region from 1.5 to 4 magnitudes below the brightest, including all evolutionary dips, were excluded, the KS statistic for a normal distribution would be D=0.16 with a P-value of 0.46 indicating a 50\% improvement in the fit. The $\chi^{2}$ test returns a statistic of 5.75 and P-value of 0.056 thereby improving the fit by a factor of almost 3 and allowing the null hypothesis to only be rejected at the average critical value (0.05). The excluded data, representing the deficit in the curve, has a KS statistic of 0.17 and P-value of 0.96 which is a very poor fit with 99\% certainty.}

%The theoretical and convolved exponential curves, \textbf{when fitted in this way to the number versus luminosity relation} over 6 magnitudes, give a direct estimate of 14 $\pm$ 0.28 for $\textsl{m}^{\ast}$ where they intersect
%with the $\textit{x}$-axis. \textbf{This represents the first time a distance estimate has been attempted by a direct fit of the theoretical truncated exponential curve. Figure~\ref{Figure 7}, however warns us that this curve is not a perfect fit. The mean agreement is 87\,$\pm$47\% over 6 magnitudes. The effects of reddening and evolutionary dips take their toll on any attempt to truly use this evolutionary path as a distance indicator.}
%\textbf{Nonetheless,} using our value $\textsl{M}^{\ast}$ --4.44 $\pm$ 0.05,
%and including the line-of-sight extinction towards the LMC, it follows that the distance modulus to the LMC using the exponential curve over 6 magnitudes is 18.44 $\pm$ 0.28.

The same objects, after correction for internal extinction, are plotted in linear space (Figure~\ref{Figure 9})
in order to fit and test the truncated exponential curve while examining the bright end with increased detail. Error bars represent poisson statistics. The same truncated exponential curve, again has a mean bin-to-bin agreement of 87\,$\pm$53\% against the observed data over 6 magnitudes. The goodness of fit tests between the observed and empirical PNLFs are summarised in Table~\ref{Table 2}. Clearly, the exponential increase in the number of PNe to be found with decreasing magnitude, first suggested by Henize \& Westerlund et al. (1963) is equally invalid once internal extinction is removed. The close statistical results are also further proof of the identical shape of the observed and internally corrected PNLFs, discussed in section 4.

%For most external galaxies, the current maximum likelihood and $\chi^{2}$ methods on foreground de-reddened data remain the most robust ways for determining distance.

%\textbf{The improved fit to the de-reddened function is especially true if we remove the two most outlying data points,
%marked as open circles in Figure~\ref{Figure 9}, and re-run the tests. The KS test then returns a statistic of D=0.29
%and P-value of 0.016 while the $\chi^{2}$ test returns a statistic of 5.20 and P-value of 0.07 thereby showing a very
%good agreement. This shows that the exponential increase in the number of PNe to be found with decreasing luminosity is
%a good approximation to the observed data once the effects of extinction are corrected. Unfortunately, with most external
%galaxies, it is almost impossible to correct each observed PN for these individual effects and so the current fitting
%methods remain the most robust ways for determining distance.}

An up to date compendium of distance estimates to the LMC is shown in Table~\ref{table 3}. These results represent a variety of calibration objects, methods and extinction estimates. The PNLF method has been shown to be a very robust method (Ciardullo et al. 2005) across all galaxies tested. Our confirmed  distance estimate of 18.46 $\pm$0.20 is in very good agreement with Cepheid and RR Lyrae distances derived from optical observations. It is also in close agreement with the previous PNLF distance estimate of 18.44 $\pm$0.18 as found by Jacoby et al. (1990) where the underlying universal luminosity
function is used as a probability distribution function. The estimated distance was given by the abscissa of graphed maximum likelihood solutions.

It is extremely encouraging that a direct fit to the bright end of the PNLF using our considerable new data over the first 6 magnitudes has produced a result in keeping with the previous LMC PNLF work using far fewer PNe. This important result shows that the bright end cut-off fit is robust to far less complete PN samples obtained for a galaxy and that its use as a standard candle is not compromised by modest sampling of the bright end of the PNLF.

The 1990 IAU LMC distance consensus of
50 kpc was presented by Mould (1990). Using an absolute
magnitude of -4.44 for M$^{\ast}$ in the LMC, the brightest PN would need to have an apparent magnitude of 14.05 to equal the 50 kpc distance. An increase in the value of M$^{\ast}$ would move the PNLF-based distance toward higher distance estimates. Our new PNLF distance indicator places the LMC at precisely 49.2\,kpc $\pm$0.2\,kpc, very close to 48.7\,kpc $\pm$2.32\,kpc, which is the mean distance and mean error found from all the distance results shown in Table~\ref{table 3}.

\begin{table*}
\begin{center}
\caption{A 20 year compendium of distance estimates to the Large
Magellanic Cloud including our new estimate using the LMC PNLF.}
\begin{tabular}{llll}
\hline\hline \noalign{\smallskip}
Method & Survey & Distance & Distance  \\
    &           &    Modulus   &  kpc     \\
  \hline\noalign{\smallskip}
  B stars & Shobbrock and Visvanathan (1987) & 18.3 $\pm$ 0.2 & 45.7$^{1}$ \\
  Cepheids & Feast and Walker (1987) & 18.47 $\pm$ 0.15 & 49.4  \\
  Cepheids & Welch et al. (1987) & 18.57 $\pm$ 0.05 & 51.7 \\
  Cepheids  & Visvanathan (1989) & 18.42 $\pm$ 0.04 &  48.3   \\
  Cepheids  &  Paturel et al. (1997) & 18.7 $\pm$ 0.02    &  54.9   \\
  Cepheids & Luri and Torra et al. (1999) &  18.35 $\pm$ 0.13 &  46.7 \\
  Cepheids & Bono et al. (2002)  &  18.53 $\pm$ 0.08  & 50.8  \\
  Cepheids & Keller and Wood (2006)  &  18.54 $\pm$ 0.018 & 51.0 \\
  LT Eclipsing Binary Systems & Pietrzy$\acute{n}$ski et al. (2009) &  18.50 $\pm$ 3\%  &  50.1 \\
  LPV \& Miras & Bergeat, Knapik and Rutily (1998) &  18.50 $\pm$ 0.17  &  50.1   \\
  Miras & Feast (1988) & 18.28 $\pm$ 0.6 & 45.3$^{1}$ \\
  Model Atmospheres & Eastman and Kirshner (1989) & 18.45 $\pm$ 0.28 &  48.9  \\
  MS fitting & Chiosi and Pigatto (1986) & 18.5 $\pm$ 0.1 & 50.1 \\
  MS fitting  & Schommer et al. (1984) & 18.2 $\pm$ 0.2 & 43.6$^{1}$ \\
  MS fitting & VandenBerg and Poll (1989) & 18.4  &  47.8  \\
  Novae & Capaccioli et al. (1990) & 18.70 $\pm$ 0.2 & 54.9 \\
  O stars & Conti, Garmany and Massey (1986) & 18.3 $\pm$ 0.3 & 45.7$^{1}$ \\
  PNLF & Jacoby, Walker and Ciardullo (1990) & 18.44 $\pm$ 0.18 & 48.7 \\
  PNLF & Dopita, Jacoby, Vassiliadis (1992) &  18.37 $\pm$ 0.15  &  47.2 \\
  \textbf{PNLF} & \textbf{Reid and Parker (this work) (2009)} & \textbf{18.46} $\pm$ \textbf{0.2} & \textbf{49.2} \\
  Red Clump stars & Grocholski et al. (2007) & 18.40 $\pm$ 0.04  &   47.8  \\
  RR Lyraes & Reid and Strugnell (1986) & 18.37 $\pm$ 0.15 & 47.2 \\
  RR Lyraes & Walker and Mack (1987) & 18.44 $\pm$ 0.05 & 48.7 \\
  RR Lyraes & Alcock, Alves, Axelrod et al. (2004) & 18.43 $\pm$ 1.6 & 48.5 \\
  RR Lyraes  &  Catelan and Cort$\acute{e}$s (2008)   &  18.44 $\pm$ 0.11  &  48.7  \\
  \hline
\end{tabular}\label{table 3}
\end{center}
$^{1}$ LMC short distances.
\end{table*}

\subsection{The shape of the PNLF}

Since the PN population in the LMC is now one of the most complete, observed samples at fixed distance available, modeling of the PNLF can be undertaken using this population. Although this study has extended the
luminosity range to far fainter limits than previously achievable, and
is therefore more complete, there still exists the
possibility that the most luminous PNe are over-represented
due to all surveys being flux-limited. The large number of PNe now available from the
 RP sample has a strong effect on the shape of the PNLF fainter than
 5 absolute magnitudes below the brightest. When examining the overall shape of the LMC PNLF, it is important to remember that the sample includes PNe that are both optically thick and optically thin, hydrogen burners and helium burners, type I and type II PNe (Peimbert 1978), young, middle-aged and old with different evolutionary characteristics and morphologies (Vassiliadis \& Wood, 1994; Dopita et al. 1992). It is beyond the scope of this paper to undertake evolutionary modeling of the PNe at this stage, however, the shape of the PNLF can now be examined in detail. Unless specified, we use the PNLF, corrected for internal extinction when discussing the shape of the PNLF and any evolutionary models.

The peak number of PNe
occurs at magnitude 19, which is half a magnitude fainter than half the magnitude range of the sample.
The distribution on both sides of the peak is rather different. The
bright end is relatively flat for the initial 3.5 mag before a steep
and almost constant rise to the peak. The faint end drops off far
more gradually with only three detections in the faintest bin.

%shape of faint end

The rapid rise at the bright end, from mag 17.5 ($\sim$4 mag below
the brightest) is a new major feature. It has not been seen before in an observed sample of PNe in any galaxy but was previously predicted due to the mean age of 3-8 Gyr for stars in the LMC (Stryker \& Butcher 1981; Frogel \& Blanco 1983; Hyland 1991). This places the largest number of PNe close to a luminosity of log (L/L$_{\odot}$) $\sim$ 3.65-3.75 in keeping with previous predictions (Dopita et al. 1992).

This rise in the function coincides with the faint end of a sudden dip or inflection
point found in some extragalactic studies (Ciardullo 2006). Importantly, we don't see a strong `dip' (also known as the `Jacoby dip') hinted at in the earlier but far more incomplete LMC PNLF (see Figure~\ref{Figure 6} and~\ref{Figure 6a}). Although there is a small decrease in the number of PNe between magnitudes 16.4 and 17.2 it is not a strong feature. Rather, the function is relatively flat until mag 17.2. The dip in the number of PNe occasionally seen at $\sim$3.5--4 mag below the bright cut-off (eg. Jacoby \& De Marco, 2002) has been attributed to the decline in luminosity of the
central star as it starts descending the White Dwarf (WD) cooling track (eg. Frew 2008). Alternatively, it may indicate that the PN evolutionary phase can occur at different luminosity levels for stars of different progenitor masses where 4 magnitudes below the brightest is representative of the largest group. It may also be evidence for the bimodal luminosity function expected from post-AGB stars (Vassiliadis \& Wood, 1994). Interestingly, the PNe populations sampled in
star forming galaxies exhibit a broad dip at 1.5-2 magnitudes below
the brightest. Both PNLFs (Figures~\ref{Figure 6}~and~\ref{Figure 6a}, corrected for foreground and internal extinction respectively, show some convincing evidence for the existence of this dip. This is characteristic of galaxies and models in
which the evolution of \OIII\,5007\AA~in PNe is governed by the
rapid evolution of a high mass core.

\begin{figure}
\begin{center}
  \includegraphics[width=0.5\textwidth]{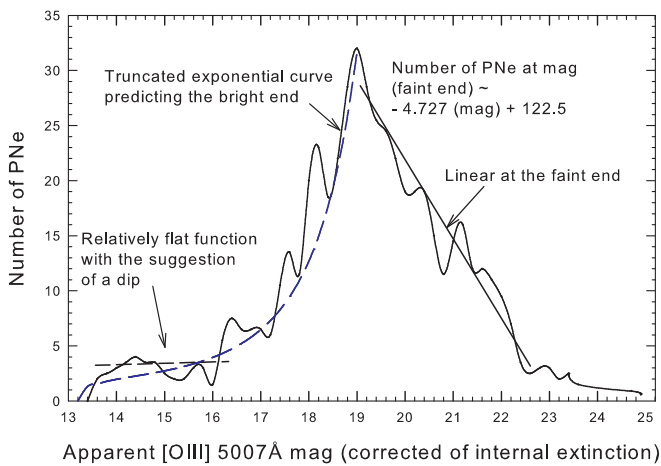}\\
  \caption{The planetary nebulae luminosity function for the central 25 deg$^{2}$ of the LMC in 0.4 magnitude bins. The overall shape of the PNLF begins to resemble that simulated by M\'{e}ndez et al (2008, Fig.7). The simulation resulted in a smooth distribution which may be close to the truth once all the bright PNe across the LMC are included in the PNLF. The truncated exponential curve (Ciadullo et al. 1989) has been included for comparison. Although we point out the broadly linear distribution at the faint end, no scientific conclusions can be drawn from this result as the faint end of the PNLF will suffer from incompleteness. It is however worth noting that our survey, to a depth of mag 25 in \OIII5007, shows this linear decline.}
  \label{Figure 10}
  \end{center}
  \end{figure}

By graphing the PNLF into 0.4 magnitude bins as a smoothed frequency distribution (Figure~\ref{Figure 10}), it becomes easier to identify any possible evolutionary dips in the shape of the PNLF. By including the truncated exponential curve we have a clearer impression of the dip between magnitudes 15 and 16. Overall, however, we see a relatively flat distribution over the brightest 3 to 4 magnitudes followed by a steep rise in the function. It agrees very well with the theoretically simulated PNLF created by M\'{e}ndez et al. (2008), where there is a sudden rise at the bright end followed by a very gradual decline for 4 magnitudes and a steep rise over 2 magnitudes to a peak in the function. The reason for the function remaining relatively flat for 4 magnitudes may be related to the bright LMC PNe being a combination of H-burning and He-burning central stars with a small proportion of low mass H-burners. Figure 7 in M\'{e}ndez at el. (2008) shows that the function steadily rises with no dip if the simulation includes too many central stars with masses as low as 0.55M$_{\odot}$. It conversely dips too low if there are too many high mass central stars. The relatively steady distribution over the brightest 4 magnitudes of the LMC PNLF found through this study adds weight to probability that central star mass plays a major role in shaping the bright end; especially if we accept that the steep rise $\sim$4 mags in represents the point at which the central star starts its descent down the WD cooling track.

 The number of PNe identified in our deep H$\alpha$~survey begins to decline at magnitudes m$_{5007}$
 $>$ 19. Is this due to incompleteness or is there a real turnover close to this point in the function? The decline begins rather steeply over the next 4 mag but clearly becomes increasingly incomplete in the faintest two
 magnitudes of the survey (mag 23 $\rightarrow$ 25). If the theoretical and empirical PNLFs (Jacoby 1989;
 Ciardullo et al. 1989) were employed beyond the peak magnitude of
 19 as found in this work, they would predict the presence of
 a large number of PNe by magnitude 25. Are they there? Is there sufficient mass in the
 LMC to permit the co-existence of tens of thousands of faint PNe at this end of the function?

 It would be very surprising if we have missed detecting hundreds of PNe more than 5 magnitudes brighter than our detection limit. Our PNe were selected using a deep H$\alpha$ filter (see RPa) which includes the \NII6548 and 6583\AA~lines. Deep PN surveys of the LMC have already been conducted using \OIII~filters with enormous success (RPa). Our survey using H$\alpha$ and \NII~has allowed us to discover a large number of PNe with extremely low \OIII~levels. Where \OIII~is low in abundance, \NII~is normally enhanced (RPa,b), permitting us to uncover a large population of PNe that would be very difficult to discover in an \OIII~survey. Even so, it is always possible that a deeper \OIII~survey and new extraction techniques may uncover more PNe. The only really hard
constraint on the number of PNe to be found in the LMC comes from stellar evolutionary theory. The
number of PNe shouldn't be greater than the product of the stellar
evolutionary flux (2E-11 stars yr$^{-1}$ L$_{\odot}$$^{-1}$) times the LMC bolometric luminosity
times the PN lifetime. Our previous work (RPb) estimates the existence of 956 $\pm$\,141 PNe in the LMC, in close agreement with an earlier estimation of 996 by Jacoby (1980). The derived luminosity-specific number of PNe ($\alpha$) in the LMC according to Buzzoni et al. (2006), is given as log $\alpha$ = -6.57 $\pm$\,0.04, representing 1040 $\pm$\,60 PNe.

 We therefore suggest that a turnover is likely to occur at some midpoint in the function. At this stage, we can safely say that equation 6 represents the bright end of the PNLF with a range of 6 magnitudes. After this initial 6 magnitudes and the peak in the PNLF, there is likely to be a flattening or even a decline in the number of PNe to be found. Only further deep surveys will be able to solve this issue conclusively.
\label{subsection 4.2}

 \subsection{Luminosity implied time}
\begin{figure}
\begin{center}
  \includegraphics[width=0.48\textwidth]{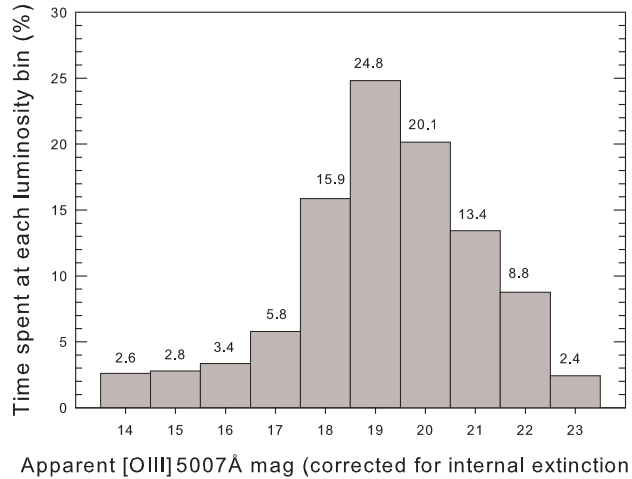}\\
  \caption{The percentage of PNe at each luminosity bin represents the percentage of time PNe spend at that luminosity according to predictions of Jacoby (1980); Ciardullo et al. (1989) and Ciardullo et al. (2004). This plot ignores the possible incompleteness of the faint end. It is also insensitive to any intrinsic variation and `birth-to-death' luminosity range for a given PN as progenitors range between 1 and 8 M$_{\odot}$. A simulated increase of the faintest 2 magnitudes would further decrease the percentages at the bright end.}
  \label{Figure 11}
  \end{center}
  \end{figure}

 If we accept that the number of PNe in each luminosity bin is generally proportional to the time those PNe spend at that luminosity (Jacoby 1980; Ciardullo et al. 1989; Ciardullo et al. 2004), then we may conclude that PNe evolve away from the brightest 3 magnitudes at a proportionally fast pace. The time spent at each luminosity bin is plotted in Figure~\ref{Figure 11} where each luminosity bin is 1 magnitude wide. This histogram ignores the possible incompleteness of the faint end. It is also insensitive to any intrinsic variation and `birth-to-death' luminosity range for a given PN as progenitors range between 1 and 8 M$_{\odot}$. Many PNe may have low-luminosity cores, and therefore join the luminosity function at fainter magnitudes. A simulated increase of the faintest 2 magnitudes would further decrease the percentages at the bright end. All this plot can effectively show is the number of PNe that may be found per luminosity bin at a single instant of time.

 Each bin has been scaled as a percentage of the number of PNe in the whole sample. Figure~\ref{Figure 14} shows that the percentage of PNe in the brightest 4 magnitudes accounts for only 14.6\% of all LMC PNe within the deep sample. In other words, 40\% of the magnitude range contains only 14.6\% of the PNe. It has been suggested that this drop in number density brighter than mag 18 is due to the negative slope of the initial mass function, especially effecting young stellar populations (Shaw, 1989) and the rapidly accelerating rate of evolution across the H-R diagram with increasing luminosity, particularly evident on the hydrogen-burning tracks (Dopita et al. 1992). Simulations conducted by M\'{e}ndez et al. (1993) suggest that these PNe at the bright end are also predominantly optically thin.

\subsection{The cumulative PNLF and comparisons with other galaxies}

In order to compare the depth of our derived PNLF with previous LMC PNLF models and PNLF in other galaxies, we create a cumulative distribution of the function. We present the cumulative function as a percentage of PNe found at each magnitude and compare this to model cumulative functions. This approach helps to reveal the positions at which incompleteness begins to set in. It also shows the way in which incompleteness affects the overall shape of the cumulative function.

\begin{figure}
\begin{center}
  \includegraphics[width=0.48\textwidth]{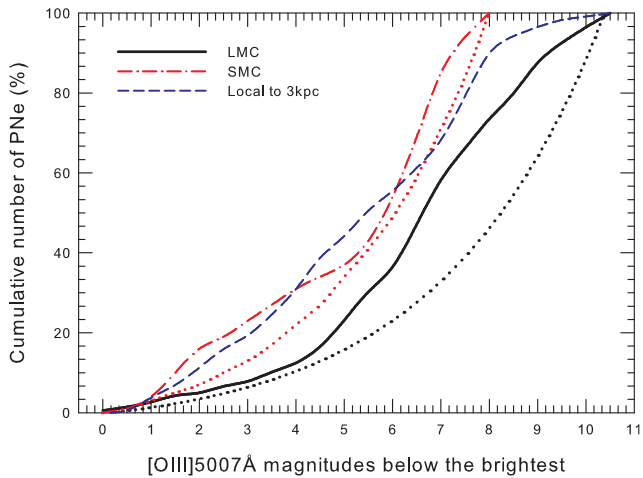}\\
  \caption{The cumulative PNLFs for \OIII5007 using the new PNLF for the central LMC (solid line - this work), the revised deep SMC PNLF of Jacoby (2006) and a deep local sample from Frew (2008), to a volume radii of 3.0kpc, centred on the sun. The dotted lines are the plots of Eq. (8) from M\'{e}ndez et al. (1993) with c$_{1}$ = 12 for the SMC, c$_{1}$ = 6 for the LMC and local PNLF to 3kpc, c$_{2}$ = 0.307 for both.}
  \label{Figure 12}
  \end{center}
  \end{figure}

The cumulative LMC PNLF for \OIII5007 is shown in Figure~\ref{Figure 12} where it is compared with the cumulative PNLFs calculated in the same way for the SMC, using data from Jacoby (2006) and for local PNe using the new highly complete local Galactic volume catalogue of Frew (2008) which includes PNe out to 3kpc. These objects are included in the MASH catalogue of Parker et al (2006). Although each PNLF is affected by incompleteness, compared to the LMC, the local and SMC samples have a larger percentage of PNe in the brightest 6 magnitude range (41\% for the LMC, compared to 54\% for the SMC and 55\% for the local PNLFs). The shape of the
bright end of the PNLF is therefore very sensitive to the relative
number of PNe included at the very brightest magnitude. It is also
sensitive to the peak number density, the flux limit of the
survey and the decreasing number of PNe at the faint end.

The large increase in the number of LMC PNe between 4 and the peak at 7 magnitudes below the brightest causes the LMC PNLF to rise away from the predicted exponential curve. With an identical magnitude range, the local PNLF shows a more constant rise. The SMC PNLF, with its smaller magnitude range, shows a small dip between 4 and 5 mag below the brightest, after which it also rises steeply. The steepest section of each cumulative function represents a rise to the peak in number density. This feature is easiest to see in the LMC PNLF, however, it's presence also in the SMC and between 6 and 7.5 magnitude ranges in the local PNLF suggests that it may be common to PN evolution in most galaxies.

%The best SMC PNLF currently available covers a 7.5 magnitude range but converges with the local PNLF between 4 and 6 magnitudes below the brightest. \textbf{This represents a dip, or decrease in the number of PNe discovered in the SMC between 4 and 5 magnitudes below the brightest.} A comparison of the LMC and local PNLFs indicates that either there are many more PNe to be found in the SMC between 3.5 and 5.4 magnitudes or the evolutionary dip occurs 1 magnitude fainter in the SMC than it does in the LMC.

 The final three magnitudes of the LMC PNLF show a gradual turnover towards the faint drop-off. A stronger turnover is also seen in the local sample, affecting the final 2 magnitudes, with only 8\% of PNe found in this range, prior to the faint drop-off. These turnovers are the result of incompleteness at the faint end, where we would expect the number of PNe to either rise or remain stable beyond the peak density ($\sim$6 mag below the brightest). We do not discount the possibility that a certain drop in number density towards the faint end may be a real feature suggesting that most PNe spend the majority of their lifecycle between 4 and 7 magnitudes below the brightest. The SMC PNLF only shows a turnover (drop in number density) in the final magnitude of its faint end. This short turnover would normally indicate that the SMC PNLF is more complete up to its final magnitude drop-off than the other two PNLFs. The SMC faint turnover, however, also coincides with the drop-off from the peak number density in the LMC PNLF. The LMC contains a further 228 PNe fainter than this magnitude. Therefore, the faint end of the SMC PNLF is actually less complete than the other two and we predict more PNe be found $>$4 mag below the brightest and 3 magnitudes deeper than currently available.

Using equation 6, M\'{e}ndez et al. (1993) have obtained the cumulative PNLF for any magnitude $\textit{M}$ fainter than the cutoff by integrating $\textit{N(M)}$ between $\textit{M}^{\ast}$ and $\textit{M}$ using the following formula:

\begin{equation}
K(M) = \frac{c^{1}}{c^{2}} [e^{c_{2}M} - e^{c_{2}M^{\ast}}] \\
- \frac{c_{1} e^{3M^{\ast}}}{(c_{2} - 3)} [e^{M(c_{2} - 3)} - e^{M\ast(c_{2} - 3)}]
\end{equation}

The dotted lines in Figure~\ref{Figure 12} represent this predicted cumulative PNLF from Eq. (8), adjusted to fit the magnitude range of PNe in the SMC and LMC. Comparing this predicted curve to the SMC PNLF suggests there is a lack of PNe in the range 4-6 below the brightest. This is affecting the overall shape of this PNLF and creating the upward `hump' from 1 to 4 mag below the brightest. The cumulative curve from Eq. (8), when fitted to the LMC, agrees very well at the brightest 4 magnitudes. After this, it does not follow the strong peak in the function but continues at a steady exponential rate. A larger number of PNe at the faintest 3 magnitudes would effectively bring the LMC PNLF closer to the predicted curve. The local PNLF also shares a 10 mag range in \OIII~and so shares the same predicted curve as our LMC PNLF, but traces out a path much further to the bright end and away from the simulated PNLF. This position for the local PNLF, when compared to the LMC PNLF either suggests that there should be a great many more faint PNe ($>$6 mag below the brightest) to be found within the local 3 kpc radii range or the local PNLF is affected by selection effects since it is only sampling a relatively small region of our Galaxy.

While the formula of Ciardullo et al. (1989) results in a constantly
increasing PNLF towards fainter magnitudes, the simulated PNLF of M\'{e}ndez et al. (1997) produces two rises and declines.
The first decline is between \OIII\,5007\AA~magnitudes
-3.5 and -2.5 (M\'{e}ndez et al. 1997, fig. 4). It then rises again from mags fainter than -2.3 until it reaches a peak at $\sim$--0.2. From this point, it begins a steady decline. This simulation however was re-worked to allow for PNe on cooling tracks by including a random distribution of the absorbing factor between 0.1 and 1. With this adjustment, the simulation fits the formula of Ciardullo et al. (1989) very well (see M\'{e}ndez et al. 1997, fig 5). This was a prediction, where severe
 incompleteness hampered any true test with observations. The simulation of M\'{e}ndez et al. (1997, fig. 5) after a correction to allow absorbing factors to tend to 0, shows a peak in the function at mag $\sim$6 and a faint end decline from that point. This is exactly what is seen in the LMC PNLF using the RP sample (Figure~\ref{Figure 12}). This in turn strengthens the case for a strong rise in the function from
 4 magnitudes below the brightest. This rise is the result of an evolutionary change where a young, high mass and luminous PN reaches a peak in its photoionisation potential. It then shifts towards the WD cooling track with increasingly lower density to mass ratios. The sudden rise in number density shows that this change occurs rapidly.

 The extension of the PNLF to fainter magnitudes describes complete
population effects as predicted by the Shklovski model (1981) and
permits us to observe what effect the faint PN population may reveal
for successful PNLF modeling. It is hoped that this will in turn
help to further refine PNLF simulations. %The faint end
%of the central LMC sample however is best described as linear. This ignores large variations in the number of PNe found in consecutive bins. The fitted equation that adequately describes the faint end has been determined as:

%\begin{equation}
%N_{PNe}(mag) = - 4.727 * M_{5007} - 122.5 .
%\end{equation}

%This equation implies that once PNe pass the point of their maximum
%number density (an absolute \OIII5007\AA~magnitude of $\sim$0.8 for the LMC, 6 mag below the brightest), the
%decline to fainter magnitudes is quite consistent despite the large bin-to-bin variations. This is in contrast
%to the more sudden increase in the number of PNe at 4 magnitudes
%below the brightest.

 \label{section4.1}

\section{The potential of the LMC H$\beta$ PNLF}

The PNLF is essentially based on the evolution of
the central ionising star (Jacoby, 1989). In particular, the H$\beta$ intensity is
a measure of the central star's ionising luminosity (Osterbrock,
1989) and can also indicate the heating of the nebula as long as it
is optically thick. This would indicate that a hydrogen emission line would also be
a natural choice for a PNLF. Since the evolution of the central
star's luminosity gives rise to the PN luminosity function, there
should be a correlation between the relative strengths of the
integrated \OIII\,5007\AA~and H$\beta$ fluxes. The H$\beta$ line is also less influenced by the metallicity of the parent galaxy, which could make it an ideal choice for distance determination. A PNLF was constructed using H$\beta$ in order to discover the bright cutoff and shape of the function. A comparison with the \OIII5007 PNLF should then reveal broad influences of the LMC's low metallicity. The H$\beta$ flux by comparison is typically $>$\,5 times fainter than \OIII5007 in most PNe making it much more difficult to spectroscopically measure in
external galaxies. This is why H$\beta$ is not generally used for distance determination. H$\beta$ has a bright cut-off at about 2.4 mag fainter than that of \OIII\,5007\AA.

\begin{figure*}
\begin{center}
  \includegraphics[width=0.82\textwidth]{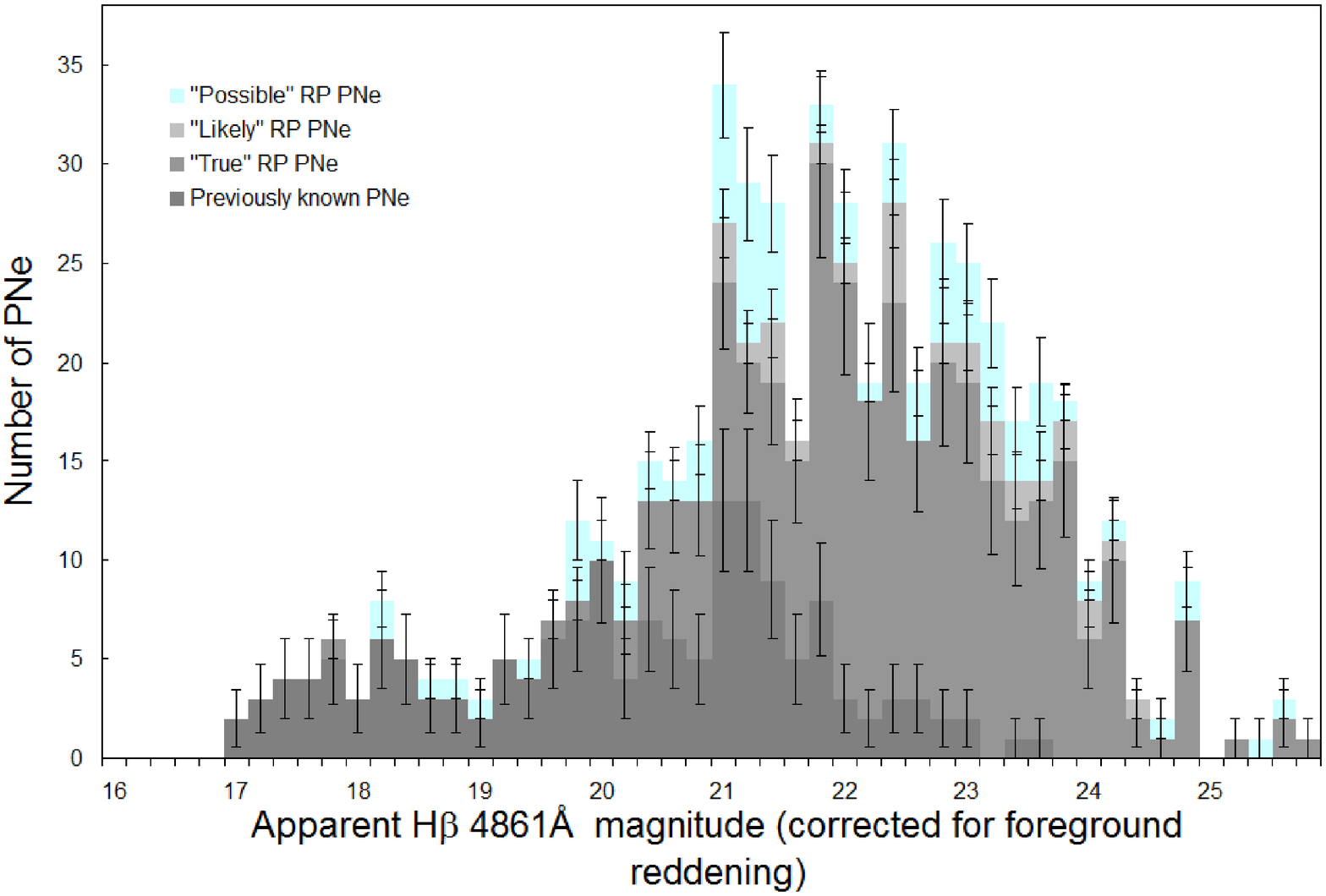}\\
  \caption{The H$\beta$ planetary nebulae luminosity function for the central 25 deg$^{2}$ of the LMC, with PNe separated into the categories of previously known, true, likely and possible PNe in the survey area for this study. Only foreground de-reddening has been applied to the raw magnitudes. The bins are 0.2 mags in width. Poisson error bars are included. The errors are calculated individually for previously known, true, likely and possible PNe as shown in Tables~\ref{table 4} \& \ref{table 5} and then combined for each magnitude bin. }
  \label{Figure 13}
  \end{center}
 % \end{figure*}
%\begin{figure*}
\begin{center}
  \includegraphics[width=0.82\textwidth]{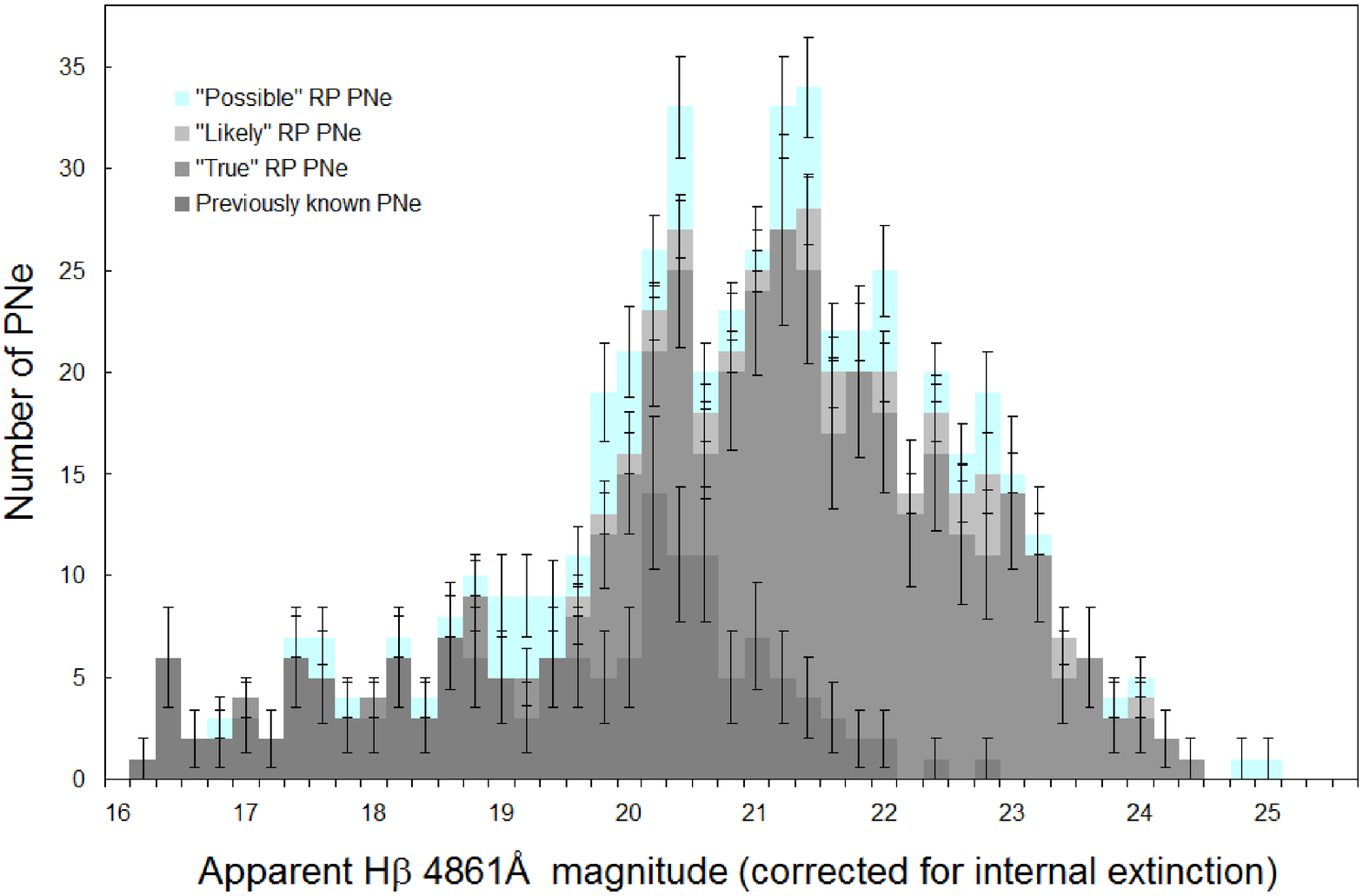}\\
  \caption{The H$\beta$ planetary nebulae luminosity function for the central 25 deg$^{2}$ of the LMC, with PNe separated into the categories of previously known, true, likely and possible PNe in the survey area for this study. A correction for internal extinction has been applied to each PNe using the Balmer decrement. The bins are 0.2 mags in width. Poisson error bars have been included. The errors are calculated individually for previously known, true, likely and possible PNe as shown in Tables~\ref{table 4} \& \ref{table 5} and then combined for each magnitude bin.}
  \label{Figure 13a}
  \end{center}
  \end{figure*}

For optically thick PNe, the H$\beta$ flux should be a reliable indicator of stellar luminosity. The maximum conversion efficiency from stellar luminosity to luminosity in the H$\beta$ line only varies by a factor of 2, even though a maximum of just under 1\% of stellar luminosity is converted into H$\beta$ photons (Dopita et al. 1992). Dopita et al. (1992) have found the variation in this conversion efficiency to be even less.

In order to convert our H$\beta$ fluxes, expressed in ergs cm$^{-2}$ s$^{-1}$, to magnitudes, we use the definition given by Jacoby (1989):

\begin{equation}
\textit{m}_{f}= -2.5~\textrm{log}~f-13.74
\end{equation}

where $\textit{f}$ represents the flux to be roughly converted to the broad V-band magnitude. The close interval between
the \OIII5007\AA~and H$\beta$~lines allows use of the same definition to convert fluxes from ergs cm$^{-2}$ s$^{-1}$, to magnitudes for both lines. This technique, previously adopted by M$\acute{e}$ndez et al. (1993), allows us to make a direct comparison between \OIII5007\AA~and H$\beta$~magnitudes on the same scale.
The luminosity function for the H$\beta$~line is shown in Figure~\ref{Figure 13} for foreground de-reddened objects where the brightest PN is found at an apparent magnitude of $\textit{m}_{4861}$ = 16.8 In Figure~\ref{Figure 13a} the PNLF is shown following correction for internal extinction where the brightest PN is found at an apparent magnitude of 16.2.

\begin{figure}
\begin{center}
  \includegraphics[width=0.48\textwidth]{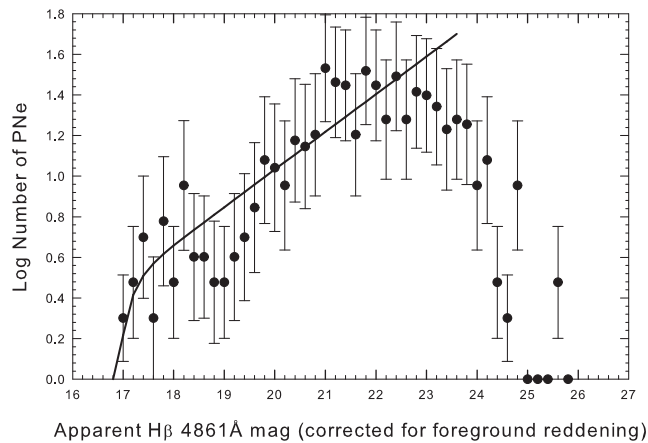}\\
  \caption{The H$\beta$ planetary nebulae luminosity function for the central 25 deg$^{2}$ of the LMC, derived from a homogeneous sample of LMC PNe with m$_{4861}$ $<$27.0. The data have been binned into 0.2 mag intervals and plotted in log space. Error bars are based on poisson statistics. The solid line is the truncated exponential curve as predicted by Ciardullo et al. (1989) convolved for errors and placed at the best fit position. It marks out magnitude 16.8 as the position of m$^{\ast}$. The curve matches the brightest 6 magnitudes of the function quite well. It is assumed that the falloff at magnitudes m$_{4861}$ $>$ 24 is due to incompleteness.}
  \label{Figure 14}
  \end{center}
  \end{figure}

In Figure~\ref{Figure 14} we show the H$\beta$~PNLF as a log plot with the truncated exponential curve of Ciardullo et al. (1989) added for comparison. The curve has been placed at the best fit to the bright end, using the Levenberg-Marquardt fit method. Using the bright intersection of the curve with the $\textit{x}$-axis, a value of M$^{\ast}_{4861}$ = 16.8$\pm$0.21 was found. This corresponds to an absolute magnitude of -1.68$\pm$0.21 for M$^{\ast}_{4861}$. Statistically, the goodness of fit tests to the theoretical exponential curve give F=1.31, a KS statistic of 0.13 and $\chi^{2}$ statistic of 2.85 providing 95\% confidence limits that the observed PNLF is unlikely to be drawn from the empirical one. At present there is little data on extragalactic H$\beta$~fluxes available with which to compare these results. Stanghellini (1995) constructed an H$\beta$ PNLF for the Magellanic Clouds based on the fluxes of Richer (private communication) together with a simulation of 1000 optically thick PNe using Salpeter's IMF and Weidemann's initial mass-final mass relation, with M$_{max}$ = 0.7\,M$_{\odot}$ (Weidemann 1987). This data produced very encouraging comparisons, however, variations in the mass truncation produced significantly different results. Using M$_{max}$ = 0.7\,M$_{\odot}$, the peak in the distribution occurred only 1 magnitude below the brightest PN. Increasing M$_{max}$ to 1.4\,M$_{\odot}$ shifted the peak 2.2 magnitudes fainter but the entire distribution only covered a 4 magnitude range.

\begin{figure}
\begin{center}
  \includegraphics[width=0.48\textwidth]{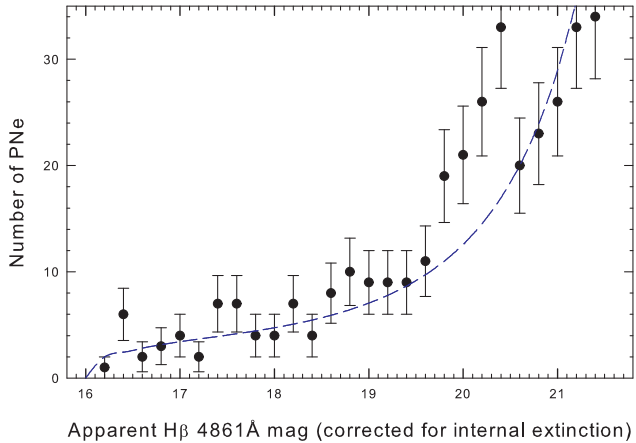}\\
  \caption{The bright end of the H$\beta$ planetary nebulae luminosity function to mag 21.4 for the central 25 deg$^{2}$ of the LMC, binned into 0.2 mag intervals and plotted in linear space. The broken line is the truncated exponential curve as predicted by Ciardullo et al. (1989). Error bars are based on poisson statistics. The curve matches the bright end of the function well. This provides the first, direct estimation of H$\beta$ $\textsl{M}^{\ast}$ for the LMC using the bright end of the LMC PNLF.  }
  \label{Figure 15}
  \end{center}
  \end{figure}

The bright end only is plotted in linear space in Figure~\ref{Figure 15} using the same data corrected for internal extinction. Analogous to the \OIII5007\AA~function, the brightest PN in the H$\beta$ function also increases 0.8 magnitudes after correction. Assuming an exponential growth in the function, and comparing the empirical PNLF of Ciardullo et al. (1989), the KS statistic is 0.122 and the $\chi^{2}$ statistic is 5.0. This indicates that the observed PNLF is unlikely to be drawn from the extrapolated one at the 90\% confidence level.

Improved simulations of the empirical PNLF were presented for both the \OIII5007\AA~and H$\beta$ lines by M\'{e}ndez and Soffner (1997). By varying the estimated fraction of stellar ionising luminosity absorbed by the nebula, better known as the absorbing factor $\mu$, and using available evolutionary tracks to produce a good representation of varying $\textit{T}_{eff}$, mass and log L for 3000 100 year intervals between 0 and 30,000 years, they produced good representations to the standard predictions of Ciardullo (1995). Their Fig.6 simulation of the H$\beta$ PNLF shows a peak in the distribution at precisely the same position of M = 3.8 found in this study using real LMC PNe. This remarkable match to model predictions proves that not all PNe on cooling tracks can have $\mu$ = 1, since that would create a strong hump at the peak and make fainter PNe extremely hard to find. The values of $\mu$ most likely vary between 0.1 and 1 between ages of 1 and 30,000 years.


  \label{section5}

\subsection{PNe with low central star temperatures}

%The next paper in this series will discuss excitation classes, stellar and nebulae temperatures.
It is important to note that we have found 3 PNe (SMP numbers 64, 31, 55) with very high H$\beta$ fluxes belonging to excitation class 1, the lowest category (Reid \& Parker, 2009). In RPb, SMP64 and SMP31 were classified as possible PNe belonging to a class known as Very Low Excitation (VLE) PNe. They have central stars with $\textit{T}$$_{eff}$ so low (31,200\,K to 38,500\,K) that the \OIII5007\AA~is extremely low by comparison to H$\beta$ (details in RPb). A further 4 PNe (SMP67, 77, 76 and MG62) also have \OIII5007/H$\beta$ flux ratios $\sim$3, low excitation classes from 2 to 3 and $\textit{T}$$_{eff}$  between 46,500\,K and 64,200\,K. The presence of VLE PNe has previously been reported by M\'{e}ndez et al. (1993) and can be seen in Fig. 4a of Doptia et al. (1992) for several Magellanic Cloud PNe.

In order for the H$\beta$ luminosity to remain constantly high in these PNe, the number of ionising photons from the central
star must increase by a factor of 2.5 as the temperature increases from $\textit{T}$$_{eff}$ 30,000\,K to 70,000\,K. If these
nebulae were optically thick, we would expect a proportional increase in the H$\beta$ luminosity with increasing central star
temperature. To keep these nebulae constantly bright on the horizontal portion of the post-AGB evolutionary tracks,
M\'{e}ndez et al. (1993) suggest they must be optically thin in the H Lyman continuum. The fraction of stellar ionising
luminosity absorbed by the nebula, defined as the absorbing factor, $\mu$, (M\'{e}ndez et al. 1992) may also play a
defining role here. Another possibility is that many of these central stars are burning He rather than H. This might account
for a decrease in luminosity as $\textit{T}$$_{eff}$ increases. %It is beyond the scope of this work to explore these options
%but our next paper will present nebula diagnostics and central star temperatures.

\subsection{\OIII5007~vs H$\beta$ magnitudes}

Comparisons have been previously drawn between the \OIII5007\AA~and H$\alpha$ + \NII~line ratios in a given population of extragalactic PNe (eg. Ciardullo et al. 2002, 2004). The H$\alpha$, being the brightest line in the Balmer series, is the easiest hydrogen line to measure in extra-galactic PNe. Unfortunately, due to filter responses, it usually includes the adjacent \NII6548\AA~+ 6583\AA~lines. The LMC is close enough that the H$\beta$ line is easy to measure using our spectroscopic observations. Since it is cleanly separated and yet relatively close to the \OIII5007~line in the optical spectrum, the ratios will be less affected by errors in flux calibration and extinction estimates.

The shape of the H$\beta$ PNLF has strong similarities to that of \OIII~Figure~\ref{Figure 6} but there some differences. Once again there is a peak to the distribution, however, it is not as strongly marked as its counterpart in \OIII. With the H$\beta$ peak occurring at magnitude 22.4 (absolute mag 2.6), 5 magnitudes below the brightest, 48\% or 287 PNe have fluxes occurring at the bright end. This may be compared with the \OIII~function where the peak occurs at magnitude 20.4 (absolute mag 0.4), however only 242 or 41\% of PNe are found at the bright end of the peak. This is despite the \OIII~peak occurring 5.2 magnitudes below the brightest, which is relatively 0.2 magnitudes fainter than the peak in H$\beta$. In general terms, across a broad range of evolution, this suggests that the H$\beta$ flux increases and fades at a steadier rate than \OIII~flux. The \OIII~flux appears to be highly sensitive to evolutionary spikes and dips whereas the H$\beta$ flux is more sensitive to the opacity of the nebula. It also suggests that the central star evolves over the dynamic lifetime of a PN as it ionises the uniformly expanding shells (Henize \& Westerlund, 1963), affecting temperature, luminosity and excitation of the nebula. These parameters will be closely examined in a further paper in this series.

\begin{figure}
\begin{center}
  \includegraphics[width=0.48\textwidth]{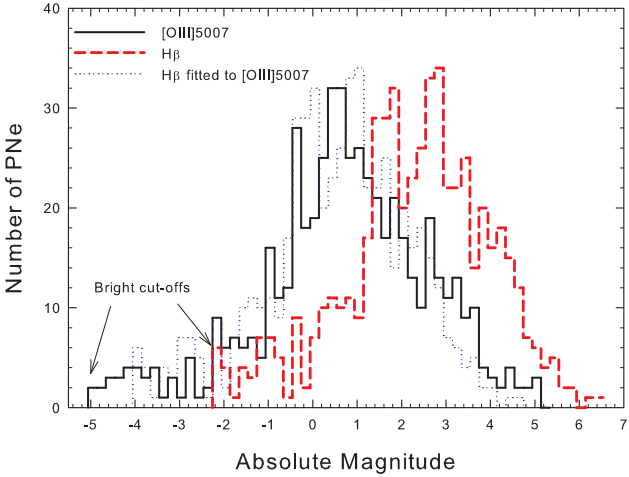}\\
  \caption{The planetary nebulae luminosity functions of H$\beta$ and \OIII5007, corrected for extinction, are directly compared using all PNe uncovered in the central 25 deg$^{2}$ of the LMC. The data have been binned into 0.2 mag intervals and plotted according to their measured and corrected magnitudes. The bright cut-off for H$\beta$ occurs at magnitude 16.8, 2.4 magnitudes below that of \OIII5007. At the peak of the distributions (shown) the $\Delta$magnitude (H$\beta$-\OIII5007) is reduced to 1.8 magnitudes and remains at approximately this level to the faint cut-off.}
  \label{Figure 16}
  \end{center}
  \end{figure}

All the previously known and new
PN in this survey are included in Figure~\ref{Figure 16} where the
\OIII\,5007\AA~magnitudes are plotted against the
H$\beta$ magnitudes, both de-reddened and corrected for extinction. The correction between the
line fluxes is generally very small. Even for a huge reddening of
$\textit{E(B--V)}$=1.0, the differential reddening between
\OIII\,5007\AA~and H$\beta$ is less than 0.06 dex (Schild, 1977).

Figure~\ref{Figure 16} shows that the \OIII\,5007\AA~line intensities span $\sim$10 magnitudes. The H$\beta$~line intensities on the other hand only span $\sim$8.5 magnitudes. A faint dotted line included in Figure~\ref{Figure 16} represents the H$\beta$ function, moved to the best fitting position with the \OIII5007 function. Statistically, on a bin-to-bin basis, there is an agreement of 82\% between them. This, however, should be viewed only in terms of overall shape of the PNLF since \OIII5007/H$\beta$ in individual PNe may vary by as much as 2.5 magnitudes. With this in place, it becomes apparent that \OIII5007 fluxes extend approximately 1 magnitude brighter than the bright end of the H$\beta$ function. Assuming an exponential distribution, the KS statistic is 0.10 and the $\chi^{2}$ statistic is 9.72 indicating a 92\% probability that the two functions do not correspond. Nonetheless, we cannot statistically exclude the possibility that the two samples are essentially the same shape.

At the bright end, the H$\beta$~PNLF like that of \OIII\,5007\AA~starts to climb almost immediately without any strongly defined dips but a series of peaks and troughs. This may indicate that the central star's progression onto the WD cooling track has little immediate impact on stellar luminosity. Although the stellar temperature begins to fall, causing a dip in the \OIII5007\AA~line, the H$\beta$~flux is maintained longer, probably due to the higher ionisation potential of H.

\begin{figure}
\begin{center}
  \includegraphics[width=0.48\textwidth]{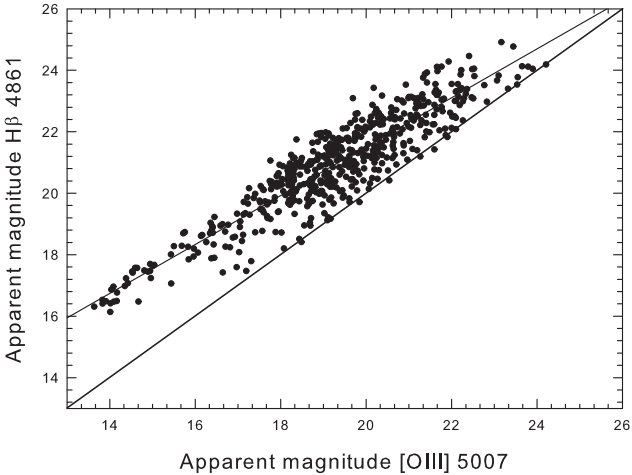}\\
  \caption{A direct comparison of \OIII5007\AA~and H$\beta$4861\AA~apparent magnitudes corrected for internal extinction shows a steady regression towards the line of equality with fainter magnitudes. The faint trend line marks out this regression while the darker line represents the line of equality where the 5007\AA~and 4861\AA~magnitudes would be equal. PNe in the brightest 4 magnitudes mainly exhibit high 5007/4861\AA~ratios.}
  \label{Figure 17}
  \end{center}
  \end{figure}

 Although the transition probability (3 to 1) between the \OIII5007\AA~and 4959\AA~lines (Osterbrock \& Ferland, 2006) remains constant, our measurements of PNe in the LMC show that the H$\beta$ flux varies widely by comparison to these lines. Combined \OIII5007\AA~+ 4959\AA~fluxes lie between 62$\times$ and 0.9$\times$ the strength of H$\beta$ PN fluxes in the LMC survey sample. Figure~\ref{Figure 17} directly compares the measured fluxes. A fitted trend line shows a steady regression towards the line of equality with fainter magnitudes. PNe at the brightest 4 magnitudes have consistently high 5007/4861\AA~ratios. The marked lack of low 5007/4861\AA~ratios in this region suggests that these PNe are young, optically thick, medium to high excitation PNe with central stars which have not yet reached the WD cooling track.

 PNe with low \OIII~fluxes, -0.9 to 2 times H$\beta$ are found in both faint, highly evolved PNe and within magnitude ranges fainter than 18. Other diagnostics such as high \NII6583/H$\alpha$~together with morphology, when available from HST images, help to distinguish these candidates as possible Very Low Excitation (VLE) PNe. The mean ratio (\OIII5007,4959/H$\beta$) across the entire central 25deg$^{2}$ survey region is 8.93$\pm$7.31. By far, the majority (94\%) of \OIII5007 + 4959~fluxes are below 20 $\times$~H$\beta$~fluxes.

%\begin{figure}
%\begin{center}
 % \includegraphics[width=0.48\textwidth]{REID_Fig17a.eps}\\
  %\caption{The \OIII5007\AA~/H$\beta$4861\AA~flux ratio is shown as a function of log \OIII5007\AA~flux. The plot shows a broad regression towards the line of equality with decreasing fluxes.  }
 % \label{Figure 17}
 % \end{center}
 % \end{figure}

The H$\beta$ bright cutoff occurs 2.4 magnitudes fainter than the equivalent position for \OIII5007\AA. Figure~\ref{Figure 17} shows that PNe with the brightest H$\beta$ fluxes also have strong \OIII5007\AA~fluxes. In fact, sampling the brightest 50 H$\beta$ fluxes, the average \OIII5007\AA~flux is 8.9 times H$\beta$. These include mostly medium and some high excitation PNe with central star temperatures between \textit{T}$_{eff}$ =31,800\,K and 136,000\,K.

\begin{figure*}
\begin{minipage}[h]{8.8cm}
\psfig{file=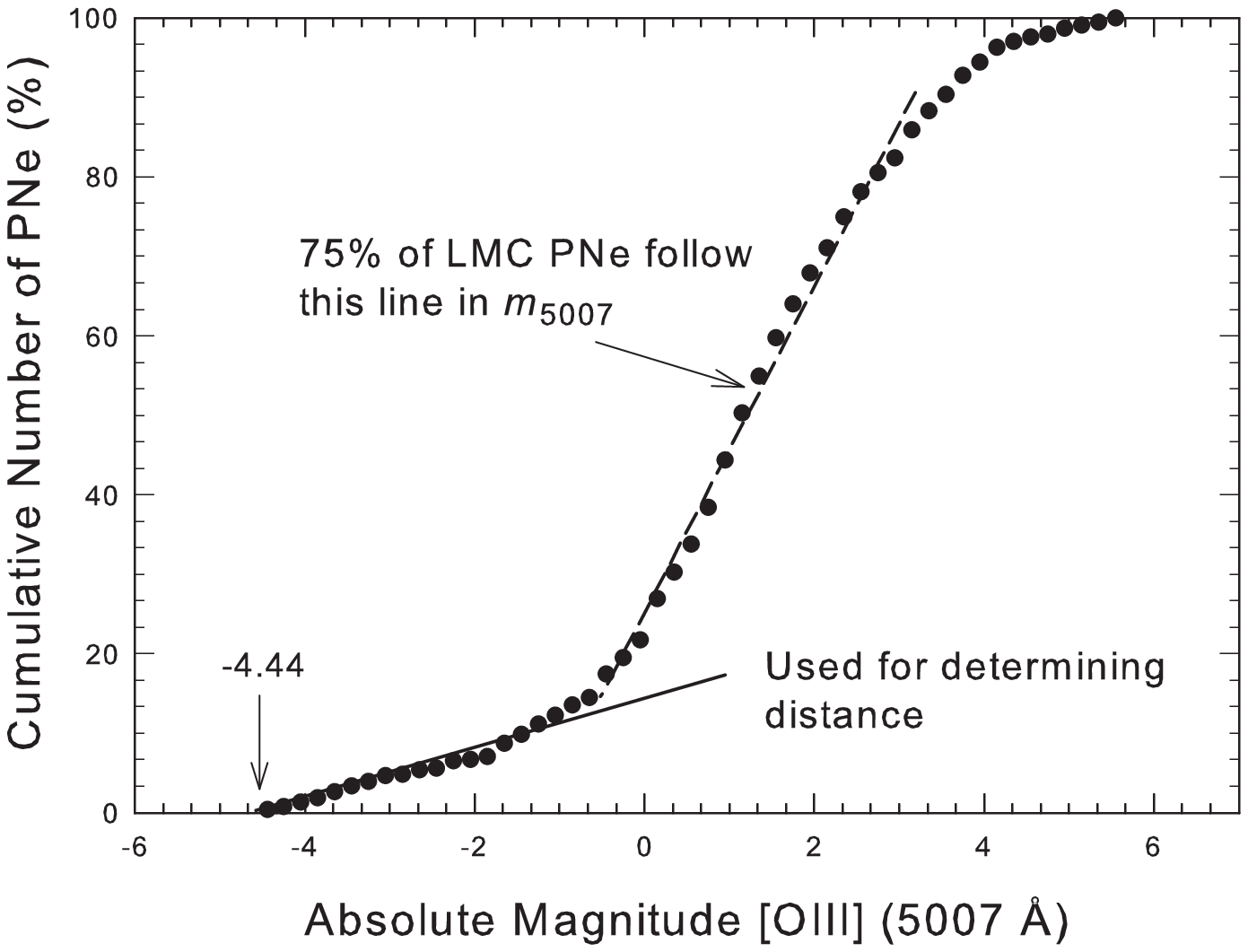,angle=0,height=5.8cm,width=7.7cm}
%\hfill
\end{minipage}
\begin{minipage}{8.7cm}
\psfig{file=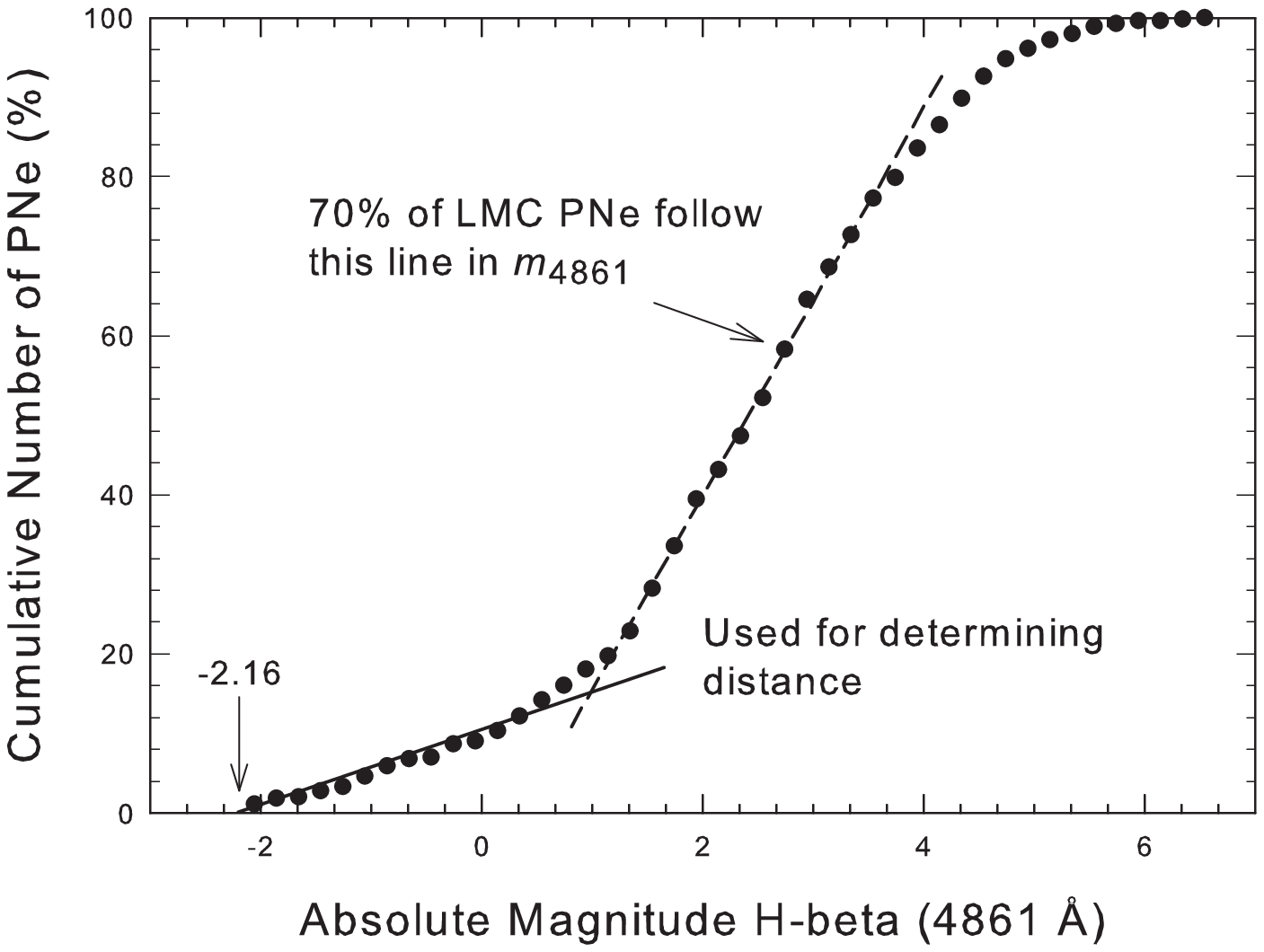,angle=0,height=5.8cm,width=7.7cm}
 \end{minipage}
\caption{The cumulative PNLF using the \OIII5007~line (left) and H$\beta$4861 line (right). The bright end of the PNLF, traditionally used for distance determination, is marked on both wavelengths. Both lines equally contribute to this section of the function, making either a good choice for modeling the bright end and distance estimations. The stronger peak in the \OIII5007\AA~function results in the oscillation seen either side of the arrow, to the right and left of the central line.}\label{Figure 18}
\end{figure*}

The H$\beta$ PNLF shows the same rather flat distribution seen across the initial 2 brightest magnitudes of the \OIII5007\AA~PNLF. The initial peak in the H$\beta$ distribution is only 1.8 magnitudes fainter than the equivalent position for \OIII5007\AA. Similarly the faintest PNe in each wavelength are generally 1.8 magnitudes in separation. Both of these PNLF estimates agree closely with the average magnitude difference of 1.7 $\pm$0.9 by directly comparing 591 LMC PNe. By comparison, if we average the magnitude difference between \OIII5007\AA~and H$\beta$~for the brightest 30 PNe in H$\beta$, the result is a much higher 2.3 $\pm$0.6 mag difference. The flat distribution at the bright end of the \OIII5007\AA~PNLF is therefore the result of a bright extension, peculiar to the bright end of the \OIII5007\AA~PNLF. This extension is an average 0.6 mag greater than the average difference for the remaining 95\% of PNe. These results, despite the evolutionary effects, indicate
that the temperature of the central star is the most dominant contributor to the PN
luminosity function (eg. Jacoby 1989), giving rise to both the H$\beta$ and \OIII5007\AA~luminosity.

The raw cumulative luminosity functions derived from the \OIII~and H$\beta$~magnitudes are shown in Figure~\ref{Figure 18}. In each case, the bright end of the functions are fitted with a straight line, indicating the upper luminosity cutoff. This line traditionally represents the observational basis for distance estimates once corrections for reddening and luminosity efficiency of the \OIII~or H$\beta$ line photons are made. The graphs clearly show however, that most PNe follow the steeper slope which represents the greatest growth, peak and fall in the function. The number of PNe on this line is equal in \OIII5007\AA~and H$\beta$ (70-75\% for both). The stronger peak in the \OIII5007\AA~function results in the oscillation seen either side of the arrow, to the right and left of the central line.

Tables~\ref{table 4} and~\ref{table 5} in the appendix provide a compendium of H$\beta$ and \OIII5007\AA~raw flux estimates for previously known and newly discovered RP PNe respectively in the Large Magellanic Cloud. The tables also include the measured values of \textit{c}H$\beta$ and the estimated errors. Table~\ref{table 5} includes the PN object probability where `T' represents `True', `L' represents `Likely' and `P' represents `Possible'.

  It is anticipated that once the RP survey is
  extended to the outer LMC, a similar shape for the PNLF to that newly
  obtained for the central 25 deg$^{2}$ would be found. The increase in total numbers should create a much tighter fit to the exponential curve, providing a very accurate distance estimate. The combination of previously known and newly discovered PNe in this study currently provide the most accurate estimate for the shape of the LMC PNLF obtained thus far.

\label{section5.1}

%%%%%%%%%%%%%%%%%%%%%%%%%%%%%%%%%%%%%%%%%%%%%%%%%%%%%%%%%%%%%%%%%%%%%%%%%%%%%%%%%%%%%%%%%%%%%%%%%%%%%%%%%%%%%%%%%%%%%%%%%%%%%%%%%%%%%%%%%%%%

\section{Conclusion}

Due to the depth of the UKST multi-stacked H$\alpha$ survey, the subsequently uncovered large RP PN sample is able to show the influence of population effects on the
shape of the PNLF for the first time. We present details of a new technique for the flux calibration of fibre-based spectra. Using this technique, we provide carefully calibrated \OIII5007\AA~and H$\beta$ flux estimates from 2dF and FLAMES spectra together with fluxes from calibrated longslit spectra for 584 LMC PNe. These fluxes are converted to magnitudes and used to construct a PNLF in \OIII~and in H$\beta$. The PNe are separated into previously known
and newly discovered PNe where a significant increase in the number
of LMC PNe below an apparent magnitude of 18.6 in \OIII5007~is possible thanks to
this work. The \OIII5007\AA~PNLF is used to directly estimate the distance to the LMC. By fitting an exponential curve to the bright end of the function over a 6 magnitude range, an m$^{\ast}$ value of 14.02 is found. Applying the accepted M$^{\ast}$ value of -4.44, we find a distance modulus of 18.46 for the LMC. This is in excellent agreement with the best recent calibrations (see table 2). The bright end cutoff is found to be robust to small samples of bright PNe since our significantly increased PN samples have no effect on this fiducial.

We find a peak in the \OIII5007~PNLF $\sim$6 mags below the brightest. This peak was previously predicted but not seen in a real sample until now. The steep rise to the peak in the function over a 2 mag range was also not seen before but was also predicted due to the position of the WD cooling track and a mixture of H and He-burning stars in the LMC. Not withstanding incompleteness, the faint end of the PNLF can now be examined for the first time. Over a 4 mag range we find a linear drop-off in the number of PNe per mag bin.

A comparison of the foreground de-reddened \OIII5007 PNLF with the same \OIII5007 PNLF after correction for internal extinction reveals almost no change in the overall shape of the PNLF. If the observed PNLF is shifted by applying 1 magnitude of extinction to each object, then the de-reddened PNLF can be almost exactly recovered. This remarkable finding means that although the PN extinction ranges from 0 $<$ \textit{c}(H$\beta$) $<$ 1.3, from a global prospective, the PNLF acts as if each one has \textit{c}(H$\beta$) = 0.4. This will clearly make it easier to model the PNLF in distant galaxies, as it implies that, for a 1 dimensional analysis, there is only a mean offset to be applied.

We have constructed the first LMC PNLF in H$\beta$ using 591 measured PN spectra. We find a value of 16.8 for the m$^{\ast}_{4861}$ bright cut-off. This leads to an estimate of M$^{\ast}$ = --1.68 for H$\beta$-based PNLF distance estimates.

A clear relation is found between the shape of the \OIII\,5007\AA~and
the H$\beta$ PNLF with a mean difference of 2.8 magnitudes at the bright cut-off decreasing to 1.8 magnitudes by the peak of the distributions or 6 magnitudes below m$^{\ast}_{{5007}}$.

The new LMC PNLF is compared to a new PNLF constructed using local
Galactic PNe to 3kpc and the newest sample from the SMC. The magnitude range is almost exactly the
same for both the LMC and local surveys. A small dip $\sim$3 and $\sim$4 mag below the brightest
is evident in the local and LMC surveys respectively. The rollover at the faint end of
the PNLF is generally interpreted as a sign of incompleteness in the
survey, where the rollover occurs across the faintest 1 or 2
magnitudes. Using 584 PNe in the RP sample from this survey, the
rollover occurs over 1.5 magnitudes, causing the shape of
the faint end to inversely reflect the shape of the PNLF at the bright
end. This would indicate that the peak distribution found in this
work is close to the true peak distribution for LMC PNe. This occurs
at an absolute magnitude of $\sim$0.8. By comparison, the peak
distribution of local Galactic PNe occurs at an absolute magnitude
of $\sim$2, 1.2 magnitudes fainter than that found for the LMC.

Once we complete our survey by including the outer LMC, we expect the estimated
extra 80 PNe at the bright end of the function will fill out much of the exponential curve. It should thereby clarify the position of the rise, 4 mags below the brightest and the position of any dip if they exist. The extra objects are not expected to alter the distance modulus of 18.46; similar to the result found by Jacoby (1989). Final results will largely depend on individual luminosities up to 6 magnitudes below m$^{\ast}$.

%The
%lower metallicity in the LMC also has the effect of making LMC PNe
%brighter than local Galactic PNe. It has been shown by Stasi\'{n}ska
%(1978) and Garnett (1992) that high metallicities induce a
%temperature drop in the O$^{++}$ zone because cooling is dominated
%by the \OIII\,52 $\mu$m and \OIII\,88 $\mu$m lines which are almost
%independent of $\textsl{T}_{e}$. Figure~\ref{Figure 21.29} suggests
%that this effect is most pronounced at the peak of the PN
%distribution and may be observed in any sample which covers a large
%magnitude range. Metallicity has minimal effect on the magnitude of
%the brightest and faintest PNe.

% One
%possible explanation for the brighter PNe in the LMC is the
%generally lower \OIII~abundance for LMC PNe compared to Galactic
%PNe, where \OIII~acts as a coolant in the nebula.

%%%%%%%%%%%%%%%%%%%%%%%%%%%%%%%%%%%%%%%%%%%%%%%%%%%%%%%%%%%%%%%%%%%%%%%%%%%%%%%%%%%%%%%%%%%%%%%%%%%%%%%%%%%%%%%%%%%%%%%%%%%%%%%%%%%%%%%%%%%%%%%%%%%%%%%%%%%%%%

\label{section 9}%%%%%%%%%%%%%%%%%%%%%%%%%%%%%%%%%%%%%%%%%%%%%%%%%%%%%%%%%%%%%%%%%%%%%%%%%%%%%%%%%%%%%%%%%%%%%%%%%%%%%%%%%%%%%

\section*{Acknowledgments}

The authors wish to thank the AAO board for observing time on the
AAT and UKST. The authors also thank the European Southern
Observatory for observing time on the VLT and Australian National University along
with their telescope time allocation committees for supporting our
programme of follow-up spectroscopy. We thank Martin Cohen for making SAGE data available and assisting in the followup analysis of LMC PNe. We also thank Robin Ciardullo for very helpful comments and suggestions while carefully reviewing the paper. WR thanks Macquarie University, Sydney, for travel grants.

\label{section 10}%%%%%%%%%%%%%%%%%%%%%%%%%%%%%%%%%%%%%%%%%%%%%%%%%%%%%%%%%%%%%%%%%%%%%%%%%%%%%%%%%%%%%%%%%%%%%%%%%%%

\section{Appendix}
\begin{table}
\scriptsize
\begin{center}
\caption{A compendium of H$\beta$ and \OIII5007\AA~raw flux estimates for Large
Magellanic Cloud PNe. * = A flux measured directly from long-slit spectra. $\dag$ = PNe with extremely high H$\alpha$/H$\beta$
$>$10 ratios which may have been affected by power-law
photoionisation or shock-heating (see section~\ref{section3.3}).}
\begin{tabular}{llccc}
\hline\hline \noalign{\smallskip}
RP  & Common  & $\textit{c}$H$\beta$ & Flux H$\beta$ & Flux \OIII  \\
Ref. & Nomenclature &  & 4861\AA~  &  5007\AA~  \\
    &     &    &    {\tiny(erg cm$^{-2}$ s$^{-1}$)}   &  {\tiny(erg cm$^{-2}$ s$^{-1}$)}     \\
 \hline\noalign{\smallskip}
10	&	SMP88		WS27			&	0.56	&	4.63E-14\,$\pm$2.3E-15	&	2.02E-13\,$\pm$1.0E-14		\\
12	&	SMP90					&	0.42	&	3.65E-14\,$\pm$1.8E-15	&	2.31E-13\,$\pm$1.1E-14		\\
74	&				Mo35		&	0.46	&	4.94E-15\,$\pm$4.9E-16	&	3.78E-14\,$\pm$3.2E-14		\\
75	&				Mo40		&	0.34	&	6.42E-15\,$\pm$5.4E-15	&	4.61E-14\,$\pm$2.5E-14		\\
106	&	LM2-40 N217					&	0.65	&	1.28E-15\,$\pm$1.2E-16	&	4.12E-15\,$\pm$3.5E-15		\\
133	&	SMP89		WS38			&	0.35	&	2.09E-13\,$\pm$1.0E-14	&	2.80E-12\,$\pm$1.4E-13		\\
134	&				Mo39		&	0.48	&	3.51E-15\,$\pm$3.5E-16	&	2.66E-14\,$\pm$2.2E-14		\\
152	&				Mo37		&	0.34	&	1.25E-14\,$\pm$6.2E-16	&	6.78E-14\,$\pm$3.3E-15		\\
179	&				Mo36		&	0.32	&	6.13E-15\,$\pm$5.2E-15	&	5.71E-14\,$\pm$3.1E-14		\\
209	&	SMP93					&	0.33	&	4.35E-14\,$\pm$2.1E-15	&	1.91E-13\,$\pm$9.5E-15		\\
213	&	SMP92		WS39			&	0.12	&	2.10E-13\,$\pm$1.0E-14	&	3.70E-12\,$\pm$1.8E-13		\\
214	&		MG73				&	0.36	&	8.81E-15\,$\pm$7.4E-15	&	4.37E-14\,$\pm$2.4E-14		\\
215	&		MG76				&	0.45	&	1.39E-14\,$\pm$6.9E-16	&	7.60E-14\,$\pm$3.8E-15		\\
267	&		MG68				&	0.19	&	2.28E-14\,$\pm$1.1E-15	&	7.06E-14\,$\pm$3.5E-15		\\
269	&	SMP86					&	0.11	&	1.94E-14\,$\pm$9.6E-16	&	1.24E-13\,$\pm$6.2E-15		\\
270	&	SMP91					&	0.02	&	2.86E-14\,$\pm$1.4E-15	&	1.98E-13\,$\pm$9.9E-15		\\
271	&		MG75				&	0.41	&	8.15E-15\,$\pm$6.9E-15	&	6.14E-14\,$\pm$3.3E-14		\\
272	&		MG77				&	0.30	&	9.05E-15\,$\pm$7.6E-15	&	4.17E-14\,$\pm$2.2E-14		\\
273	&		MG78				&	0.32	&	2.48E-15\,$\pm$2.4E-16	&	1.61E-14\,$\pm$1.3E-14		\\
316	&	Sa124					&	0.41	&	7.65E-15\,$\pm$6.5E-15	&	8.72E-14\,$\pm$4.3E-15		\\
317	&	SMP83		WS35			&	0.01	&	1.70E-13\,$\pm$8.5E-15	&	1.85E-12\,$\pm$9.2E-14		\\
318	&	SMP85					&	0.06	&	2.16E-13\,$\pm$1.0E-14	&	9.21E-13\,$\pm$4.6E-14	*	\\
346	&		MG72				&	0.33	&	6.07E-15\,$\pm$5.1E-15	&	4.93E-14\,$\pm$2.7E-14	*	\\
350	&		MG69				&	0.39	&	5.98E-15\,$\pm$5.0E-15	&	5.60E-14\,$\pm$3.0E-14		\\
356	&		MG71				&	0.35	&	1.00E-14\,$\pm$5.0E-16	&	7.13E-14\,$\pm$3.5E-15		\\
398	&	SMP55					&	0.41	&	6.36E-13\,$\pm$6.8E-15	&	1.40E-12\,$\pm$1.4E-14		\\
399	&	SMP59					&	0.23	&	6.96E-14\,$\pm$3.4E-15	&	2.22E-13\,$\pm$1.1E-14	*	\\
400	&	SMP60					&	0.18	&	6.89E-14\,$\pm$9.5E-16	&	4.62E-13\,$\pm$9.9E-15		\\
401	&	SMP62		WS25			&	0.36	&	4.95E-13\,$\pm$1.8E-14	&	2.64E-12\,$\pm$2.4E-13		\\
402	&	SMP65					&	0.65	&	3.19E-14\,$\pm$1.5E-15	&	2.59E-13\,$\pm$1.2E-14	*	\\
403	&	SMP68					&	0.02	&	1.30E-13\,$\pm$6.5E-15	&	4.49E-13\,$\pm$2.2E-14		\\
404	&	SMP71					&	0.31	&	1.48E-13\,$\pm$7.3E-15	&	1.38E-12\,$\pm$6.8E-14		\\
405	&	SMP72					&	0.16	&	1.23E-14\,$\pm$6.1E-16	&	1.21E-13\,$\pm$6.0E-15		\\
406	&	SMP73					&	0.32	&	3.72E-13\,$\pm$1.8E-14	&	5.52E-12\,$\pm$2.7E-13		\\
407	&	SMP80		WS24			&	0.16	&	7.16E-14\,$\pm$3.5E-15	&	3.44E-13\,$\pm$1.7E-14		\\
408	&	Sa120					&	0.51	&	9.81E-15\,$\pm$8.3E-15	&	9.32E-14\,$\pm$4.6E-15		\\
409	&	Sa121					&	0.29	&	1.40E-14\,$\pm$6.9E-16	&	1.16E-13\,$\pm$5.7E-15		\\
410	&	Sa123 LM2-36					&	0.11	&	1.56E-13\,$\pm$7.8E-15	&	7.22E-13\,$\pm$3.6E-14		 \\
411	&		MG51				&	0.32	&	9.79E-15\,$\pm$8.3E-15	&	7.34E-14\,$\pm$3.6E-15		\\
412	&				Mo24		&	0.41	&	3.14E-14\,$\pm$5.7E-15	&	3.12E-13\,$\pm$3.2E-15		\\
413	&				Mo28		&	0.44	&	5.53E-15\,$\pm$4.7E-15	&	6.42E-14\,$\pm$3.2E-15		\\
414	&				Mo33		&	0.51	&	1.66E-14\,$\pm$8.3E-16	&	9.94E-14\,$\pm$4.9E-15		\\
539	&		MG43				&	0.37	&	2.96E-14\,$\pm$1.4E-15	&	1.16E-13\,$\pm$5.8E-15		\\
642	&	SMP56					&	0.33	&	1.07E-13\,$\pm$5.3E-15	&	4.28E-13\,$\pm$2.1E-14		\\
643	&	SMP57					&	0.45	&	2.67E-14\,$\pm$1.3E-15	&	2.42E-13\,$\pm$1.2E-14		\\
644	&	SMP58		WS23			&	0.08	&	2.97E-13\,$\pm$1.4E-14	&	1.93E-12\,$\pm$9.6E-14		\\
646	&	SMP77					&	0.33	&	1.58E-13\,$\pm$7.9E-15	&	6.34E-13\,$\pm$3.1E-14	*	\\
647	&	SMP78		WS33			&	0.41	&	3.17E-13\,$\pm$1.5E-14	&	3.36E-12\,$\pm$1.6E-13		\\
648	&	SMP82					&	0.46	&	3.77E-14\,$\pm$1.8E-15	&	3.33E-13\,$\pm$1.6E-14		\\
649	&	Sa117					&	0.45	&	1.98E-14\,$\pm$9.8E-16	&	1.28E-13\,$\pm$6.3E-15		\\
650	&	Sa122					&	0.38	&	2.29E-14\,$\pm$1.1E-15	&	1.22E-13\,$\pm$6.1E-15		\\
651	&	Sa116 LM2-24					&	0.32	&	2.97E-14\,$\pm$1.4E-15	&	2.19E-13\,$\pm$1.1E-14		 \\
652	&	Sa118					&	0.35	&	1.18E-14\,$\pm$5.9E-16	&	9.97E-14\,$\pm$4.9E-15		\\
654	&		MG49				&	0.27	&	2.93E-14\,$\pm$1.4E-15	&	1.03E-13\,$\pm$5.1E-15		\\
655	&		MG52				&	0.60	&	5.33E-15\,$\pm$4.5E-15	&	5.88E-14\,$\pm$3.2E-14		\\
656	&		MG53				&	0.36	&	1.90E-14\,$\pm$9.5E-16	&	1.43E-13\,$\pm$7.1E-15		\\
657	&		MG56				&	0.52	&	2.55E-15\,$\pm$2.5E-16	&	7.66E-15\,$\pm$6.5E-15		\\
658	&		MG60				&	0.33	&	1.21E-14\,$\pm$6.0E-16	&	8.07E-14\,$\pm$4.0E-15		\\
659	&		MG65				&	0.37	&	3.04E-14\,$\pm$1.5E-15	&	1.04E-13\,$\pm$5.2E-15		\\
660	&				Mo23		&	0.05	&	1.16E-14\,$\pm$5.7E-16	&	3.68E-14\,$\pm$3.1E-14		\\
661	&				Mo27		&	0.20	&	1.05E-14\,$\pm$5.2E-16	&	4.67E-14\,$\pm$2.5E-14		\\
662	&				Mo30		&	0.42	&	1.22E-14\,$\pm$6.1E-16	&	6.02E-14\,$\pm$3.3E-14		\\
663	&				Mo34		&	0.52	&	3.75E-14\,$\pm$1.8E-15	&	1.69E-13\,$\pm$8.4E-15		\\
664	&				Mo32		&	0.46	&	1.62E-14\,$\pm$8.1E-16	&	9.82E-14\,$\pm$4.9E-15		\\
668	&				Mo26		&	0.48	&	7.82E-15\,$\pm$6.6E-15	&	7.02E-14\,$\pm$3.5E-15		\\
890	&	SMP63		WS26			&	0.40	&	3.24E-13\,$\pm$1.6E-14	&	3.90E-12\,$\pm$1.9E-13		\\
891	&	SMP75		WS31			&	0.39	&	2.62E-13\,$\pm$1.3E-14	&	2.91E-12\,$\pm$1.4E-13		\\
892	&	SMP76		WS32			&	0.38	&	3.55E-13\,$\pm$1.7E-14	&	1.85E-12\,$\pm$9.2E-14		\\
893	&		MG40				&	0.31	&	2.92E-14\,$\pm$1.4E-15	&	2.18E-13\,$\pm$1.0E-14		\\
894	&		MG54				&	0.32	&	9.75E-15\,$\pm$8.2E-15	&	8.42E-14\,$\pm$4.2E-15		\\
895	&		MG62				&	0.31	&	2.81E-15\,$\pm$2.8E-16	&	1.06E-14\,$\pm$9.0E-15		\\
1047	&	SMP66					&	0.01	&	2.89E-13\,$\pm$1.4E-14	&	2.19E-12\,$\pm$1.1E-13	*	\\
1048	&	SMP67					&	0.14	&	1.61E-13\,$\pm$8.0E-15	&	5.15E-13\,$\pm$2.5E-14		\\
1049	&	SMP69					&	0.29	&	4.24E-14\,$\pm$2.1E-15	&	3.44E-13\,$\pm$1.7E-14		\\
1050	&		MG42				&	0.29	&	1.78E-15\,$\pm$1.7E-16	&	9.33E-15\,$\pm$7.9E-15		\\
1051	&		MG50				&	0.27	&	2.31E-14\,$\pm$1.1E-15	&	8.66E-14\,$\pm$4.3E-15		\\
\end{tabular}\label{table 4}
\end{center}
\end{table}
\begin{table}
\scriptsize
\begin{center}
\begin{tabular}{llccc}
\hline\hline \noalign{\smallskip}
RP  & Common  & $\textit{c}$H$\beta$ & Flux H$\beta$ & Flux \OIII  \\
Ref. & Nomenclature &  & 4868\AA~  &  5007\AA~  \\
    &     &    &    {\tiny(erg cm$^{-2}$ s$^{-1}$)}   &  {\tiny(erg cm$^{-2}$ s$^{-1}$)}     \\
 \hline\noalign{\smallskip}
1052	&		MG58				&	0.35	&	1.27E-14\,$\pm$6.3E-16	&	7.15E-14\,$\pm$3.5E-15		\\
1053	&		MG64				&	0.46	&	3.66E-15\,$\pm$3.6E-16	&	3.20E-14\,$\pm$2.7E-14		\\
1114	&	SMP32		WS10			&	0.29	&	5.83E-13\,$\pm$3.4E-14	&	4.16E-12\,$\pm$4.5E-13		 \\
1115	&	SMP41					&	0.28	&	1.61E-13\,$\pm$8.0E-15	&	2.39E-12\,$\pm$1.2E-13		\\
1116	&	SMP49					&	0.29	&	7.22E-14\,$\pm$2.2E-15	&	5.25E-13\,$\pm$2.9E-14		\\
1117	&		MG30				&	0.26	&	9.15E-14\,$\pm$3.0E-15	&	3.36E-13\,$\pm$1.6E-14		\\
1118	&				Mo20		&	0.16	&	3.88E-14\,$\pm$9.9E-16	&	1.49E-13\,$\pm$5.7E-15		\\
1212	&	SMP38		WS15			&	0.22	&	2.06E-13\,$\pm$1.0E-14	&	3.03E-12\,$\pm$1.5E-13		 \\
1213	&	SMP43					&	0.46	&	9.53E-14\,$\pm$4.7E-15	&	1.20E-12\,$\pm$6.0E-14		\\
1214	&	SMP47		WS18			&	0.55	&	2.57E-13\,$\pm$1.2E-14	&	2.77E-12\,$\pm$1.3E-13		 \\
1215	&	SMP48		WS19			&	0.50	&	2.41E-13\,$\pm$1.2E-14	&	1.88E-12\,$\pm$9.4E-14		 \\
1216	&	SMP51					&	0.44	&	6.08E-14\,$\pm$3.0E-15	&	9.21E-13\,$\pm$4.6E-14		\\
1217	&					J-5	&	0.59	&	3.99E-14\,$\pm$2.0E-15	&	2.28E-13\,$\pm$1.1E-14		\\
1218	&					J-04	&	0.27	&	6.59E-15\,$\pm$5.6E-15	&	3.52E-14\,$\pm$2.9E-14		\\
1219	&					J-7	&	0.57	&	1.49E-16\,$\pm$1.4E-17	&	9.83E-16\,$\pm$8.3E-16		\\
1220	&					J-12	&	0.54	&	7.53E-15\,$\pm$6.4E-15	&	3.96E-14\,$\pm$3.3E-14		\\
1221	&					J-14	&	0.62	&	3.46E-15\,$\pm$3.4E-16	&	5.19E-14\,$\pm$2.8E-14		\\
1222	&					J-15	&	0.85	&	3.50E-15\,$\pm$3.5E-16	&	3.69E-14\,$\pm$3.1E-14		\\
1223	&					J-16	&	0.64	&	3.40E-15\,$\pm$3.4E-16	&	3.64E-14\,$\pm$3.0E-14		\\
1224	&					J-17	&	0.70	&	3.43E-14\,$\pm$1.7E-15	&	6.98E-14\,$\pm$3.4E-15		\\
1225	&					J-18	&	0.57	&	1.98E-15\,$\pm$1.9E-16	&	2.91E-14\,$\pm$2.4E-14		\\
1227	&					J-20	&	0.33	&	1.94E-14\,$\pm$9.6E-16	&	1.12E-13\,$\pm$5.6E-15		\\
1228	&					J-21	&	0.43	&	7.70E-15\,$\pm$6.5E-15	&	3.36E-14\,$\pm$2.8E-14		\\
1229	&					J-22	&	0.39	&	8.64E-15\,$\pm$7.3E-15	&	4.92E-14\,$\pm$2.7E-14		\\
1230	&					J-23	&	0.32	&	7.68E-14\,$\pm$3.8E-15	&	3.23E-13\,$\pm$1.6E-14		\\
1231	&					J-24	&	0.53	&	9.13E-15\,$\pm$7.7E-15	&	4.71E-14\,$\pm$2.5E-14		\\
1232	&					J-26	&	0.32	&	1.19E-14\,$\pm$5.9E-16	&	1.19E-13\,$\pm$5.9E-15		\\
1234	&					J-32	&	0.57	&	5.38E-15\,$\pm$4.5E-15	&	6.12E-14\,$\pm$3.3E-14		\\
1235	&					J-33	&	0.06	&	2.59E-14\,$\pm$1.2E-15	&	1.95E-13\,$\pm$9.74E-15		\\
1236	&		MG28				&	0.32	&	1.07E-14\,$\pm$5.3E-16	&	7.48E-14\,$\pm$3.74E-15		\\
1237	&				Mo19		&	0.26	&	1.44E-14\,$\pm$7.2E-16	&	1.75E-13\,$\pm$8.77E-15		\\
1395	&	SMP28					&	0.39	&	3.15E-14\,$\pm$1.5E-15	&	2.06E-13\,$\pm$1.0E-14		\\
1396	&	SMP29		WS9			&	0.34	&	1.59E-13\,$\pm$7.9E-15	&	1.67E-12\,$\pm$8.3E-14		\\
1398	&	SMP33		WS11			&	0.47	&	1.24E-13\,$\pm$6.2E-15	&	1.74E-12\,$\pm$8.7E-14		 \\
1399	&	SMP34					&	0.36	&	1.05E-13\,$\pm$5.2E-15	&	6.58E-13\,$\pm$3.2E-14		\\
1400	&	SMP36		WS13			&	0.26	&	5.73E-14\,$\pm$2.8E-15	&	8.88E-13\,$\pm$4.4E-14		 \\
1401	&	SMP37		WS14			&	0.37	&	8.72E-14\,$\pm$4.3E-15	&	1.44E-12\,$\pm$7.1E-14		 \\
1402	&	SMP39					&	0.48	&	5.35E-14\,$\pm$2.6E-15	&	6.92E-13\,$\pm$3.4E-14		\\
1403	&	SMP42					&	0.00	&	6.46E-14\,$\pm$3.2E-15	&	8.25E-13\,$\pm$4.1E-14		\\
1404	&	SMP46					&	0.51	&	3.43E-14\,$\pm$1.7E-15	&	4.55E-13\,$\pm$2.2E-14		\\
1405	&	SMP52		WS21		&	0.22	&	3.27E-13\,$\pm$1.6E-14	&	4.59E-12\,$\pm$2.2E-13		\\
1406	&	SMP54		J-35	&	0.36	&	3.85E-14\,$\pm$1.9E-15	&	3.32E-13\,$\pm$1.6E-14		\\
1407	&	Sa110					&	0.28	&	1.46E-14\,$\pm$7.3E-16	&	2.15E-13\,$\pm$1.0E-14		\\
1408	&					J-10	&	0.49	&	2.33E-15\,$\pm$2.3E-16	&	2.52E-14\,$\pm$2.1E-14		\\
1409	&		MG19				&	0.37	&	1.10E-14\,$\pm$5.4E-16	&	8.75E-14\,$\pm$4.3E-15		\\
1410	&		MG20				&	0.33	&	1.00E-14\,$\pm$5.0E-16	&	7.90E-14\,$\pm$3.9E-15		\\
1411	&		MG23				&	0.46	&	2.24E-15\,$\pm$2.2E-16	&	2.07E-14\,$\pm$1.7E-14		\\
1412	&		MG29				&	0.38	&	3.33E-14\,$\pm$1.6E-15	&	3.94E-13\,$\pm$1.9E-14		\\
1413	&		MG31				&	0.40	&	9.48E-15\,$\pm$8.0E-15	&	8.81E-14\,$\pm$4.4E-15		\\
1552	&	SMP30					&	0.44	&	3.11E-14\,$\pm$1.5E-15	&	2.35E-13\,$\pm$1.1E-14		\\
1553	&	SMP44					&	0.33	&	3.91E-14\,$\pm$1.9E-15	&	3.40E-13\,$\pm$1.7E-14		\\
1554	&	SMP45		WS17			&	0.53	&	8.25E-14\,$\pm$4.1E-15	&	1.17E-12\,$\pm$5.8E-14		 \\
1555	&	SMP50		WS20			&	0.39	&	1.68E-13\,$\pm$8.4E-15	&	1.38E-12\,$\pm$6.9E-14		 \\
1556	&	SMP53		WS22			&	0.25	&	5.96E-13\,$\pm$2.9E-14	&	3.30E-12\,$\pm$1.6E-13		 \\
1557	&		MG34				&	0.33	&	8.05E-15\,$\pm$6.8E-15	&	2.05E-14\,$\pm$1.7E-14		\\
1558	&		MG35				&	0.45	&	1.72E-14\,$\pm$8.6E-16	&	4.20E-14\,$\pm$2.3E-14		\\
1602	&	SMP13		WS4			&	0.33	&	1.57E-13\,$\pm$7.8E-15	&	1.64E-12\,$\pm$8.2E-14		\\
1603	&	SMP14		WS2			&	0.35	&	1.85E-14\,$\pm$9.2E-16	&	1.29E-13\,$\pm$6.4E-15		\\
1604	&	SMP15		WS5			&	0.41	&	1.66E-13\,$\pm$8.2E-15	&	2.28E-12\,$\pm$1.1E-13		\\
1605	&	SMP19		WS6			&	0.33	&	3.36E-13\,$\pm$1.6E-14	&	1.50E-12\,$\pm$7.5E-14		\\
1606	&		MG4				&	0.19	&	7.03E-15\,$\pm$5.9E-15	&	5.65E-14\,$\pm$3.1E-14	*	\\
1607	&				Mo9		&	0.51	&	9.97E-15\,$\pm$8.4E-15	&	3.05E-14\,$\pm$2.5E-14		\\
1608	&				Mo11		&	0.30	&	7.05E-15\,$\pm$5.9E-15	&	5.56E-14\,$\pm$3.0E-14		\\
1609	&				Mo14		&	0.24	&	5.96E-15\,$\pm$5.0E-15	&	3.81E-14\,$\pm$3.2E-14		\\
1677	&	SMP16					&	0.43	&	4.30E-14\,$\pm$2.1E-15	&	4.19E-13\,$\pm$2.0E-14		\\
1678	&	SMP17					&	0.04	&	8.63E-15\,$\pm$7.3E-15	&	5.69E-14\,$\pm$3.1E-14		\\
1679	&	SMP18					&	0.31	&	5.27E-14\,$\pm$2.6E-15	&	4.37E-13\,$\pm$2.1E-14		\\
1680	&	SMP20					&	0.35	&	4.06E-14\,$\pm$2.0E-15	&	3.95E-13\,$\pm$1.9E-14		\\
1681	&	SMP22					&	0.42	&	1.61E-14\,$\pm$8.0E-16	&	1.75E-13\,$\pm$8.7E-15		\\
1682	&	SMP24					&	0.02	&	1.15E-14\,$\pm$5.7E-16	&	1.54E-13\,$\pm$7.6E-15		\\
1683	&	SMP25					&	0.31	&	4.03E-13\,$\pm$2.0E-14	&	3.47E-12\,$\pm$1.7E-13		\\
1684	&	Sa105					&	0.52	&	9.41E-15\,$\pm$8.0E-15	&	9.91E-14\,$\pm$4.9E-15		\\
1685	&	Sa107					&	0.45	&	1.61E-14\,$\pm$8.0E-16	&	1.51E-13\,$\pm$7.5E-15		\\
1686	&		MG14				&	0.28	&	3.12E-14\,$\pm$1.5E-15	&	3.17E-13\,$\pm$1.5E-14		\\
1687	&				Mo12		&	0.41	&	5.92E-15\,$\pm$5.0E-15	&	4.75E-14\,$\pm$2.6E-14		\\
1688	&				Mo13		&	0.26	&	5.41E-15\,$\pm$4.5E-15	&	3.42E-14\,$\pm$2.9E-14		\\
1689	&				Mo16		&	0.14	&	4.90E-15\,$\pm$4.9E-16	&	4.15E-14\,$\pm$2.2E-14		\\
1797	&	SMP21		WS7			&	0.46	&	1.68E-13\,$\pm$8.4E-15	&	1.95E-12\,$\pm$9.7E-14		\\
1798	&	SMP23		WS8			&	0.11	&	2.14E-13\,$\pm$1.0E-14	&	1.67E-12\,$\pm$8.3E-14		\\
1799	&	Sa106					&	0.33	&	1.66E-14\,$\pm$8.3E-16	&	1.38E-13\,$\pm$6.8E-15		\\
\end{tabular}
\end{center}
\end{table}
\begin{table}
\scriptsize
\begin{center}
\begin{tabular}{llccc}
\hline\hline \noalign{\smallskip}
RP  & Common  & $\textit{c}$H$\beta$ & Flux H$\beta$ & Flux \OIII  \\
Ref. & Nom. &  & 4868\AA~  &  5007\AA~  \\
    &     &    &    {\tiny(erg cm$^{-2}$ s$^{-1}$)}   &  {\tiny(erg cm$^{-2}$ s$^{-1}$)}     \\
 \hline\noalign{\smallskip}
1800	&		MG8				&	0.38	&	1.52E-14\,$\pm$7.6E-16	&	1.24E-13\,$\pm$6.2E-15		\\
1801	&		MG10				&	0.43	&	1.25E-14\,$\pm$6.2E-16	&	1.34E-13\,$\pm$6.7E-15		\\
1802	&		MG15				&	0.34	&	4.46E-15\,$\pm$4.4E-16	&	1.78E-14\,$\pm$1.5E-14		\\
1894	&	SMP27					&	0.26	&	1.72E-14\,$\pm$8.6E-16	&	2.24E-13\,$\pm$1.1E-14	*	\\
1895	&		MG9				&	0.39	&	9.07E-15\,$\pm$7.7E-15	&	7.32E-14\,$\pm$3.6E-15		\\
1896	&		MG11				&	0.41	&	1.23E-14\,$\pm$6.1E-16	&	5.88E-14\,$\pm$3.2E-14		\\
1897	&		MG12				&	0.41	&	9.59E-15\,$\pm$8.1E-15	&	4.42E-14\,$\pm$2.4E-14		\\
1898	&				Mo17		&	0.28	&	5.89E-15\,$\pm$5.0E-15	&	3.97E-14\,$\pm$3.3E-14		\\
 \hline\hline \noalign{\smallskip}
\end{tabular}
\end{center}
%\\[1cm]
Explanation of abbreviations used: J: Jacoby (1980),~LM2: Lindsay \& Mullan (1963), ~MG: Morgan \& Good (1992), ~Mo:
Morgan (1994),~N: Henize (1956),~SMP: Sanduleak et al. (1978), ~Sa: Sanduleak (1984), ~WS:
Westerlund \& Smith (1964)
\end{table}

\begin{table}
\scriptsize
\begin{center}
\caption{A compendium of H$\beta$ and \OIII5007\AA~raw flux estimates for RP PNe in the Large
Magellanic Cloud. P = Probability where T = True, L = Likely, P = Possible. A detailed review of T, L and P is given in [RPb]. * = A flux measured directly from long-slit spectra. $\dag$ = PNe with extremely high H$\alpha$/H$\beta$
$>$10 ratios which may have been affected by power-law
photoionisation or shock-heating (see section~\ref{section3.3}).}
\begin{tabular}{lcccccc}
\hline\hline \noalign{\smallskip}
RP  & P  & $\textit{c}$H$\beta$ & Flux H$\beta$ & Flux \OIII  \\
Ref. &  & &   4861\AA~  &  5007\AA~  \\
    &    &      &    {\tiny(erg cm$^{-2}$ s$^{-1}$)}   &  {\tiny(erg cm$^{-2}$ s$^{-1}$)}     \\
 \hline\noalign{\smallskip}
1	&	L	&	0.56	&	1.46E-15\,$\pm$7.29E-17	&	2.52E-15\,$\pm$1.26E-16		\\
9	&	T	&	0.49	&	2.60E-15\,$\pm$1.30E-16	&	2.32E-14\,$\pm$1.16E-15		\\
18	&	T	&	0.28	&	7.11E-15\,$\pm$3.56E-16	&	1.39E-14\,$\pm$6.96E-16		\\
25	&	T	&	0.10	&	9.66E-16\,$\pm$8.21E-16	&	7.76E-15\,$\pm$3.88E-16		\\
26	&	T	&	0.21	&	3.39E-15\,$\pm$1.69E-16	&	1.72E-14\,$\pm$8.58E-16		\\
35	&	T	&	0.06	&	5.12E-15\,$\pm$2.56E-16	&	5.60E-15\,$\pm$2.80E-16		\\
44	&	T	&	0.45	&	1.06E-14\,$\pm$5.32E-16	&	2.07E-14\,$\pm$1.04E-15		\\
46	&	P	&	0.29	&	2.64E-13\,$\pm$1.32E-14	&	3.52E-13\,$\pm$1.76E-14	*	\\
61	&	T	&	0.33	&	5.42E-15\,$\pm$2.71E-16	&	2.28E-14\,$\pm$1.14E-15		\\
62	&	T	&	0.51	&	1.74E-15\,$\pm$8.69E-17	&	1.01E-14\,$\pm$5.04E-16		\\
70	&	T	&	1.01	&	3.79E-16\,$\pm$3.79E-17	&	3.32E-15\,$\pm$1.66E-16		\\
72	&	L	&	0.09	&	5.30E-15\,$\pm$2.65E-16	&	3.05E-14\,$\pm$1.53E-15	*	\\
77	&	T	&	0.78	&	3.28E-16\,$\pm$3.28E-17	&	2.28E-15\,$\pm$1.14E-16		\\
85	&	L	&	0.84	&	7.68E-15\,$\pm$3.84E-16	&	1.64E-14\,$\pm$8.20E-16		\\
86	&	P	&	0.86	&	1.43E-15\,$\pm$7.16E-17	&	6.20E-15\,$\pm$3.10E-16		\\
87	&	L	&	0.32	&	7.24E-16\,$\pm$7.24E-17	&	9.28E-16\,$\pm$9.28E-17		\\
89	&	T	&	0.63	&	6.03E-15\,$\pm$3.02E-16	&	1.11E-14\,$\pm$5.54E-16		\\
90	&	T	&	0.63	&	1.29E-15\,$\pm$6.46E-17	&	3.52E-15\,$\pm$1.76E-16		\\
93	&	T	&	0.56	&	2.32E-15\,$\pm$1.16E-16	&	1.57E-14\,$\pm$7.84E-16		\\
95	&	P	&	0.72	&	2.08E-15\,$\pm$1.04E-16	&	7.02E-15\,$\pm$3.51E-16		\\
97	&	P	&	0.69	&	2.04E-15\,$\pm$1.02E-16	&	2.27E-15\,$\pm$1.13E-16		\\
99	&	T	&	0.43	&	3.45E-15\,$\pm$1.73E-16	&	4.26E-14\,$\pm$2.13E-15		\\
102	&	T	&	0.07	&	2.23E-15\,$\pm$1.12E-16	&	7.00E-15\,$\pm$3.50E-16		\\
103	&	T	&	0.55	&	5.08E-15\,$\pm$2.54E-16	&	2.06E-14\,$\pm$1.03E-15		\\
111	&	T	&	0.83	&	4.74E-15\,$\pm$2.37E-16	&	1.26E-14\,$\pm$6.28E-16		\\
116	&	L	&	0.53	&	3.03E-14\,$\pm$1.51E-15	&	8.84E-14\,$\pm$4.42E-15	*	\\
120	&	P	&	0.36	&	7.47E-15\,$\pm$3.74E-16	&	1.73E-14\,$\pm$8.67E-16		\\
122	&	T	&	0.15	&	1.31E-15\,$\pm$6.55E-17	&	4.29E-15\,$\pm$2.15E-16		\\
125	&	T	&	0.93	&	3.60E-16\,$\pm$3.60E-17	&	1.57E-15\,$\pm$1.34E-15		\\
127	&	L	&	0.43	&	3.26E-15\,$\pm$1.63E-16	&	7.80E-15\,$\pm$3.90E-16		\\
129	&	T	&	0.35	&	7.19E-15\,$\pm$3.60E-16	&	1.74E-14\,$\pm$8.72E-16		\\
130	&	P	&	0.26	&	4.82E-15\,$\pm$2.41E-16	&	7.93E-15\,$\pm$3.97E-16		\\
135	&	T	&	0.23	&	6.77E-15\,$\pm$3.38E-16	&	2.24E-14\,$\pm$1.12E-15		\\
142	&	T	&	0.48	&	4.62E-15\,$\pm$2.31E-16	&	7.49E-15\,$\pm$3.75E-16		\\
143	&	T	&	0.40	&	4.36E-15\,$\pm$2.18E-16	&	2.50E-14\,$\pm$1.25E-15		\\
144	&	T	&	0.76	&	1.13E-15\,$\pm$5.65E-17	&	8.83E-15\,$\pm$4.42E-16		\\
145	&	T	&	0.06	&	2.55E-15\,$\pm$1.28E-16	&	2.84E-15\,$\pm$1.42E-16		\\
147	&	T	&	0.51	&	6.09E-16\,$\pm$6.09E-17	&	3.59E-15\,$\pm$1.79E-16		\\
162	&	T	&	0.34	&	3.67E-15\,$\pm$1.84E-16	&	1.21E-14\,$\pm$6.05E-16		\\
163	&	P	&	0.28	&	4.95E-15\,$\pm$2.47E-16	&	8.55E-15\,$\pm$4.28E-16		\\
171	&	T	&	0.13	&	3.73E-15\,$\pm$1.87E-16	&	1.47E-14\,$\pm$7.34E-16	*	\\
172	&	T	&	0.19	&	1.40E-15\,$\pm$6.98E-17	&	3.96E-15\,$\pm$1.98E-16		\\
178	&	T	&	0.24	&	6.36E-15\,$\pm$3.18E-16	&	2.31E-14\,$\pm$1.16E-15		\\
180	&	L	&	0.18	&	1.50E-15\,$\pm$7.49E-17	&	1.61E-15\,$\pm$8.04E-17		\\
182	&	L	&	0.09	&	1.99E-15\,$\pm$9.93E-17	&	2.26E-15\,$\pm$1.13E-16		\\
187	&	P	&	0.59	&	2.31E-15\,$\pm$1.15E-16	&	2.27E-14\,$\pm$1.13E-15		\\
188	&	P	&	0.41	&	2.11E-15\,$\pm$1.05E-16	&	3.30E-15\,$\pm$1.65E-16		\\
189	&	P	&	0.68	&	1.70E-15\,$\pm$8.52E-17	&	3.98E-15\,$\pm$1.99E-16		\\
194	&	T	&	0.53	&	2.57E-15\,$\pm$1.28E-16	&	9.78E-15\,$\pm$4.89E-16		\\
198	&	P	&	0.96	&	4.12E-15\,$\pm$2.06E-16	&	8.13E-15\,$\pm$4.07E-16	*	\\
202	&	T	&	0.39	&	2.83E-15\,$\pm$1.42E-16	&	1.75E-14\,$\pm$8.75E-16		\\
203	&	T	&	0.07	&	1.83E-15\,$\pm$9.17E-17	&	1.71E-14\,$\pm$8.53E-16		\\
207	&	L	&	0.28	&	2.17E-15\,$\pm$1.08E-16	&	3.37E-15\,$\pm$1.68E-16		\\
217	&	P	&	0.37	&	5.91E-15\,$\pm$2.96E-16	&	7.20E-15\,$\pm$3.60E-16		\\
219	&	P	&	0.49	&	9.00E-15\,$\pm$4.50E-16	&	1.49E-14\,$\pm$7.45E-16		\\
223	&	P	&	0.43	&	1.73E-14\,$\pm$8.66E-16	&	4.65E-14\,$\pm$2.32E-15		\\
227	&	P	&	0.37	&	2.27E-14\,$\pm$1.14E-15	&	3.65E-14\,$\pm$1.82E-15		\\
228	&	P	&	0.65	&	2.78E-15\,$\pm$1.39E-16	&	7.84E-15\,$\pm$3.92E-16		\\
231	&	L	&	0.44	&	1.12E-14\,$\pm$5.58E-16	&	2.74E-14\,$\pm$1.37E-15		\\
232	&	L	&	0.28	&	1.24E-14\,$\pm$6.19E-16	&	4.01E-14\,$\pm$2.00E-15		\\
234	&	P	&	0.22	&	1.15E-14\,$\pm$5.74E-16	&	4.75E-14\,$\pm$2.38E-15		\\
240	&	P	&	0.48	&	1.02E-14\,$\pm$5.08E-16	&	2.72E-14\,$\pm$1.36E-15		\\
241	&	P	&	0.49	&	8.60E-15\,$\pm$4.30E-16	&	2.93E-14\,$\pm$1.46E-15		\\
242	&	P	&	0.44	&	3.82E-14\,$\pm$1.91E-15	&	3.07E-14\,$\pm$1.54E-15		\\
246	&	P	&	0.59	&	8.71E-15\,$\pm$4.36E-16	&	3.96E-14\,$\pm$1.98E-15		\\
247	&	P	&	0.62	&	9.85E-15\,$\pm$4.92E-16	&	1.80E-14\,$\pm$8.98E-16		\\
250	&	P	&	0.52	&	2.50E-14\,$\pm$1.25E-15	&	4.98E-14\,$\pm$2.49E-15		\\
251	&	P	&	0.52	&	1.43E-14\,$\pm$7.15E-16	&	1.22E-14\,$\pm$6.12E-16		\\
252	&	P	&	0.79$\dag$	&	4.16E-15\,$\pm$2.08E-16	&	5.11E-15\,$\pm$2.56E-16		\\
%\hline\hline\noalign{\smallskip}
  \end{tabular}\label{table 5}
\end{center}
\end{table}
\begin{table}
\scriptsize
\begin{center}
\begin{tabular}{lcccccc}
\hline\hline \noalign{\smallskip}
RP  & P  & $\textit{c}$H$\beta$ & Flux H$\beta$ & Flux \OIII  \\
Ref. &  & &   4868\AA~  &  5007\AA~  \\
    &    &      &    {\tiny(erg cm$^{-2}$ s$^{-1}$)}   &  {\tiny(erg cm$^{-2}$ s$^{-1}$)}     \\
 \hline\noalign{\smallskip}
254	&	P	&	0.47	&	9.39E-15\,$\pm$4.70E-16	&	1.80E-14\,$\pm$9.01E-16		\\
259	&	P	&	0.55	&	7.73E-15\,$\pm$3.86E-16	&	1.20E-14\,$\pm$5.98E-16		\\
261	&	P	&	0.54	&	3.32E-14\,$\pm$1.66E-15	&	3.04E-14\,$\pm$1.52E-15		\\
263	&	P	&	0.55	&	9.27E-15\,$\pm$4.64E-16	&	1.42E-14\,$\pm$7.08E-16		\\
264	&	P	&	0.68	&	1.16E-14\,$\pm$5.80E-16	&	1.19E-14\,$\pm$5.96E-16		\\
265	&	T	&	0.12	&	1.36E-14\,$\pm$6.80E-16	&	4.39E-14\,$\pm$2.20E-15		\\
266	&	P	&	0.35	&	7.81E-14\,$\pm$3.90E-15	&	8.69E-14\,$\pm$4.35E-15		\\
268	&	P	&	0.00	&	3.39E-14\,$\pm$1.69E-15	&	7.75E-14\,$\pm$3.87E-15		\\
277	&	P	&	0.12	&	1.80E-15\,$\pm$9.02E-17	&	3.00E-15\,$\pm$1.50E-16		\\
283	&	P	&	1.00	&	3.31E-15\,$\pm$1.66E-16	&	6.40E-15\,$\pm$3.20E-16		\\
295	&	P	&	0.58	&	1.01E-14\,$\pm$5.07E-16	&	1.25E-14\,$\pm$6.23E-16		\\
296	&	P	&	0.59	&	1.25E-15\,$\pm$6.25E-17	&	1.60E-15\,$\pm$7.99E-17		\\
299	&	T	&	0.32	&	4.85E-15\,$\pm$2.42E-16	&	1.95E-14\,$\pm$9.73E-16		\\
303	&	P	&	1.14	&	7.70E-16\,$\pm$7.70E-17	&	1.79E-15\,$\pm$8.95E-17	*	\\
307	&	P	&	0.54	&	9.12E-15\,$\pm$4.56E-16	&	2.51E-14\,$\pm$1.26E-15		\\
312	&	L	&	0.32	&	7.22E-15\,$\pm$3.61E-16	&	2.57E-14\,$\pm$1.29E-15		\\
%313	&	L	&		&			&				\\
315	&	P	&	0.39	&	9.22E-15\,$\pm$4.61E-16	&	2.76E-14\,$\pm$1.38E-15		\\
326	&	P	&	0.71	&	4.39E-15\,$\pm$2.19E-16	&	9.02E-15\,$\pm$4.51E-16		\\
328	&	P	&	0.32	&	4.44E-16\,$\pm$4.44E-17	&	2.05E-15\,$\pm$1.03E-16		\\
331	&	P	&	0.44	&	1.02E-15\,$\pm$5.12E-17	&	2.28E-15\,$\pm$1.14E-16		\\
347	&	T	&	0.51	&	4.03E-15\,$\pm$2.01E-16	&	2.24E-14\,$\pm$1.12E-15		\\
359	&	T	&	0.58	&	8.30E-16\,$\pm$8.30E-17	&	5.02E-15\,$\pm$2.51E-16		\\
363	&	P	&	0.74	&	1.65E-16\,$\pm$1.65E-17	&	1.66E-15\,$\pm$8.29E-17		\\
366	&	T	&	0.02	&	1.16E-15\,$\pm$5.79E-17	&	2.00E-15\,$\pm$1.00E-16		\\
384	&	T	&	0.20	&	9.86E-16\,$\pm$8.38E-16	&	2.93E-15\,$\pm$1.46E-16		\\
393	&	T	&	0.20	&	1.04E-15\,$\pm$5.20E-17	&	4.91E-15\,$\pm$2.46E-16	*	\\
394	&	T	&	0.07	&	1.00E-15\,$\pm$5.02E-17	&	1.13E-15\,$\pm$9.61E-16	*	\\
395	&	L	&	0.37	&	6.53E-16\,$\pm$6.53E-17	&	1.90E-15\,$\pm$9.48E-17		\\
396	&	T	&	0.34	&	1.71E-15\,$\pm$8.57E-17	&	8.07E-15\,$\pm$4.03E-16		\\
397	&	T	&	0.16	&	8.81E-16\,$\pm$8.81E-17	&	8.64E-16\,$\pm$8.64E-17		\\
415	&	T	&	0.33	&	2.21E-14\,$\pm$1.10E-15	&	8.00E-14\,$\pm$4.00E-15		\\
427	&	T	&	0.31	&	1.61E-14\,$\pm$8.05E-16	&	4.16E-14\,$\pm$2.08E-15		\\
440	&	T	&	0.22	&	1.46E-14\,$\pm$7.28E-16	&	6.26E-14\,$\pm$3.13E-15	*	\\
441	&	T	&	0.40	&	1.33E-15\,$\pm$6.65E-17	&	9.60E-15\,$\pm$4.80E-16		\\
442	&	T	&	0.19	&	1.96E-14\,$\pm$9.78E-16	&	6.69E-14\,$\pm$3.34E-15		\\
445	&	L	&	0.57	&	1.47E-15\,$\pm$7.37E-17	&	2.33E-14\,$\pm$1.16E-15		\\
460	&	P	&	0.39	&	9.67E-16\,$\pm$8.22E-16	&	2.29E-15\,$\pm$1.15E-16		\\
463	&	P	&	0.76	&	1.74E-15\,$\pm$8.68E-17	&	3.82E-15\,$\pm$1.91E-16		\\
473	&	L	&	0.51	&	2.22E-15\,$\pm$1.11E-16	&	5.87E-15\,$\pm$2.94E-16	*	\\
474	&	P	&	0.22	&	2.14E-16\,$\pm$2.14E-17	&	1.08E-15\,$\pm$9.16E-16		\\
478	&	T	&	0.55	&	1.59E-15\,$\pm$7.96E-17	&	1.21E-14\,$\pm$6.05E-16		\\
491	&	T	&	0.56	&	9.74E-15\,$\pm$4.87E-16	&	5.30E-14\,$\pm$2.65E-15		\\
492	&	P	&	0.99	&	9.75E-15\,$\pm$4.88E-16	&	6.42E-14\,$\pm$3.21E-15	*	\\
492	&	P	&		&	9.75E-15\,$\pm$4.88E-16	&	6.42E-14\,$\pm$3.21E-15		\\
493	&	T	&	0.84	&	2.56E-14\,$\pm$1.28E-15	&	6.65E-14\,$\pm$3.32E-15		\\
499	&	T	&	0.57	&	1.79E-15\,$\pm$8.96E-17	&	1.35E-14\,$\pm$6.76E-16		\\
506	&	P	&	0.28	&	1.42E-15\,$\pm$7.10E-17	&	3.52E-15\,$\pm$1.76E-16	*	\\
523	&	T	&	0.44	&	1.21E-15\,$\pm$6.07E-17	&	3.96E-15\,$\pm$1.98E-16		\\
524	&	L	&	0.21	&	5.07E-15\,$\pm$2.54E-16	&	1.38E-14\,$\pm$6.89E-16		\\
525	&	T	&	0.27	&	1.08E-15\,$\pm$5.40E-17	&	3.45E-15\,$\pm$1.72E-16		\\
530	&	T	&	0.35	&	1.11E-14\,$\pm$5.56E-16	&	1.15E-13\,$\pm$5.73E-15		\\
548	&	T	&	0.27	&	1.18E-14\,$\pm$5.91E-16	&	5.92E-14\,$\pm$2.96E-15		\\
553	&	P	&	0.40	&	1.99E-15\,$\pm$9.93E-17	&	2.68E-15\,$\pm$1.34E-16		\\
561	&	T	&	0.31	&	2.94E-15\,$\pm$1.47E-16	&	1.77E-14\,$\pm$8.85E-16		\\
565	&	P	&		&			&	6.28E-15\,$\pm$3.14E-16	*	\\
577	&	T	&	0.39	&	1.40E-15\,$\pm$7.00E-17	&	1.96E-14\,$\pm$9.79E-16		\\
580	&	T	&	0.30	&	9.97E-16\,$\pm$8.48E-16	&	6.29E-15\,$\pm$3.14E-16		\\
589	&	T	&	0.38	&	1.76E-15\,$\pm$8.80E-17	&	1.20E-14\,$\pm$5.98E-16		\\
590	&	T	&	0.43	&	7.90E-15\,$\pm$3.95E-16	&	2.81E-14\,$\pm$1.40E-15		\\
594	&	T	&	0.44	&	9.88E-15\,$\pm$4.94E-16	&	3.66E-14\,$\pm$1.83E-15		\\
603	&	T	&	0.71	&	1.85E-14\,$\pm$9.27E-16	&	1.62E-13\,$\pm$8.09E-15		\\
604	&	T	&	0.08	&	1.74E-15\,$\pm$8.68E-17	&	6.90E-15\,$\pm$3.45E-16		\\
607	&	T	&	0.29	&	4.44E-15\,$\pm$2.22E-16	&	1.85E-14\,$\pm$9.26E-16		\\
613	&	P	&	0.58	&	1.46E-15\,$\pm$7.31E-17	&	4.27E-15\,$\pm$2.14E-16		\\
614	&	T	&	1.02	&	1.61E-16\,$\pm$1.61E-17	&	6.84E-16\,$\pm$6.84E-17		\\
615	&	T	&	0.62	&	2.01E-14\,$\pm$1.00E-15	&	6.09E-14\,$\pm$3.05E-15		\\
618	&	P	&	0.83	&	1.53E-15\,$\pm$7.64E-17	&	1.56E-15\,$\pm$9.88E-16		\\
619	&	T	&	0.20	&	1.47E-15\,$\pm$7.36E-17	&	2.71E-15\,$\pm$1.36E-16		\\
620	&	T	&	0.62	&	1.12E-15\,$\pm$5.58E-17	&	4.20E-15\,$\pm$2.10E-16		\\
621	&	P	&	0.39	&	2.37E-15\,$\pm$1.18E-16	&	2.38E-15\,$\pm$1.19E-16		\\
622	&	T	&	0.33	&	4.75E-15\,$\pm$2.38E-16	&	3.85E-14\,$\pm$1.93E-15		\\
624	&	T	&	0.41	&	6.56E-16\,$\pm$6.56E-17	&	1.53E-15\,$\pm$1.30E-15		\\
637	&	T	&	0.45	&	2.31E-15\,$\pm$1.16E-16	&	6.46E-15\,$\pm$3.23E-16		\\
645	&	P	&	0.69	&	1.56E-13\,$\pm$7.82E-15	&	6.18E-14\,$\pm$3.09E-15		\\
666	&	T	&	0.41	&	1.41E-14\,$\pm$7.03E-16	&	6.33E-14\,$\pm$3.16E-15		\\
671	&	T	&	0.70	&	1.08E-15\,$\pm$5.41E-17	&	1.24E-14\,$\pm$6.18E-16		\\
676	&	T	&	0.20	&	1.47E-15\,$\pm$7.35E-17	&	3.50E-15\,$\pm$1.75E-16		\\
678	&	T	&	0.28	&	2.87E-15\,$\pm$1.43E-16	&	1.22E-14\,$\pm$6.11E-16		\\
%\hline\hline\noalign{\smallskip}
  \end{tabular}
\end{center}
\end{table}
\begin{table}
\scriptsize
\begin{center}
\begin{tabular}{lcccccc}
\hline\hline \noalign{\smallskip}
RP  & P  & $\textit{c}$H$\beta$ & Flux H$\beta$ & Flux \OIII  \\
Ref. &  & &   4868\AA~  &  5007\AA~  \\
    &    &      &    {\tiny(erg cm$^{-2}$ s$^{-1}$)}   &  {\tiny(erg cm$^{-2}$ s$^{-1}$)}     \\
 \hline\noalign{\smallskip}
679	&	T	&	0.18	&	7.45E-15\,$\pm$3.73E-16	&	1.37E-14\,$\pm$6.87E-16		\\
680	&	T	&	0.37	&	1.25E-14\,$\pm$6.27E-16	&	4.50E-14\,$\pm$2.25E-15	*	\\
681	&	T	&	0.23	&	5.32E-15\,$\pm$2.66E-16	&	1.67E-14\,$\pm$8.34E-16	*	\\
682	&	P	&	1.06	&	6.06E-16\,$\pm$6.06E-17	&	1.76E-15\,$\pm$8.79E-17		\\
683	&	P	&	0.31	&	5.32E-15\,$\pm$2.66E-16	&	4.13E-15\,$\pm$2.06E-16		\\
686	&	T	&	0.49	&	3.02E-15\,$\pm$1.51E-16	&	3.43E-14\,$\pm$1.72E-15		\\
687	&	T	&	0.23	&	5.45E-15\,$\pm$2.72E-16	&	3.38E-14\,$\pm$1.69E-15		\\
691	&	T	&	0.56	&	1.12E-15\,$\pm$5.62E-17	&	1.29E-14\,$\pm$6.43E-16	*	\\
692	&	T	&	0.44	&	4.74E-15\,$\pm$2.37E-16	&	3.31E-14\,$\pm$1.66E-15		\\
693	&	L	&	0.68	&	1.85E-15\,$\pm$9.24E-17	&	3.85E-15\,$\pm$1.92E-16		\\
695	&	T	&	0.03	&	4.73E-15\,$\pm$2.37E-16	&	3.16E-14\,$\pm$1.58E-15	*	\\
700	&	T	&	0.37	&	3.77E-15\,$\pm$1.88E-16	&	3.24E-15\,$\pm$1.62E-16		\\
701	&	T	&	0.19	&	1.34E-14\,$\pm$6.69E-16	&	1.64E-14\,$\pm$8.19E-16		\\
711	&	P	&	0.21	&	1.44E-15\,$\pm$7.21E-17	&	5.73E-15\,$\pm$2.87E-16		\\
714	&	T	&	0.33	&	4.97E-15\,$\pm$2.49E-16	&	3.50E-14\,$\pm$1.75E-15		\\
717	&	T	&	0.37	&	2.74E-15\,$\pm$1.37E-16	&	4.78E-15\,$\pm$2.39E-16		\\
719	&	P	&	2.13	&	1.53E-15\,$\pm$7.63E-17	&	4.23E-15\,$\pm$2.11E-16	*	\\
721	&	T	&	0.63	&	3.24E-15\,$\pm$1.62E-16	&	4.06E-14\,$\pm$2.03E-15		\\
722	&	T	&	1.15$\dag$	&	5.33E-16\,$\pm$5.33E-17	&	6.01E-15\,$\pm$3.00E-16		\\
723	&	T	&	0.36	&	1.10E-14\,$\pm$5.49E-16	&	7.01E-14\,$\pm$3.50E-15		\\
727	&	L	&	0.82$\dag$	&	3.10E-15\,$\pm$1.55E-16	&	3.82E-15\,$\pm$1.91E-16		\\
732	&	T	&	0.79	&	2.62E-15\,$\pm$1.31E-16	&	3.10E-14\,$\pm$1.55E-15		\\
733	&	T	&	0.97$\dag$	&	5.76E-16\,$\pm$5.76E-17	&	4.12E-15\,$\pm$2.06E-16		\\
735	&	T	&	0.06	&	7.87E-15\,$\pm$3.93E-16	&	2.85E-14\,$\pm$1.42E-15		\\
736	&	T	&	0.41	&	9.51E-15\,$\pm$4.75E-16	&	3.38E-14\,$\pm$1.69E-15		\\
737	&	L	&	0.96	&	3.19E-15\,$\pm$1.59E-16	&	1.12E-14\,$\pm$5.62E-16		\\
748	&	T	&	0.02	&	1.79E-14\,$\pm$8.95E-16	&	5.19E-14\,$\pm$2.59E-15		\\
753	&	P	&	0.58	&	8.46E-14\,$\pm$4.23E-15	&	1.31E-13\,$\pm$6.55E-15		\\
756	&	P	&	0.94	&	1.02E-14\,$\pm$5.09E-16	&	4.83E-14\,$\pm$2.41E-15	*	\\
757	&	T	&	0.52	&	3.00E-15\,$\pm$1.50E-16	&	1.96E-14\,$\pm$9.79E-16		\\
758	&	T	&	0.03	&	3.93E-14\,$\pm$1.96E-15	&	5.68E-14\,$\pm$2.84E-15		\\
764	&	T	&	0.50	&	5.06E-15\,$\pm$2.53E-16	&	2.77E-14\,$\pm$1.38E-15		\\
771	&	T	&	0.31	&	4.09E-15\,$\pm$2.04E-16	&	1.20E-14\,$\pm$5.99E-16		\\
774	&	P	&	0.58	&	2.13E-14\,$\pm$1.06E-15	&	2.96E-14\,$\pm$1.48E-15		\\
775	&	P	&	0.90	&	2.91E-15\,$\pm$1.45E-16	&	1.51E-14\,$\pm$7.54E-16		\\
776	&	T	&	0.66	&	7.46E-15\,$\pm$3.73E-16	&	1.24E-14\,$\pm$6.19E-16		\\
777	&	P	&	0.29	&	3.61E-14\,$\pm$1.81E-15	&	2.75E-14\,$\pm$1.38E-15		\\
%782	&	L	&		&			&				\\
789	&	T	&	0.71	&	2.27E-15\,$\pm$1.13E-16	&	3.13E-14\,$\pm$1.56E-15		\\
790	&	L	&	0.40	&	7.63E-15\,$\pm$3.81E-16	&	5.52E-14\,$\pm$2.76E-15		\\
791	&	L	&	0.37	&	3.40E-15\,$\pm$1.70E-16	&	1.16E-14\,$\pm$5.82E-16		\\
793	&	T	&	0.29	&	3.64E-15\,$\pm$1.82E-16	&	9.39E-15\,$\pm$4.69E-16		\\
803	&	T	&	0.23	&	5.53E-15\,$\pm$2.76E-16	&	5.98E-14\,$\pm$2.99E-15		\\
806	&	T	&	0.31	&	8.48E-16\,$\pm$8.48E-17	&	1.02E-15\,$\pm$8.67E-16		\\
815	&	T	&	0.26	&	9.23E-15\,$\pm$4.61E-16	&	3.02E-14\,$\pm$1.51E-15		\\
828	&	P	&	0.73	&	1.24E-14\,$\pm$6.22E-16	&	1.83E-14\,$\pm$9.16E-16		\\
850	&	T	&	0.84	&	8.85E-16\,$\pm$8.85E-17	&	3.28E-15\,$\pm$1.64E-16		\\
856	&	P	&	0.40	&	1.42E-14\,$\pm$7.09E-16	&	2.12E-14\,$\pm$1.06E-15		\\
857	&	L	&	0.23	&	4.58E-15\,$\pm$2.29E-16	&	1.40E-14\,$\pm$7.01E-16		\\
865	&	T	&	0.48	&	2.22E-15\,$\pm$1.11E-16	&	1.84E-14\,$\pm$9.21E-16		\\
875	&	T	&	0.55	&	2.20E-15\,$\pm$1.10E-16	&	1.87E-14\,$\pm$9.35E-16		\\
883	&	P	&	1.22	&	2.09E-14\,$\pm$1.04E-15	&	3.54E-14\,$\pm$1.77E-15		\\
885	&	T	&	0.42	&	1.19E-14\,$\pm$5.93E-16	&	5.86E-14\,$\pm$2.93E-15		\\
887	&	P	&	1.05	&	1.64E-15\,$\pm$8.21E-17	&	6.50E-15\,$\pm$3.25E-16		\\
889	&	T	&	0.34	&	3.33E-14\,$\pm$1.67E-15	&	1.49E-13\,$\pm$7.45E-15		\\
896	&	T	&	0.28	&	1.61E-14\,$\pm$8.03E-16	&	1.02E-13\,$\pm$5.11E-15		\\
900	&	T	&	0.50	&	2.63E-16\,$\pm$2.63E-17	&	1.36E-15\,$\pm$1.16E-15		\\
903	&	T	&	0.29	&	1.12E-14\,$\pm$5.60E-16	&	5.33E-14\,$\pm$2.67E-15		\\
907	&	T	&	0.25	&	2.35E-14\,$\pm$1.18E-15	&	7.77E-14\,$\pm$3.89E-15		\\
908	&	L	&	0.80	&	1.34E-15\,$\pm$6.70E-17	&	1.84E-15\,$\pm$9.18E-17		\\
913	&	P	&	0.29	&	1.15E-14\,$\pm$5.75E-16	&	1.03E-14\,$\pm$5.17E-16		\\
916	&	T	&	0.53	&	1.31E-14\,$\pm$6.53E-16	&	3.34E-14\,$\pm$1.67E-15		\\
917	&	T	&	0.25	&	1.80E-14\,$\pm$9.00E-16	&	9.28E-14\,$\pm$4.64E-15		\\
919	&	T	&	0.40	&	1.38E-15\,$\pm$6.89E-17	&	3.61E-15\,$\pm$1.81E-16		\\
926	&	T	&	0.45	&	8.10E-15\,$\pm$4.05E-16	&	3.19E-14\,$\pm$1.59E-15		\\
931	&	P	&	0.75	&	1.04E-15\,$\pm$5.22E-17	&	3.92E-16\,$\pm$3.92E-17		\\
932	&	P	&	0.35	&	3.14E-15\,$\pm$1.57E-16	&	5.56E-15\,$\pm$2.78E-16		\\
945	&	T	&	0.57	&	3.96E-16\,$\pm$3.96E-17	&	1.18E-15\,$\pm$1.01E-15		\\
958	&	L	&	0.24	&	5.12E-16\,$\pm$5.12E-17	&	1.00E-15\,$\pm$8.51E-16		\\
972	&	T	&	0.42	&	5.79E-15\,$\pm$2.89E-16	&	2.56E-14\,$\pm$1.28E-15		\\
977	&	P	&	0.81	&	1.09E-14\,$\pm$5.47E-16	&	1.19E-14\,$\pm$5.97E-16		\\
979	&	T	&	0.52	&	2.00E-14\,$\pm$1.00E-15	&	8.47E-14\,$\pm$4.24E-15		\\
980	&	T	&	0.38	&	2.05E-13\,$\pm$1.03E-14	&	9.10E-13\,$\pm$4.55E-14		\\
992	&	P	&	0.43	&	9.96E-14\,$\pm$4.98E-15	&	5.83E-14\,$\pm$2.91E-15		\\
1006	&	T	&	0.35	&	3.24E-15\,$\pm$1.62E-16	&	2.84E-14\,$\pm$1.42E-15		\\
1012	&	T	&	0.80	&	1.57E-14\,$\pm$7.86E-16	&	5.25E-14\,$\pm$2.62E-15		\\
1018	&	P	&	0.57	&	2.48E-15\,$\pm$1.24E-16	&	4.46E-15\,$\pm$2.23E-16		\\
1024	&	T	&	0.47	&	1.82E-15\,$\pm$9.09E-17	&	4.10E-15\,$\pm$2.05E-16		\\
1037	&	T	&	0.47	&	4.44E-15\,$\pm$2.22E-16	&	1.60E-14\,$\pm$8.01E-16		\\
%\hline\hline\noalign{\smallskip}
  \end{tabular}
\end{center}
\end{table}
\begin{table}
\scriptsize
\begin{center}
\begin{tabular}{lcccccc}
\hline\hline \noalign{\smallskip}
RP  & P  & $\textit{c}$H$\beta$ & Flux H$\beta$ & Flux \OIII  \\
Ref. &  & &   4868\AA~  &  5007\AA~  \\
    &    &      &    {\tiny(erg cm$^{-2}$ s$^{-1}$)}   &  {\tiny(erg cm$^{-2}$ s$^{-1}$)}     \\
 \hline\noalign{\smallskip}
1038	&	T	&	0.19	&	2.30E-15\,$\pm$1.15E-16	&	6.72E-15\,$\pm$3.36E-16		\\
1040	&	T	&	0.38	&	1.97E-15\,$\pm$9.86E-17	&	1.52E-14\,$\pm$7.61E-16		\\
1045	&	L	&	0.04	&	2.96E-16\,$\pm$2.96E-17	&	1.45E-15\,$\pm$1.23E-15	*	\\
1046	&	T	&	0.35	&	3.53E-16\,$\pm$3.53E-17	&	1.45E-15\,$\pm$1.24E-15		\\
1065	&	P	&	0.49	&	5.79E-15\,$\pm$2.89E-16	&	1.67E-14\,$\pm$8.37E-16	*	\\
1066	&	L	&	0.28	&	4.69E-15\,$\pm$2.35E-16	&	1.16E-14\,$\pm$5.80E-16	*	\\
1067	&	T	&	0.43	&	1.30E-15\,$\pm$6.48E-17	&	4.27E-15\,$\pm$2.14E-16		\\
1068	&	T	&	0.67	&	6.98E-16\,$\pm$6.98E-17	&	2.73E-15\,$\pm$1.37E-16		\\
1069	&	P	&	0.42	&	3.39E-14\,$\pm$1.69E-15	&	6.90E-14\,$\pm$3.45E-15	*	\\
1071	&	P	&		&	2.58E-13\,$\pm$1.29E-14	&	8.94E-15\,$\pm$4.47E-16	*	\\
1072	&	T	&	0.26	&	3.26E-15\,$\pm$1.63E-16	&	2.23E-14\,$\pm$1.12E-15		\\
1078	&	T	&	0.36	&	3.06E-15\,$\pm$1.53E-16	&	4.11E-14\,$\pm$2.06E-15		\\
1080	&	T	&	0.44	&	2.75E-15\,$\pm$1.37E-16	&	1.14E-14\,$\pm$5.70E-16		\\
1081	&	T	&	0.39	&	2.80E-15\,$\pm$1.40E-16	&	1.94E-14\,$\pm$9.72E-16		\\
1084	&	T	&	0.56	&	3.51E-16\,$\pm$3.51E-17	&	1.45E-15\,$\pm$1.23E-15		\\
1088	&	T	&	0.03	&	4.44E-17\,$\pm$6.67E-18	&	4.89E-16\,$\pm$4.89E-17		\\
1089	&	T	&	1.21	&	3.90E-15\,$\pm$1.95E-16	&	3.60E-14\,$\pm$1.80E-15	*	\\
1090	&	T	&	0.48	&	5.92E-16\,$\pm$5.92E-17	&	7.84E-15\,$\pm$3.92E-16		\\
1092	&	T	&	0.34	&	9.76E-15\,$\pm$4.88E-16	&	1.99E-14\,$\pm$9.93E-16		\\
1093	&	L	&	0.45	&	1.17E-15\,$\pm$5.85E-17	&	1.84E-15\,$\pm$9.19E-17		\\
1095	&	T	&	0.44	&	3.41E-16\,$\pm$3.41E-17	&	3.38E-15\,$\pm$1.69E-16		\\
1106	&	L	&	0.55	&	1.17E-14\,$\pm$5.83E-16	&	9.36E-14\,$\pm$4.68E-15		\\
1148	&	T	&	0.23	&	3.88E-15\,$\pm$7.63E-17	&	4.00E-15\,$\pm$1.30E-15		\\
1168	&	T	&	0.31	&	7.22E-15\,$\pm$3.61E-16	&	4.40E-14\,$\pm$2.20E-15		\\
1179	&	T	&	0.19	&	1.59E-15\,$\pm$7.94E-17	&	4.07E-15\,$\pm$2.03E-16		\\
1183	&	T	&	0.42	&	3.62E-15\,$\pm$1.81E-16	&	2.73E-14\,$\pm$1.37E-15		\\
1184	&	T	&	0.40	&	5.86E-16\,$\pm$5.86E-17	&	2.85E-15\,$\pm$1.43E-16		\\
1185	&	L	&	1.07	&	1.26E-15\,$\pm$6.30E-17	&	1.17E-14\,$\pm$5.84E-16	*	\\
1186	&	T	&	0.00	&	1.27E-15\,$\pm$6.36E-17	&	1.19E-14\,$\pm$5.93E-16		\\
1188	&	T	&	0.32	&	7.70E-15\,$\pm$3.85E-16	&	4.68E-14\,$\pm$2.34E-15		\\
1189	&	T	&	0.38	&	3.00E-15\,$\pm$1.50E-16	&	9.72E-15\,$\pm$4.86E-16		\\
1190	&	T	&	0.36	&	4.54E-15\,$\pm$2.27E-16	&	3.25E-14\,$\pm$1.63E-15		\\
1191	&	T	&	0.22	&	3.19E-14\,$\pm$7.64E-16	&	1.81E-13\,$\pm$7.44E-15		\\
1196	&	T	&	0.06	&	1.19E-14\,$\pm$2.11E-16	&	4.15E-14\,$\pm$1.08E-15		\\
1197	&	T	&	0.23	&	1.39E-14\,$\pm$1.06E-16	&	8.43E-14\,$\pm$1.04E-15		\\
1201	&	L	&	1.56	&	2.81E-16\,$\pm$8.48E-17	&	2.95E-15\,$\pm$3.17E-16	*	\\
1207	&	T	&	0.74	&	5.77E-16\,$\pm$5.77E-17	&	7.76E-15\,$\pm$3.88E-16		\\
1208	&	T	&	0.44	&	9.76E-15\,$\pm$1.62E-16	&	2.45E-14\,$\pm$5.42E-16		\\
1233	&	L	&	0.40	&	9.28E-16\,$\pm$7.89E-16	&	1.14E-14\,$\pm$5.72E-16		\\
1240	&	T	&	0.32	&	2.79E-15\,$\pm$1.39E-16	&	1.14E-14\,$\pm$5.72E-16		\\
1241	&	L	&	0.15	&	2.07E-14\,$\pm$4.35E-16	&	2.78E-14\,$\pm$6.39E-16		\\
1242	&	P	&		&	7.11E-17\,$\pm$1.07E-17	&	3.26E-16\,$\pm$3.26E-17		\\
1245	&	P	&		&	7.76E-16\,$\pm$7.76E-17	&	4.04E-15\,$\pm$2.02E-16		\\
1246	&	T	&	0.55	&	6.23E-15\,$\pm$3.11E-16	&	2.99E-14\,$\pm$1.50E-15		\\
1259	&	T	&	0.05	&	2.67E-15\,$\pm$1.33E-16	&	5.89E-15\,$\pm$2.94E-16		\\
1267	&	T	&	0.28	&	1.90E-15\,$\pm$9.50E-17	&	2.19E-14\,$\pm$1.09E-15		\\
1270	&	T	&	0.02	&	3.92E-15\,$\pm$1.96E-16	&	2.12E-14\,$\pm$1.06E-15		\\
1281	&	P	&	0.56	&	9.46E-16\,$\pm$8.04E-16	&	1.33E-15\,$\pm$1.13E-15		\\
1284	&	T	&	0.27	&	2.91E-15\,$\pm$1.45E-16	&	1.72E-14\,$\pm$8.62E-16		\\
1288	&	T	&	0.51	&	3.14E-15\,$\pm$1.57E-16	&	2.53E-14\,$\pm$1.27E-15		\\
1289	&	T	&	0.36	&	6.20E-15\,$\pm$3.10E-16	&	4.78E-14\,$\pm$2.39E-15		\\
1296	&	T	&	0.10	&	3.24E-15\,$\pm$1.62E-16	&	5.90E-15\,$\pm$2.95E-16		\\
1300	&	P	&	0.88	&	1.22E-14\,$\pm$6.09E-16	&	1.26E-13\,$\pm$6.32E-15		\\
1303	&	T	&	0.23	&	1.83E-15\,$\pm$9.13E-17	&	1.23E-14\,$\pm$6.13E-16		\\
1304	&	T	&	0.33	&	3.05E-15\,$\pm$1.52E-16	&	3.71E-14\,$\pm$1.85E-15		\\
1308	&	T	&	0.45	&	3.01E-15\,$\pm$1.50E-16	&	9.06E-15\,$\pm$4.53E-16		\\
1309	&	T	&	0.29	&	1.50E-15\,$\pm$7.50E-17	&	9.82E-15\,$\pm$4.91E-16		\\
1310	&	T	&	0.62	&	8.65E-16\,$\pm$8.65E-17	&	5.44E-15\,$\pm$2.72E-16		\\
1314	&	T	&	0.28	&	9.99E-16\,$\pm$8.49E-16	&	5.69E-15\,$\pm$2.84E-16		\\
1315	&	T	&	0.54	&	1.03E-15\,$\pm$5.14E-17	&	5.76E-15\,$\pm$2.88E-16		\\
1317	&	T	&	0.72	&	3.41E-16\,$\pm$3.41E-17	&	1.77E-15\,$\pm$8.86E-17		\\
1323	&	T	&	0.38	&	1.02E-15\,$\pm$5.10E-17	&	2.02E-15\,$\pm$1.01E-16		\\
1324	&	T	&	0.48	&	2.12E-15\,$\pm$1.06E-16	&	1.45E-14\,$\pm$7.25E-16		\\
1336	&	T	&	0.51	&	4.31E-15\,$\pm$2.15E-16	&	1.31E-14\,$\pm$6.55E-16		\\
1337	&	T	&	0.54	&	6.86E-15\,$\pm$3.43E-16	&	2.08E-14\,$\pm$1.04E-15		\\
1338	&	T	&	0.32	&	1.50E-14\,$\pm$7.49E-16	&	4.24E-14\,$\pm$2.12E-15		\\
1341	&	P	&	0.40	&	1.27E-15\,$\pm$6.35E-17	&	3.00E-15\,$\pm$1.50E-16		\\
1345	&	T	&	0.40	&	9.79E-15\,$\pm$4.89E-16	&	2.75E-14\,$\pm$1.37E-15		\\
1352	&	T	&	0.57	&	1.96E-14\,$\pm$9.82E-16	&	7.06E-14\,$\pm$3.53E-15		\\
1353	&	T	&	0.26	&	2.18E-15\,$\pm$1.09E-16	&	7.48E-15\,$\pm$3.74E-16		\\
1354	&	T	&	0.58	&	7.24E-16\,$\pm$7.24E-17	&	3.02E-15\,$\pm$1.51E-16		\\
1357	&	T	&	0.73	&	1.66E-15\,$\pm$8.29E-17	&	1.81E-14\,$\pm$9.07E-16		\\
1358	&	T	&	0.43	&	5.09E-15\,$\pm$2.54E-16	&	4.25E-14\,$\pm$2.12E-15		\\
1371	&	T	&	0.58	&	1.10E-14\,$\pm$5.50E-16	&	4.66E-14\,$\pm$2.33E-15		\\
1375	&	P	&	0.42	&	1.14E-14\,$\pm$5.68E-16	&	2.11E-14\,$\pm$1.05E-15		\\
1376	&	T	&	0.47	&	4.93E-15\,$\pm$2.47E-16	&	2.93E-14\,$\pm$1.46E-15		\\
1378	&	T	&	0.70	&	4.84E-15\,$\pm$2.42E-16	&	1.30E-14\,$\pm$6.52E-16		\\
1387	&	T	&	0.41	&	7.93E-15\,$\pm$3.97E-16	&	6.31E-14\,$\pm$3.16E-15		\\
1397	&	P	&	0.38	&	1.41E-13\,$\pm$7.07E-15	&	1.82E-13\,$\pm$9.08E-15		\\
%\hline\hline\noalign{\smallskip}
  \end{tabular}
\end{center}
\end{table}
\begin{table}
\scriptsize
\begin{center}
\begin{tabular}{lcccccc}
\hline\hline \noalign{\smallskip}
RP  & P  & $\textit{c}$H$\beta$ & Flux H$\beta$ & Flux \OIII  \\
Ref. &  & &   4868\AA~  &  5007\AA~  \\
    &    &      &    {\tiny(erg cm$^{-2}$ s$^{-1}$)}   &  {\tiny(erg cm$^{-2}$ s$^{-1}$)}     \\
 \hline\noalign{\smallskip}
1415	&	T	&	0.47	&	1.44E-15\,$\pm$7.20E-17	&	8.34E-15\,$\pm$4.17E-16		\\
1416	&	T	&	0.33	&	2.24E-15\,$\pm$1.12E-16	&	1.56E-14\,$\pm$7.78E-16		\\
1418	&	T	&	0.26	&	1.21E-15\,$\pm$6.07E-17	&	5.74E-15\,$\pm$2.87E-16		\\
1426	&	T	&	0.36	&	4.21E-15\,$\pm$2.10E-16	&	1.60E-14\,$\pm$8.01E-16		\\
1429	&	T	&	0.26	&	3.38E-15\,$\pm$1.69E-16	&	2.65E-14\,$\pm$1.33E-15		\\
1431	&	L	&	0.46	&	1.08E-15\,$\pm$5.42E-17	&	3.19E-15\,$\pm$1.60E-16		\\
1432	&	T	&	0.30	&	9.18E-16\,$\pm$7.81E-16	&	2.24E-15\,$\pm$1.12E-16		\\
1438	&	T	&	0.66	&	3.56E-15\,$\pm$1.78E-16	&	3.60E-14\,$\pm$1.80E-15		\\
1440	&	T	&	0.57	&	1.28E-15\,$\pm$6.40E-17	&	1.48E-14\,$\pm$7.42E-16		\\
1443	&	T	&	0.39	&	5.36E-15\,$\pm$2.68E-16	&	1.13E-14\,$\pm$5.63E-16	*	\\
1444	&	T	&	0.43	&	6.36E-15\,$\pm$3.18E-16	&	4.37E-14\,$\pm$2.18E-15		\\
1446	&	T	&	0.70	&	2.55E-15\,$\pm$1.28E-16	&	2.77E-14\,$\pm$1.39E-15		\\
1447	&	T	&	0.01	&	1.58E-15\,$\pm$7.90E-17	&	5.86E-15\,$\pm$2.93E-16		\\
1456	&	T	&	0.28	&	3.99E-15\,$\pm$1.99E-16	&	2.44E-14\,$\pm$1.22E-15		\\
1462	&	T	&	0.44	&	2.26E-15\,$\pm$1.13E-16	&	9.24E-15\,$\pm$4.62E-16		\\
1463	&	T	&	0.23	&	3.88E-15\,$\pm$1.94E-16	&	1.27E-14\,$\pm$6.33E-16		\\
1465	&	T	&	0.80$\dag$	&	8.08E-16\,$\pm$8.08E-17	&	4.46E-15\,$\pm$2.23E-16		\\
1466	&	P	&	0.50	&	2.76E-14\,$\pm$1.38E-15	&	3.46E-14\,$\pm$1.73E-15		\\
1467	&	T	&	0.74	&	6.31E-16\,$\pm$6.31E-17	&	2.17E-15\,$\pm$1.09E-16		\\
1474	&	T	&	0.01	&	6.30E-16\,$\pm$6.30E-17	&	5.52E-15\,$\pm$2.76E-16		\\
1488	&	T	&	0.39	&	7.67E-16\,$\pm$7.67E-17	&	5.34E-15\,$\pm$2.67E-16		\\
1502	&	T	&	0.37	&	1.81E-14\,$\pm$9.03E-16	&	9.16E-15\,$\pm$1.58E-16		\\
1508	&	L	&	0.10	&	2.95E-15\,$\pm$1.47E-16	&	2.32E-14\,$\pm$1.16E-15		\\
1519	&	T	&	0.71	&	1.60E-16\,$\pm$1.60E-17	&	1.31E-15\,$\pm$1.11E-15		\\
1523	&	T	&	0.44	&	3.89E-15\,$\pm$1.94E-16	&	2.14E-14\,$\pm$1.07E-15		\\
1528	&	T	&	0.66	&	8.42E-15\,$\pm$4.21E-16	&	8.23E-14\,$\pm$4.12E-15		\\
1532	&	T	&	0.48	&	1.25E-14\,$\pm$6.27E-16	&	8.01E-14\,$\pm$4.00E-15		\\
%1541	&	P	&		&			&				\\
1550	&	T	&	0.26	&	2.23E-14\,$\pm$1.11E-15	&	8.59E-14\,$\pm$4.29E-15		\\
1559	&	P	&	0.32	&	1.11E-15\,$\pm$5.57E-17	&	2.10E-15\,$\pm$1.05E-16		\\
1579	&	T	&	0.37	&	2.39E-15\,$\pm$1.20E-16	&	1.14E-14\,$\pm$5.69E-16		\\
1580	&	T	&	0.88	&	3.49E-15\,$\pm$1.75E-16	&	1.04E-14\,$\pm$5.22E-16		\\
1584	&	T	&	0.86$\dag$	&	1.63E-15\,$\pm$8.16E-17	&	8.21E-15\,$\pm$4.11E-16		\\
1587	&	T	&	0.34	&	7.41E-15\,$\pm$3.70E-16	&	7.02E-14\,$\pm$3.51E-15		\\
1595	&	T	&	0.12	&	6.71E-15\,$\pm$3.36E-16	&	8.20E-15\,$\pm$4.10E-16		\\
1601	&	P	&	0.86	&	8.78E-15\,$\pm$4.39E-16	&	1.18E-14\,$\pm$5.92E-16		\\
1615	&	P	&	0.27	&	8.66E-15\,$\pm$4.33E-16	&	1.31E-14\,$\pm$6.57E-16		\\
1624	&	T	&	0.28	&	5.98E-15\,$\pm$8.53E-17	&	1.53E-14\,$\pm$2.91E-16		\\
1631	&	T	&	0.06	&	3.37E-15\,$\pm$1.69E-16	&	7.62E-15\,$\pm$3.81E-16		\\
1634	&	P	&	0.91	&	4.96E-15\,$\pm$6.67E-17	&	1.50E-14\,$\pm$2.84E-16		\\
1636	&	T	&	0.33	&	5.62E-15\,$\pm$2.81E-16	&	4.41E-14\,$\pm$2.20E-15		\\
1638	&	T	&	0.19	&	2.14E-15\,$\pm$1.07E-16	&	1.12E-14\,$\pm$5.58E-16	*	\\
1649	&	P	&	0.43	&	6.95E-16\,$\pm$6.95E-17	&	1.26E-15\,$\pm$1.07E-15		\\
1659	&	T	&	0.31	&	9.27E-16\,$\pm$7.88E-16	&	5.78E-15\,$\pm$2.89E-16		\\
1660	&	T	&	0.30	&	4.39E-15\,$\pm$2.19E-16	&	3.24E-14\,$\pm$1.62E-15		\\
1664	&	L	&	0.46	&	9.42E-16\,$\pm$8.01E-16	&	2.49E-15\,$\pm$1.25E-16		\\
1676	&	T	&	0.33	&	1.24E-14\,$\pm$6.20E-16	&	5.08E-14\,$\pm$2.54E-15		\\
1694	&	T	&	0.44	&	1.74E-15\,$\pm$8.72E-17	&	1.44E-14\,$\pm$7.22E-16		\\
1695	&	T	&	0.36	&	5.18E-15\,$\pm$2.59E-16	&	4.09E-14\,$\pm$2.04E-15		\\
1696	&	T	&	0.41	&	1.86E-15\,$\pm$9.30E-17	&	1.46E-14\,$\pm$7.29E-16		\\
1697	&	T	&	0.50	&	7.12E-16\,$\pm$7.12E-17	&	3.57E-15\,$\pm$1.78E-16		\\
1705	&	T	&	0.37	&	1.50E-15\,$\pm$7.48E-17	&	5.24E-15\,$\pm$2.62E-16		\\
1709	&	T	&	0.40	&	1.93E-15\,$\pm$9.65E-17	&	1.92E-15\,$\pm$9.62E-17		\\
1712	&	P	&	0.50	&	1.04E-15\,$\pm$5.21E-17	&	1.56E-15\,$\pm$1.33E-15		\\
1714	&	T	&	0.46	&	5.19E-15\,$\pm$2.60E-16	&	3.25E-14\,$\pm$1.63E-15		\\
1718	&	T	&	0.09	&	5.87E-16\,$\pm$5.87E-17	&	9.08E-16\,$\pm$9.08E-17		\\
1719	&	T	&	0.52	&	8.48E-16\,$\pm$8.48E-17	&	1.44E-14\,$\pm$7.20E-16		\\
1720	&	T	&	0.48	&	2.95E-15\,$\pm$1.47E-16	&	2.07E-14\,$\pm$1.04E-15		\\
1721	&	T	&	0.56	&	5.11E-16\,$\pm$5.11E-17	&	3.47E-15\,$\pm$1.73E-16		\\
1725	&	T	&	0.31	&	3.03E-15\,$\pm$1.51E-16	&	2.05E-14\,$\pm$1.03E-15		\\
1726	&	T	&	0.29	&	1.34E-15\,$\pm$6.69E-17	&	1.35E-14\,$\pm$6.74E-16		\\
1727	&	T	&	0.23	&	5.73E-15\,$\pm$2.86E-16	&	1.91E-14\,$\pm$9.53E-16		\\
1731	&	T	&	0.30	&	1.85E-15\,$\pm$9.25E-17	&	8.28E-15\,$\pm$4.14E-16		\\
1739	&	T	&	0.47	&	1.30E-15\,$\pm$6.50E-17	&	6.85E-15\,$\pm$3.42E-16		\\
1740	&	T	&	0.27	&	5.66E-15\,$\pm$2.83E-16	&	1.80E-14\,$\pm$9.00E-16		\\
1742	&	T	&	0.21	&	6.06E-16\,$\pm$6.06E-17	&	1.95E-15\,$\pm$9.77E-17		\\
1743	&	T	&	0.36	&	4.58E-15\,$\pm$2.29E-16	&	2.99E-14\,$\pm$1.49E-15		\\
1748	&	T	&	0.30	&	4.04E-15\,$\pm$2.02E-16	&	2.70E-14\,$\pm$1.35E-15		\\
1749	&	T	&	0.17	&	9.26E-16\,$\pm$7.87E-16	&	7.58E-16\,$\pm$7.58E-17		\\
1753	&	T	&	0.33	&	8.82E-16\,$\pm$8.82E-17	&	2.04E-14\,$\pm$1.02E-15		\\
1756	&	T	&	0.37	&	7.37E-16\,$\pm$7.37E-17	&	3.08E-15\,$\pm$1.54E-16		\\
1758	&	T	&	0.33	&	5.66E-15\,$\pm$2.83E-16	&	2.39E-14\,$\pm$1.20E-15		\\
1759	&	T	&	0.34	&	4.31E-15\,$\pm$2.16E-16	&	1.45E-14\,$\pm$7.26E-16		\\
1771	&	T	&	0.58	&	3.74E-15\,$\pm$1.87E-16	&	1.26E-14\,$\pm$6.31E-16		\\
1773	&	T	&	0.54	&	5.88E-15\,$\pm$2.94E-16	&	1.46E-14\,$\pm$7.31E-16		\\
1791	&	T	&	0.49	&	1.06E-14\,$\pm$5.29E-16	&	2.09E-14\,$\pm$1.04E-15		\\
1796	&	P	&	0.84	&	3.16E-15\,$\pm$1.58E-16	&	2.29E-15\,$\pm$1.14E-16		\\
1803	&	T	&	0.31	&	1.52E-15\,$\pm$7.62E-17	&	5.37E-15\,$\pm$2.69E-16		\\
1805	&	T	&	0.43	&	1.88E-15\,$\pm$9.38E-17	&	7.51E-15\,$\pm$3.76E-16		\\
%\hline\hline\noalign{\smallskip}
  \end{tabular}
\end{center}
\end{table}
\begin{table}
\scriptsize
\begin{center}
\begin{tabular}{lcccccc}
\hline\hline \noalign{\smallskip}
RP  & P  & $\textit{c}$H$\beta$ & Flux H$\beta$ & Flux \OIII  \\
Ref. &  & &   4868\AA~  &  5007\AA~  \\
    &    &      &    {\tiny(erg cm$^{-2}$ s$^{-1}$)}   &  {\tiny(erg cm$^{-2}$ s$^{-1}$)}     \\
 \hline\noalign{\smallskip}
1807	&	T	&	0.21	&	1.67E-15\,$\pm$8.35E-17	&	4.68E-15\,$\pm$2.34E-16		\\
1808	&	L	&	0.11	&	3.39E-15\,$\pm$1.69E-16	&	8.74E-15\,$\pm$4.37E-16	*	\\
1811	&	P	&	0.99	&	8.59E-16\,$\pm$8.59E-17	&	1.59E-15\,$\pm$1.35E-15		\\
1813	&	T	&	0.19	&	9.21E-16\,$\pm$7.83E-16	&	7.10E-15\,$\pm$3.55E-16		\\
1819	&	T	&	0.60	&	2.02E-15\,$\pm$1.01E-16	&	1.86E-14\,$\pm$9.28E-16		\\
1822	&	L	&	0.43	&	9.77E-15\,$\pm$4.88E-16	&	2.57E-14\,$\pm$1.28E-15		\\
1823	&	T	&	0.42	&	8.40E-16\,$\pm$8.40E-17	&	5.68E-15\,$\pm$2.84E-16		\\
1835	&	T	&	0.27	&	5.70E-15\,$\pm$2.85E-16	&	3.44E-14\,$\pm$1.72E-15		\\
1848	&	T	&	0.41	&	1.57E-14\,$\pm$7.87E-16	&	4.59E-14\,$\pm$2.29E-15		\\
1853	&	T	&	0.51	&	3.71E-15\,$\pm$1.86E-16	&	1.46E-14\,$\pm$7.32E-16		\\
1862	&	L	&	0.62	&	7.34E-16\,$\pm$7.34E-17	&	3.52E-15\,$\pm$1.76E-16		\\
1864	&	T	&	0.31	&	1.35E-14\,$\pm$6.74E-16	&	2.24E-14\,$\pm$1.12E-15		\\
1868	&	T	&	0.38	&	4.20E-15\,$\pm$2.10E-16	&	5.91E-14\,$\pm$2.96E-15		\\
1872	&	T	&	0.42	&	1.25E-14\,$\pm$6.25E-16	&	1.01E-13\,$\pm$5.03E-15	*	\\
1876	&	T	&	0.25	&	2.22E-15\,$\pm$1.11E-16	&	4.20E-15\,$\pm$2.10E-16		\\
1878	&	T	&	0.51	&	6.07E-15\,$\pm$3.04E-16	&	3.01E-14\,$\pm$1.50E-15		\\
1886	&	T	&	0.07	&	1.68E-15\,$\pm$8.42E-17	&	1.25E-14\,$\pm$6.26E-16		\\
1900	&	T	&	0.28	&	8.13E-15\,$\pm$4.06E-16	&	1.80E-14\,$\pm$8.98E-16		\\
1904	&	T	&	0.11	&	5.60E-15\,$\pm$2.80E-16	&	5.32E-14\,$\pm$2.66E-15		\\
1921	&	T	&	0.48	&	4.69E-15\,$\pm$2.35E-16	&	1.75E-14\,$\pm$8.74E-16		\\
1922	&	T	&	0.42	&	2.38E-15\,$\pm$1.19E-16	&	1.92E-14\,$\pm$9.60E-16		\\
1930	&	P	&	0.09	&	3.26E-16\,$\pm$3.26E-17	&	1.10E-15\,$\pm$9.36E-16		\\
1934	&	T	&	0.31	&	5.20E-15\,$\pm$2.60E-16	&	1.53E-14\,$\pm$7.66E-16		\\
1938	&	T	&	0.18	&	1.10E-14\,$\pm$5.52E-16	&	5.18E-14\,$\pm$2.59E-15		\\
%1951	&	L	&		&	\,$\pm$0.00E+00	&	\,$\pm$0.00E+00		\\
1954	&	T	&	1.17	&	8.78E-16\,$\pm$8.78E-17	&	4.82E-15\,$\pm$2.41E-16		\\
1955	&	T	&	0.46	&	5.08E-15\,$\pm$2.54E-16	&	2.04E-14\,$\pm$1.02E-15		\\
1956	&	T	&	0.78	&	2.14E-15\,$\pm$1.07E-16	&	4.44E-14\,$\pm$2.22E-15		\\
1957	&	T	&	0.32	&	4.82E-15\,$\pm$2.41E-16	&	1.34E-14\,$\pm$6.68E-16		\\
1958	&	P	&	0.73	&	3.72E-16\,$\pm$3.72E-17	&	1.17E-15\,$\pm$9.93E-16		\\
1962	&	P	&	0.06	&	4.96E-14\,$\pm$2.48E-15	&	2.00E-14\,$\pm$1.00E-15		\\
2193	&	P	&	0.34	&	1.32E-14\,$\pm$6.59E-16	&	1.42E-14\,$\pm$7.11E-16		\\
\hline\hline\noalign{\smallskip}
  \end{tabular}
\end{center}
\end{table}

\bsp

\label{lastpage}

\end{document}